\documentclass[fleqn,usenatbib]{mnras}

\usepackage{newtxtext,newtxmath}
\usepackage[T1]{fontenc}

\DeclareRobustCommand{\VAN}[3]{#2}
\let\VANthebibliography\thebibliography
\def\thebibliography{\DeclareRobustCommand{\VAN}[3]{##3}\VANthebibliography}

\usepackage{graphicx}	% Including figure files
\usepackage{amsmath}	% Advanced maths commands
\usepackage[dvipsnames]{xcolor}
\usepackage[version=4]{mhchem}
\usepackage{graphicx}
\usepackage{longtable}

\usepackage{orcidlink}

\newcommand{\ffuv}{$F_\mathrm{FUV}$}

\newcommand{\rhofuv}{$\rho_\text{grid}$}
\newcommand{\rhopartial}{$\rho_\text{fixed}$}

\title[External photoevaporation diagnostics]{Observational diagnostics of external photoevaporation at millimetre wavelengths}

\author[L. Keyte et al.]{
Luke Keyte\,\textsuperscript{\orcidlink{0000-0001-5849-577X}},$^{1}$\thanks{E-mail: l.keyte@qmul.ac.uk}
Thomas J. Haworth\,\textsuperscript{\orcidlink{0000-0002-9593-7618}},$^{1}$
Tyger Peake\,\textsuperscript{\orcidlink{0009-0007-4567-7735}},$^{1}$
Álvaro Ribas\,\textsuperscript{\orcidlink{0000-0003-3133-3580}}$^{1}$
\\
$^{1}$Astronomy Unit, School of Physics and Astronomy, Queen Mary University of London, London E1 4NS, UK\\
}
\date{Accepted XXX. Received YYY; in original form ZZZ}

\pubyear{\the\year{}}

\begin{document}
\label{firstpage}
\pagerange{\pageref{firstpage}--\pageref{lastpage}}
\maketitle

%%%%%%%%%%%%
% ABSTRACT %
%%%%%%%%%%%%
\begin{abstract}
External photoevaporation is increasingly recognised as an important influence on protoplanetary disk evolution. However, there are no established molecular line diagnostics that can reliably distinguish external irradiation from variations in intrinsic stellar and disk properties. We use a large grid of thermochemical models to predict the fluxes of 48 transitions from eight species observable with ALMA, spanning a wide range of system properties and external FUV fields of $10^2$--$10^5\,$G$_0$. At fixed stellar and disk properties, all transitions generally brighten with increasing FUV field strength. HCO$^+$ and CO fluxes increase monotonically in virtually all disks, whereas lower-excitation lines, particularly those of C$_2$H and N$_2$H$^+$, respond weakly and non-monotonically. However, variations in stellar mass and disk size introduce substantial degeneracies, limiting the usefulness of individual line fluxes as tracers of external irradiation. We therefore assess all pairwise line flux ratios, which can significantly reduce these degeneracies, evaluating their correlation with the external field, the magnitude of their change across the FUV range, and predicted ALMA detectability. We identify 135 promising ratios, including several combining strong FUV sensitivity and high detectability with efficient single-tuning ALMA observations. The strongest diagnostics either contrast high- and low-excitation lines of the same or co-spatial species (e.g. CO $J$=8-7/$J$=1-0), probing different regions of the photoevaporative wind, or pair an FUV-sensitive wind tracer with comparatively insensitive emission arising closer to the disk (e.g. HCO$^+$ $J$=3-2/C$_2$H $N$=3-2). Our results provide a framework for identifying externally photoevaporating systems from spatially unresolved emission and constraining their incident FUV field strengths.
\end{abstract}

\begin{keywords}
protoplanetary discs -- planets and satellites: formation -- astrochemistry -- accretion, accretion discs
\end{keywords}

%%%%%%%%%%%%%%%%%%%%%%%%%%%%%%%%%%%%%%%%%%%%%%%%%%
%%%%%%%%%%%%%%%%% BODY OF PAPER %%%%%%%%%%%%%%%%%%
%%%%%%%%%%%%%%%%%%%%%%%%%%%%%%%%%%%%%%%%%%%%%%%%%%

%%%%%%%%%%%%%%%%%%%
% 1. INTRODUCTION %
%%%%%%%%%%%%%%%%%%%

\section{Introduction}

Nearly everything we know about the chemistry of planet-forming disks has been learned from systems evolving in isolation. Observational studies have focused on individual benchmark sources such as TW Hya and disk populations in low-mass star-forming regions such as Taurus and Lupus, which are nearby, relatively bright, and generally show little evidence of environmental influence \citep[e.g.][]{qi_2013, ansdell_2016, bergin_2016, guilloteau_2016, isella_2016, miotello_2017, oberg_2021_maps_1}. Such quiescent environments, however, are the exception. Most stars instead form in massive clusters containing OB stars \citep{lada_lada_2003}, where disks are exposed to far-ultraviolet (FUV) radiation fields thousands of times stronger than the interstellar background \citep[e.g.][]{fatuzzo_adams_2008, winter_haworth_2022, planet_formation_collab_2025}.  If that irradiation leaves an imprint on disk chemistry, and by extension on the planets that disks produce, then the sample on which our understanding is built is poorly placed to reveal it.

The physical consequences of external irradiation are now well established. Disks in strongly irradiated environments are smaller \citep{eisner_2018, boyden_eisner_2020, portilla_revelo_2025}, less massive \citep{mann_2014, ansdell_2017, van_terwisga_2020, van_terwisga_2023, mauco_2025}, and shorter lived than their isolated counterparts \citep{bally_1998b, haworth_2018_fried, concha-ramirez2019, nicholson_2019, winter_2020, coleman_haworth_2022}, with many displaying the classical `proplyd' cometary morphology indicative of ongoing mass loss \citep{odell_wen_hu_1993, odell_wen_1994, bally_2000}. 

How external irradiation shapes disk chemistry is far less clear, and the relatively small number of observations to date present a mixed picture. Several irradiated systems appear chemically similar to isolated disks, both in the inner disk probed by JWST \citep{ramirez_tannus_2023} and in the outer disk probed by ALMA \citep{diaz_berrios_2024}. Others carry signatures that are difficult to explain without environmental effects. In the outer disk, these include a lack of CO depletion, in contrast to the significant depletion routinely inferred in isolated systems \citep{boyden_eisner_2023}, and bright, extended [CI] emission tracing warm atomic gas \citep{goicoechea_2024}. JWST has additionally revealed photochemically produced H$_3^+$, CH, and CH$_3^+$ in the irradiated disk surface \citep{berne_2023, schroetter_2025}, and a CO$_2$-enriched inner disk around a Herbig star \citep{frediani_2025}. These results suggest that external irradiation can alter disk chemistry, although the strength and spatial extent of its influence may vary between systems.

Chemical models generally predict that the radiation environment leaves a measurable imprint on the composition of the outer disk \citep[][]{walsh_2013, keyte_haworth_2025, keyte_ran_2025, gross_cleeves_2025, calahan_2025}, although the effect scales with field strength and may only become clearly detectable in the most extreme environments. The inner disk is expected to be less sensitive, since its radiation field is dominated by the host star \citep{ndugu_2024}, and models of the terrestrial planet-forming region predict mid-infrared tracers that are largely insensitive to the external field \citep{frediani_2026}. Reconciling these observational and theoretical results therefore requires robust diagnostics capable of identifying which systems are affected, where the effects occur, and how strongly their chemistry has been altered.

A key difficulty is that molecular line fluxes depend not only on the external FUV field, but also on a wide range of intrinsic stellar and disk properties. Stellar mass, disk mass and size, dust-to-gas ratio, dust opacity, and turbulence all shape the disk density and temperature structure and regulate the attenuation of radiation, and can therefore influence line emission as strongly as, or more strongly than, external irradiation. The influence of individual disk and stellar properties on molecular abundances and line fluxes in quiescent systems has been explored in great detail \citep[e.g.][]{aresu_2011, vasyunin_2011, cleeves_2013, walsh_2015, woitke_2016, kamp_2017, agundez_2018, wakelam_2019, gavino_2021}, where it is well established that no single line maps cleanly to any individual physical parameter. External irradiation introduces an additional source of degeneracy. The problem is further complicated by the fact that irradiated systems typically lie at much further distances than the bright, well-studied isolated disks, so their emission is only marginally resolved, if at all. A bright line measured in an irradiated disk might therefore be attributed to a strong external FUV field, or could equally arise from the disk being larger than assumed and so presenting a greater emitting area. Comparing individual line fluxes between an irradiated and isolated system cannot, on its own, separate these possibilities.

Line flux ratios offer a way to deal with these degeneracies. If two transitions depend similarly on a given stellar/disk parameter, then that shared dependence partially cancels when one is divided by the other, leaving the response to the external field. Such diagnostics have long been used in studies of photodissociation regions (PDRs), where line ratios constrain the incident FUV field and gas density \citep[e.g.][]{fuente_1993, kaufman_1999, boger_sternberg_2005, wolfire_2022}, and in isolated disks, where ratios such as CN/HCN have been proposed as probes of photochemistry \citep[e.g.][]{chapillon_2012, guzman_2015, visser_2018, bergner_2021_maps_xi}. For externally irradiated systems, comparable diagnostics have been developed only at optical wavelengths, which predominantly trace the ionised flow \citep{peake_2025}. No equivalent framework exists at millimetre wavelengths, despite this being the regime in which the bulk molecular content of disks is typically observed \citep[e.g.][]{ansdell_2016, long_2017, long_2019, oberg_2021_maps_1, booth_2024a, zhang_2025_agepro_i}.

Constructing such a framework requires models that capture the full structure of the irradiated system, not just the disk. Above a threshold external field strength, FUV photons drive a thermal wind from the disk surface, so that the system must be treated as a coupled disk--wind structure rather than a disk in isolation \citep[e.g.][]{johnstone_1998, adams_2004, facchini_2016}. The wind has its own density, temperature, and opacity structure, with dust properties that differ from the disk, since only small grains are entrained \citep{facchini_2016, owen_altaf_2021, paine_2025}. Depending on system properties, emission from the wind can contribute a significant fraction of the total line flux, and the wind can also alter the composition of the disk itself through the way in which it reprocesses the external FUV field inward \citep{keyte_haworth_2025}. A model that does not include the wind may therefore reproduce an observed flux, but wrongly attribute it to the disk, and infer the wrong composition as a result. Any diagnostic intended for irradiated systems must therefore be calibrated against models which include both the disk and wind components.

Until recently, this was not practical. Radiation-hydrodynamic simulations resolve the disk-wind structure accurately, but cost weeks-to-months to run per model, which restricts them to a handful of representative cases \citep{haworth_clarke_2019, ballabio_2023}, ruling out the broad parameter survey required to identify robust diagnostics. To overcome this limitation, \citet{keyte_haworth_2026} developed the \textsc{puffin} framework, which parametrically approximates disk-wind density structures at much lower computational cost. Coupling \textsc{puffin} to established thermochemical codes makes grids of thousands of irradiated disk models feasible, and allows a systematic search of the parameter space for diagnostics.

We carry out such a search here, computing a large grid of models with \textsc{dali} \citep{Bruderer2012, Bruderer2013} that span a range of stellar masses, disk sizes and masses, and external FUV field strengths. From these we synthesise integrated fluxes for a broad set of transitions across eight species accessible to ALMA. We then quantify the response of each transition to the external FUV field, separating this from the effects of the other model parameters, and rank all pairwise flux ratios by their FUV sensitivity, dynamic range, and detectability. The resulting diagnostics are designed to identify systems in which the photoevaporative wind contributes substantially to the observed flux, and to constrain the strength of the incident field.

This paper is organised as follows. Section \ref{sec:modelling} describes the model grid and synthetic observations. Section \ref{sec:results} presents the response of the individual line fluxes to the external FUV field and the impact of degeneracies, and finally assesses the robustness of the line flux ratios. Section \ref{sec:discussion} examines the chemical origins of the observed trends and provides guidance on applying the diagnostics with ALMA. We summarise our conclusions in Section \ref{sec:conclusions}.

\begin{table*}
    \centering
    \caption{Parameter grid used to construct the thermochemical model grid. All combinations of the listed values are computed, yielding 1470 models in total.}
    \label{table:model_grid}
    \begin{tabular}{lll}
        \hline
        \hline
        Parameter & Description & Values \\
        \hline
        $M_*$ & Stellar mass [$M_\odot$] & 0.3, 0.6, 1.0, 2.0, 3.0 \\
        $r_\mathrm{d}$ & Disk outer radius [au] & 20, 40, 60, 80, 100, 150 \\
        $\Sigma_\mathrm{1au}$ & Surface density at 1\,au [g\,cm$^{-2}$] & $10^1$, $5 \times 10^1$, $10^2$, $5 \times 10^2$,  $10^3$, $5 \times 10^3$, $10^4$ \\
        \ffuv & External FUV field [G$_0$] & $10^2$, $5 \times 10^2$, $10^3$, $5 \times 10^3$, $10^4$, $5 \times 10^4$, $10^5$ \\
        \hline
    \end{tabular}
\end{table*}

%%%%%%%%%%%%%%%%
% 2. MODELLING %
%%%%%%%%%%%%%%%%
\section{Modelling}
\label{sec:modelling}

We construct a grid of thermochemical models spanning a broad range of stellar masses, disk properties, and external FUV field strengths (Table \ref{table:model_grid}). Each model comprises a two-dimensional density structure generated with the \textsc{puffin} parametric framework \citep{keyte_haworth_2026}, which serves as input to thermochemical calculations performed with the \textsc{dali} code \citep{Bruderer2012, Bruderer2013}. The physical setup, stellar parameters, dust treatment, and chemical network follow the approach of \citet{keyte_haworth_2026}, which we summarise below. We refer the reader to that work for a complete description of the modelling procedure.

\subsection{Physical structure}
\label{sec:physical_structure}

For each combination of input parameters, we generate a two-dimensional gas density structure using the \textsc{puffin} Python package. \textsc{puffin} is a parametric framework for computing the density structures of protoplanetary disks undergoing external photoevaporation. Given a set of physical inputs and a mass-loss rate, it produces density structures that approximate the results of full radiation-hydrodynamic simulations, but can be evaluated in minutes rather than the weeks-to-months typically required. This efficiency makes it well-suited to the systematic exploration of large parameter spaces.

The density structure comprises three components: a disk in vertical hydrostatic equilibrium (accounting for external FUV heating of the disk surface), a spherically diverging photoevaporative wind, consistent with the outflow geometry adopted in one-dimensional models of external photoevaporation \citep[][]{facchini_2016, haworth_2018_fried}, and a smooth transition region that bridges the two, approximating the disk-wind structure found in two-dimensional radiation-hydrodynamic simulations \citep{haworth_clarke_2019}. The wind density profile depends on the mass-loss rate, which is interpolated from the \textsc{fried} grid of hydrodynamic simulations \citep{haworth_2018_fried, haworth_2023_fried2} and multiplied by a factor of two to account for the systematic underprediction of mass-loss rates by one-dimensional models relative to two-dimensional simulations found by \citet{haworth_clarke_2019}. The density structure therefore depends on $F_{\rm FUV}$ through both the mass-loss rate and the heating of the disk surface, and is computed once with \textsc{puffin} and held fixed during the thermochemical calculation.

The dust distribution follows the approach of \citet{haworth_clarke_2019}, in which the photoevaporative wind has a smaller maximum grain size and a lower dust-to-gas mass ratio than the disk because only sufficiently small grains can be entrained in the outflow \citep{facchini_2016, owen_altaf_2021, paine_2025}. We adopt a dust-to-gas mass ratio of $\Delta_\mathrm{d/g} = 0.01$ in the disk, consistent with the canonical interstellar value \citep{bohlin_1978}, while the wind is dust-depleted with $\Delta_\mathrm{d/g} = 3 \times 10^{-4}$ following \citet{haworth_clarke_2019}. Two grain populations are included in the disk: small grains ($0.005$--$1\,\mu$m), which share the gas scale height, and large settled grains ($0.005$--$1000\,\mu$m), whose scale height is reduced by a factor of $\chi = 0.2$ following the standard \textsc{dali} formulation based on the prescription of \citet{dalessio_2006}. The large grain mass fraction is $f = 0.9$. The wind contains only the small grain population, following a standard ISM size distribution \citep{mathis_1977}.

The computational grid is initially constructed with 1000 logarithmically spaced cells in both the radial and vertical directions, extending from $r = 0.1$ au to $r_\mathrm{out} = 8\,r_\mathrm{d}$ and $z = 0.01$ au to $z_\mathrm{out} = 8\,r_\mathrm{d}$. This resolution is necessary to obtain accurate solutions from the hydrostatic equilibrium solver within \textsc{puffin} and to capture the full spatial extent of the photoevaporative wind. The structure is subsequently regridded onto a coarser $200 \times 200$ grid to ensure that the thermochemical calculations are computationally feasible, while retaining sufficient resolution to resolve chemical abundances on scales of a few au or smaller. The resulting density structures are passed to \textsc{dali}, which computes the UV radiation field and dust temperature via Monte Carlo radiative transfer before iteratively solving for the gas temperature, molecular excitation, and time-dependent chemistry.

\subsection{Stellar parameters}
\label{sec:stellar_params}

For each stellar mass, we generate a blackbody spectrum with effective temperature and luminosity determined from the empirical scaling relations of \citet{eker_2018}. X-ray emission is modelled as a thermal spectrum with temperature $T_\mathrm{X} = 7 \times 10^{7}\,$K over the $1$--$100\,$keV energy range, adopting a fixed X-ray luminosity of $L_\mathrm{X} = 7.94 \times 10^{28}\,$erg\,s$^{-1}$ across all stellar masses. Although X-ray luminosity is observed to scale with stellar mass \citep{flaischlen_2021}, holding $L_\mathrm{X}$ constant isolates the effects of external FUV irradiation on disk chemistry and simplifies the interpretation of trends across the parameter grid. We also neglect accretion-driven UV emission, which can dominate the FUV output of low-mass stars \citep[e.g.][]{france_2014}. This additional FUV may warm the inner disk, particularly around the lowest-mass stars, whose photospheric FUV emission is weakest, although its impact is typically confined to the inner few au \citep[e.g.][]{calahan_2025}. A treatment of accretion is beyond the scope of this work, and, as with $L_{\text{X}}$, omitting accretion luminosity as a grid parameter allows us to more easily isolate the effects of the external FUV field.

\subsection{Chemical network and initial conditions}
\label{sec:chemistry}

The chemical evolution is calculated using a network based on a subset of UMIST~06 \citep{woodall2007}, as described in \citet{visser_2018}. This network comprises 134 species connected through 1855 reactions, encompassing gas-phase chemistry, freeze-out and thermal desorption, photodissociation and photoionisation, X-ray and cosmic-ray-induced processes, polycyclic aromatic hydrocarbon (PAH) charge exchange, and reactions involving vibrationally excited H$_2$. Non-thermal desorption is included for CO, CO$_2$, H$_2$O, CH$_4$, NH$_3$, and N$_2$, while grain surface chemistry is restricted to hydrogenation of simple species. A complete description of these processes is provided in \citet{Bruderer2012}. We adopt initial molecular abundances based on \citet{ballering_2021}, representing a typical protoplanetary disk with Solar-like C/O\,$=$\,0.5 and composition inherited from the parent molecular cloud (Table~\ref{table:initial_abundances}). The chemistry is evolved for 1\,Myr.

\subsection{Parameter grid}
\label{sec:grid}

Our model grid spans stellar masses from $M_*$ = 0.3 to $3.0\,M_\odot$, disk radii from $r_\text{d}$ = 20 to 150\,au, surface density normalisations from $\Sigma_\text{1au}$ = $10^1$ to $10^4\,$g\,cm$^{-2}$, and external FUV field strengths from $F_\text{FUV}$ = $10^2$ to $10^5$ G$_0$ (where G$_0 = 1.6 \times 10^{-3}$ erg cm$^{-2}$ s$^{-1}$ is the average interstellar radiation field in the solar neighbourhood; \citealt{habing_1968}). The range of $F_{\rm FUV}$ is set by the validity of \textsc{puffin}, whose density structures have been calibrated against radiation-hydrodynamic simulations only up to $10^5$ G$_0$ \citep{keyte_haworth_2026}. Structures generated at stronger fields cannot be assumed to be reliable and we do not extrapolate our results beyond this limit. The ranges of the remaining parameters were chosen to span the stellar masses and compact disk sizes and masses representative of the externally irradiated disk populations accessible to current observations \citep[e.g.][]{mann_2014, eisner_2018, boyden_eisner_2020}. The full set of parameters is listed in Table~\ref{table:model_grid}. The resulting grid contains 1470 models in total.

%--- FIGURE: Fluxes (ALL) -----------------------------------------------%
\begin{figure*}
\centering
\includegraphics[clip=,width=1\linewidth]{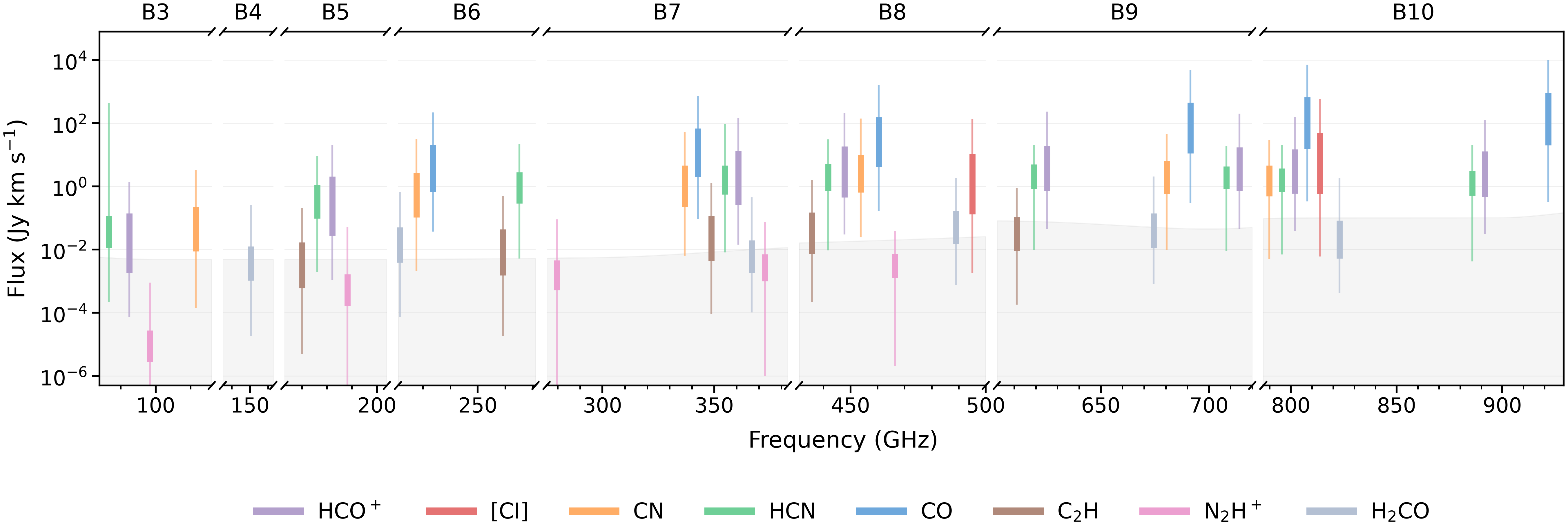}
\caption{Disk-integrated line fluxes for all models in our grid, assuming a face-on geometry and a source distance of 400 pc. Each marker represents a single transition, coloured by species and plotted at the corresponding rest frequency across ALMA Bands 3-10. For each transition, the thick bar spans the interquartile range across the model grid and the thin line covers the full range. A small horizontal jitter is applied for clarity. The shaded grey region indicates fluxes below an approximate $3\sigma$ ALMA detection threshold, estimated for a 10-hour integration, assuming all emission is unresolved within a single synthesised beam.}
\label{fig_line_fluxes_all}
\end{figure*}
%------------------------------------------------------------------------%

\subsection{Synthetic observations}
\label{sec:synth_obs}

For each model in the grid, we extract disk-integrated line fluxes for a set of species and transitions observable with ALMA at (sub-)millimetre wavelengths. We focus on eight species that are typically abundant in protoplanetary disks and readily detectable in this regime: CO, HCO$^+$, CN, HCN, C$_2$H, N$_2$H$^+$, H$_2$CO, and atomic C. For each species, we select transitions that collectively span the frequency range of ALMA Bands 3-10 (84–950 GHz), limiting the selection to at most two transitions per ALMA band per species and sampling a range of upper energy levels. In total, this yields 48 transitions that probe the broad range of temperatures and chemical conditions expected under external FUV fields of varying strength. The full list of transitions is provided in Table \ref{table:line_transitions}. For brevity, throughout this paper CN and C$_2$H transitions are labelled by their rotational quantum number $N$ alone, omitting the fine- and hyperfine-structure designations, and H$_2$CO transitions by $J$ alone, omitting the $K_a$ and $K_c$ sublevels. Full transition designations are given in Table \ref{table:line_transitions}.

Disk-integrated fluxes are extracted using the ray-tracing module in \textsc{dali}, assuming a face-on disk ($i=0^\circ$) at a distance of 400 pc, representative of the Orion star-forming region, which hosts many of the best-studied externally irradiated systems \citep[e.g.][]{odell_wen_1994,bally_2000,mann_2014}. We focus on spatially unresolved, disk-integrated emission because our aim is to identify diagnostics that can be applied independently of angular resolution. This is particularly important because externally irradiated disks are also studied in massive star-forming regions at distances of $\sim1-3$\,kpc, such as NGC~2244, Cygnus~OB2, and the Carina Nebula, where spatially resolving individual disks is substantially more challenging than in the ONC \citep[e.g.][]{balog_2007,mesa_delgado_2016,champion_2017, guarcello_2023}. Disk-integrated measurements also offer the practical advantage of not requiring a specific high-resolution array configuration, allowing observations to be conducted across a wider range of ALMA configurations and observing periods.

A small number of line fluxes ($\sim 1.6\%$) return unphysical values in excess of $10^{10}$\,Jy\,km\,s$^{-1}$. These arise in models combining a high stellar mass with a dense, high-surface-density disk. In these cases, the deep gravitational potential limits the mass removed by external photoevaporation, so the wind is very tenuous, while the high disk surface density sets up a steep density contrast between the disk surface and the wind. Under these conditions the level populations of some transitions invert within the wind, giving a negative absorption coefficient. Since \textsc{dali} solves the radiative transfer assuming this coefficient is positive, and includes no saturation mechanism, the intensities can be amplified without limit where the populations invert. The affected fluxes therefore reflect a failure of the solver in a regime it does not accurately model. To mitigate this, we exclude from our analysis any individual line flux exceeding $10^{5}$\,Jy\,km\,s$^{-1}$, together with any non-finite values. This threshold lies well above the brightest physical fluxes in the grid ($\sim$$10^{4}$\,Jy\,km\,s$^{-1}$) and far below the affected values, so our results are insensitive to its precise choice.

%%%%%%%%%%%%%%
% 3. RESULTS %
%%%%%%%%%%%%%%

\section{Results}
\label{sec:results}

\subsection{Response of individual line fluxes to model parameters}
\label{sec:individual}

We begin by examining the disk-integrated line fluxes computed across the full model grid for all transitions listed in Table \ref{table:line_transitions}. Figure \ref{fig_line_fluxes_all} shows the flux distributions as a function of their rest frequencies across the ALMA wavelength regime, coloured by species. For each transition, the thick bar spans the interquartile range of the flux distribution across all models, while the thin line extends to the minimum and maximum values.

The line fluxes span approximately 13 orders of magnitude across the full model grid, reflecting the wide range of stellar and disk properties explored. For any individual transition, the total range typically covers a few orders of magnitude. However, the interquartile ranges are much narrower, generally spanning only $\sim1$–$2$ orders of magnitude, indicating that the most extreme flux values arise from only a small subset of models.

To assess the observability of each transition, we estimate a representative ALMA $3\sigma$ detection threshold, shown by the shaded grey region in Figure \ref{fig_line_fluxes_all}. We assume a total integration time of 10 hours, broadly representative of a standard ALMA spectral-line programme, and a line width of 10 km s$^{-1}$, with the emission distributed uniformly across the line profile. The source is assumed to be spatially unresolved, such that all of the flux is contained within a single synthesised beam. We also adopt the default atmospheric water-vapour level corresponding to the third octile. Using these assumptions, we calculated the detection threshold with the ALMA Sensitivity Calculator\footnote{\url{https://almascience.eso.org/proposing/sensitivity-calculator}} at intervals of 100 GHz and smoothly interpolated between these values across the full frequency range. The resulting threshold increases from approximately 7 mJy km s$^{-1}$ at 100 GHz to 104 mJy km s$^{-1}$ at 900 GHz. Although intended only as an approximate guide, this threshold provides a useful first-order measure of detectability and is used in later sections to identify the most promising molecular line-flux ratio diagnostics.

\subsubsection{Sensitivity of line fluxes to the external FUV field}
\label{sec:fuv_sensitivity}

Our primary aim is to identify observational signatures of external photoevaporation, so we first assess how the disk-integrated line fluxes of individual species respond to the external FUV field strength \ffuv. To quantify this, we use the Spearman rank correlation coefficient, which measures the strength of monotonic association between two variables on a scale from $-1$ (perfectly decreasing) through 0 (no trend) to $+1$ (perfectly increasing), and is computed by replacing each variable with its rank before evaluating the Pearson correlation of the resulting rank sequences. We compute this statistic in two complementary ways, which answer different questions. The first, the grid-wide correlation \rhofuv, is evaluated between $\log(F_{\rm FUV})$ and the logarithm of the disk-integrated flux across all models in the grid, in which $M_*$, $r_{\rm d}$ and $\Sigma_{\rm 1au}$ vary alongside the external field. It asks whether a line flux tracks $F_{\rm FUV}$ across a population of disks with diverse, and generally unknown, properties, as in an observational survey. The second, the fixed-structure correlation \rhopartial, asks whether a line responds to the external field when nothing else changes. For each combination of $(M_*, r_{\rm d}, \Sigma_{\rm 1au})$ the corresponding models differ only in $F_{\rm FUV}$, and we compute the Spearman coefficient separately within each such set. We summarise each transition by the median of the resulting distribution and its central quartile range. A median \rhopartial\ of unity therefore means that, in a typical disk, the line flux increases at every step in $F_{\rm FUV}$. Where \rhopartial\ is high but \rhofuv\ is low, the line responds consistently to the external field, but this response is masked by variations in the stellar and disk properties. The difference between the two therefore quantifies the degeneracy that motivates the use of line flux ratios in Section \ref{sec:results_ratios}. Where both are low, the line is intrinsically insensitive to the external field. We emphasise that both statistics measure the consistency of a monotonic response rather than the magnitude of the underlying flux change.

\begin{figure*}
\centering
\includegraphics[clip=,width=0.97\linewidth]{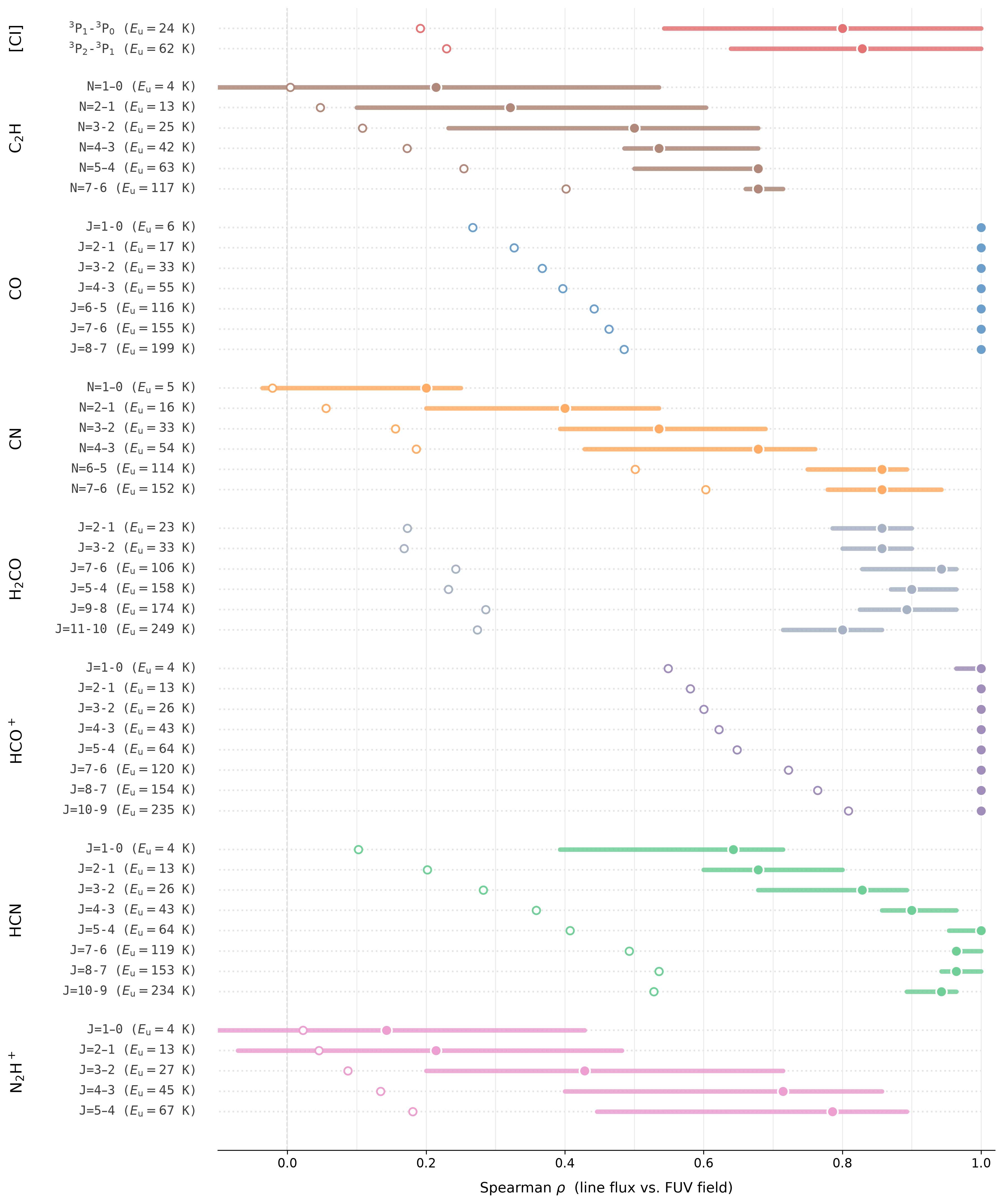}
\caption{Monotonic sensitivity of molecular line fluxes to the external FUV field, quantified by the Spearman rank correlation $\rho$ between line flux and FUV field strength \ffuv. Each row is a single transition, grouped by species (colour). For each line, a higher Spearman $\rho$ value indicates a stronger correlation between the FUV field and the line flux. Open circles show the grid-wide correlation (\rhofuv), evaluated across all models, in which $M_*$, $r_{\rm d}$, $\Sigma_{\rm 1au}$ and $F_{\rm FUV}$ all vary. Filled circles show the median fixed-structure correlation (\rhopartial), computed separately within each set of models that share the same $(M_*, r_{\rm d}, \Sigma_{\rm 1au})$ and differ only in $F_{\rm FUV}$. Coloured bars span the central quartile range of that distribution, indicating how consistently a line responds to FUV across different disk structures. Note that hyperfine transition labels have been shortened for clarity, see Table \ref{table:line_transitions} for full transition designations.}
\label{fig_spearman_megaplot}
\end{figure*}

Figure \ref{fig_spearman_megaplot} shows both statistics for every transition, grouped by species. Open circles denote \rhofuv, filled circles the median \rhopartial, and coloured bars the corresponding central quartile range. For every transition, the median \rhopartial\ exceeds \rhofuv. This is expected, since the stellar and disk parameters vary independently of $F_{\rm FUV}$ in our grid, and therefore dilute rather than strengthen a consistent response to the field. The size of this gap is nonetheless informative, since it reflects how strongly variations in stellar and disk properties mask the intrinsic FUV response. HCO$^+$ and CO exhibit the strongest correlations, with median \rhopartial\ $=1$ and negligible central quartile spread across all transitions. Their grid-wide correlations are considerably lower, with \rhofuv\ increasing with excitation from $\sim0.55$ to $\sim0.8$ for HCO$^+$ and from $\sim0.3$ to $\sim0.5$ for CO, reflecting the dilution of the perfectly monotonic response to external irradiation by structure-to-structure flux variations. HCN approaches similarly strong behaviour at moderate and high excitation, with its median \rhopartial\ rising from $\sim0.65$ for $J=1{-}0$ to ${\approx}1$ for $J=5{-}4$, and remaining above $\sim0.95$ through $J=10{-}9$ with narrow central quartile ranges. CN shows a weaker but similarly excitation-dependent trend, with the median \rhopartial\ increasing from $\sim0.15$ for $N=1{-}0$ to $\sim0.85$ for $N=7{-}6$, accompanied by a modest central quartile spread. C$_2$H and N$_2$H$^+$ show qualitatively similar increases with excitation, but with weaker overall correlations. Their median \rhopartial\ reaches only $\sim0.7$--$0.8$ for the highest-excitation transitions, while at low excitation their central quartile ranges extend below zero, indicating that for an appreciable fraction of disk structures these line fluxes are insensitive to, or even anti-correlated with, the external field. 

[CI] and H$_2$CO exhibit the strongest divergence between the two statistics. Both [CI] transitions have \rhofuv\ $\sim0.2$ but a median \rhopartial\ $\sim0.8$, although their central quartile ranges are very broad, spanning $\sim0.6$ to unity. So, while atomic carbon typically responds strongly to the external field, the strength of that response depends sensitively on the underlying disk structure. H$_2$CO exhibits the most extreme contrast, with low \rhofuv\ values across all transitions ($\lesssim0.3$) but a uniformly high median \rhopartial\ ($\sim0.8$--$0.9$). The H$_2$CO flux therefore rises consistently with \ffuv\ for any given disk structure, but this response is masked in the overall model grid by a strong dependence on the stellar and disk parameters, which we quantify in the following section. More generally, both statistics tend to increase with excitation energy within each species, such that higher-$E_\mathrm{u}$ transitions correlate more strongly with the external field.

Overall, the line responses vary substantially between species and transitions, both in their strength and consistency across disk structures. We discuss the chemical and physical origins of these trends in Section \ref{subsec:disc_connection_to_chemistry}.

%--- FIGURE: SPEARMAN MEGAPLOT -------------%
%-------------------------------------------%

%--- FIGURE: Spearman (all model parameters) ----------------------------%
\begin{figure*}
\centering
\includegraphics[clip=,width=1\linewidth]{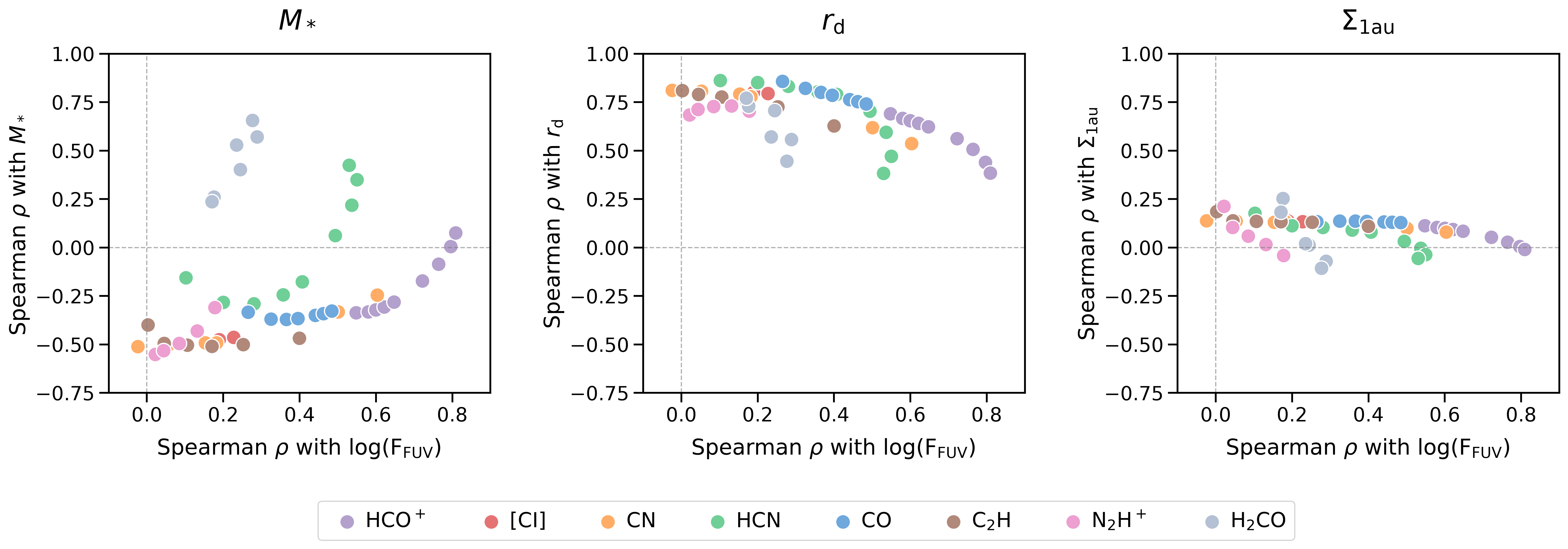}
\caption{Spearman rank correlation coefficients between the disk-integrated line fluxes and the stellar and disk parameters, for all 48 molecular transitions in the model grid. Each point represents a single transition, coloured by species. From left to right, panels show the correlation of each line flux with the stellar mass ($M_*$), disk outer radius ($r_\mathrm{d}$), and surface density at 1\,au ($\Sigma_\mathrm{1au}$), plotted against its correlation with the external FUV field strength, \rhofuv\ (Section~\ref{sec:fuv_sensitivity}). All coefficients are evaluated across the full model grid, in which all four parameters vary simultaneously, and they are therefore directly comparable to \rhofuv\ in Figure \ref{fig_spearman_megaplot} (open circles), rather than \rhopartial, which is computed at fixed disk structure. Dashed lines indicate $\rho = 0$ in each dimension. Transitions towards the upper right of each panel are positively correlated with both the external FUV field and the given stellar or disk parameter.}
\label{fig_spearman_all}
\end{figure*}
%------------------------------------------------------------------------%

\subsubsection{Degeneracies with stellar and disk parameters}
\label{sec:degeneracies}

The gap between \rhofuv\ and \rhopartial\ varies substantially between transitions, indicating that variations in the stellar or disk properties obscure the intrinsic response to the external field to differing degrees. We now examine which of the remaining grid parameters are responsible. We compute the grid-wide Spearman correlations between the disk-integrated fluxes and the stellar mass ($\rho_{M_*}$), disk radius ($\rho_{r_{\rm d}}$), and surface density normalisation ($\rho_{\Sigma_{\rm 1au}}$), evaluated across all models in the same manner as \rhofuv. Figure \ref{fig_spearman_all} presents the results as three panels, each showing the correlation with a given parameter against \rhofuv\ for every transition, coloured by species.

Most transitions display a negative correlation with stellar mass (left panel), with $\rho_{M_*}$ typically ranging from $-$0.5 to 0. The anti-correlation is strongest for the lowest-$E_\text{u}$ lines and weakens with increasing $E_\text{u}$. This is because more luminous stars heat a larger fraction of the disk, depopulating the low-$E_\text{u}$ levels that trace colder gas. High-$E_\text{u}$ transitions, on the other hand, are more efficiently excited in warmer gas, which offsets the loss of cold emitting material and weakens their anti-correlation with $M_*$. The main exceptions are H$_2$CO, whose transitions show weak to moderate positive correlations ($\rho_{M_*}\sim0.2 - 0.7$), and HCN, for which the high-$J$ transitions are weakly positively correlated with stellar mass while the low-$J$ transitions remain weakly anti-correlated.

Strong positive correlations are found between disk-integrated fluxes and $r_\mathrm{d}$ for nearly all species and transitions (centre panel of Figure \ref{fig_spearman_all}). This is expected simply because a larger disk provides a greater emitting area. However, because changing $r_\mathrm{d}$ also modifies the temperature and chemical structure in a non-linear fashion, the degree of correlation is species- and transition-dependent, with $\rho_{r_\mathrm{d}}$ varying from $\sim\,0.4$ to $\sim\,0.85$. As with the stellar mass, the correlation is strongest for low-$E_\text{u}$ transitions that trace the extended outer disk, and decreases with increasing $E_\text{u}$.

Only a weak positive correlation exists between the disk-integrated fluxes and the surface density normalisation $\Sigma_\mathrm{1au}$ (right panel of Figure \ref{fig_spearman_all}). For all species and transitions, $\rho_{\Sigma_\mathrm{1au}}$ clusters around ${\sim}\,0.15$, indicating that increasing the disk surface density, and hence disk mass, generally produces only a negligible to modest change in the emergent flux. This does not imply that the disk itself makes no contribution, but rather that the emitting gas is not set by the total disk mass. Because the incident FUV radiation is attenuated rapidly at the disk surface, the irradiated layers responsible for much of the emission are comparatively unaffected by changes in $\Sigma_\mathrm{1au}$, and the same is true of optically thick lines, which trace the temperature and area of the $\tau = 1$ surface rather than the column beneath it. The disk can nevertheless contribute substantially to the emission, particularly for low-$E_\mathrm{u}$ transitions at weak external fields, where the emission arises predominantly from the disk surface. As $F_\mathrm{FUV}$ increases, a growing fraction of the emission arises from the warm photoevaporative wind, whose temperature and excitation structure are regulated primarily by the external field rather than the underlying disk. We examine this in greater detail in Section~\ref{subsec:disc_connection_to_chemistry}.

The degeneracies discussed here demonstrate that absolute (sub-)millimetre line fluxes alone are unlikely to provide a clean diagnostic of the external FUV field. The disk radius and stellar mass, in particular, introduce correlations of comparable strength to \ffuv\ for many transitions. This motivates the use of line flux ratios as diagnostics, where the dependence on parameters that affect two lines similarly can be reduced or cancelled. We explore this approach in the following section.

\subsection{Line flux ratios as diagnostics of external FUV field strength}
\label{sec:results_ratios}

Several species exhibit strong Spearman correlations between their disk-integrated line fluxes and the external FUV field strength, but these correlations can be significantly degenerate with stellar mass and disk radius. Line flux ratios offer a natural way to reduce these degeneracies. If two transitions respond similarly to a given parameter, their shared dependence partially cancels in the ratio. However, cancellation is exact only when both lines scale by the same factor. Where their responses differ, a residual sensitivity survives and the raw Spearman correlation of the ratio with $F_\mathrm{FUV}$ may still partly reflect covariance with other parameters.

To assess the dependence of line ratios on the external FUV field, we calculated all pairwise combinations of the 48 transitions in our grid, yielding 2256 ratios (including reciprocal pairs but excluding self-ratios). We characterise each ratio using three quantities. The first is the fixed-structure correlation, \rhopartial, which, as for individual transitions, measures the relationship between the ratio and $F_{\rm FUV}$ after accounting for variations in stellar and disk properties. We calculate this using a partial Spearman correlation controlling for $M_*$, $r_{\rm d}$, and $\Sigma_{\rm 1au}$ \citep[e.g.][]{kendall_1942}. However, a high \rhopartial\ does not by itself ensure that a ratio is useful observationally. A ratio may follow $F_{\rm FUV}$ monotonically but vary only weakly across several orders of magnitude in irradiation, making it difficult to distinguish different FUV environments given realistic observational uncertainties. We therefore also compute a matched enhancement factor, $\eta_{\rm FUV}$, which measures the total change in a ratio across the full range of external fields considered. For each combination of $M_*$, $r_{\rm d}$ and $\Sigma_{\rm 1au}$, we divide the value of the ratio at $10^5$\,G$_0$ by its value at $10^2$\,G$_0$, and take $\eta_{\rm FUV}$ as the median of this quantity across all combinations. A value of $\eta_{\rm FUV} = 1$ therefore indicates no net change, while $\eta_{\rm FUV} = 10$ and $100$ indicate that the ratio increases by one and two orders of magnitude, respectively, across the three decades in $F_{\rm FUV}$. Because it depends only on the end points, $\eta_{\rm FUV}$ does not by itself guarantee a monotonic response, which is instead captured by \rhopartial. Finally, to assess observability, we record the detection fraction, $f_{\rm det}$, defined as the fraction of models in which both lines exceed the $3\sigma$ threshold of Section~\ref{sec:individual}. Together, these metrics test whether a ratio changes consistently and measurably with FUV irradiation while both lines remain detectable.

Because reciprocal ratios carry identical information with the sign of \rhopartial\ reversed, we consider each pair only once in our analysis below. Across the full set, |\rhopartial| ranges from $0.003$ ([CI]\,$^3P_2$--$^3P_1$/HCN\,$J=10-9$) to $0.936$ ([CI]\,$^3P_1$--$^3P_0$/[CI]\,$^3P_2$--$^3P_1$), and $\eta_\text{FUV}$ reaches values of up to $\sim125$, confirming that a substantial number of ratios combine a strong, structure-independent correlation with a large dynamic range. To isolate the most practical diagnostics we apply three cuts: |\rhopartial| $\geq 0.8$, ensuring at least a strong correlation with $F_{\rm FUV}$; $\eta_{\rm FUV} \geq 5$, ensuring the ratio varies by a factor large enough to be measured against realistic uncertainties (see Section \ref{sec:disc_alma_guidance}); and $f_{\rm det} \geq 0.5$, ensuring both lines are detectable across at least half the model grid. These criteria yield a shortlist of 135 ratios that are well-suited to identifying and quantifying the impact of external photoevaporation, and are listed in Table~\ref{table:ratio_shortlist} ordered by \rhopartial.

%--- FIGURE: Heatmap (3-species subset) ------%
\begin{figure}
\centering
\includegraphics[clip=,width=1.0\linewidth]{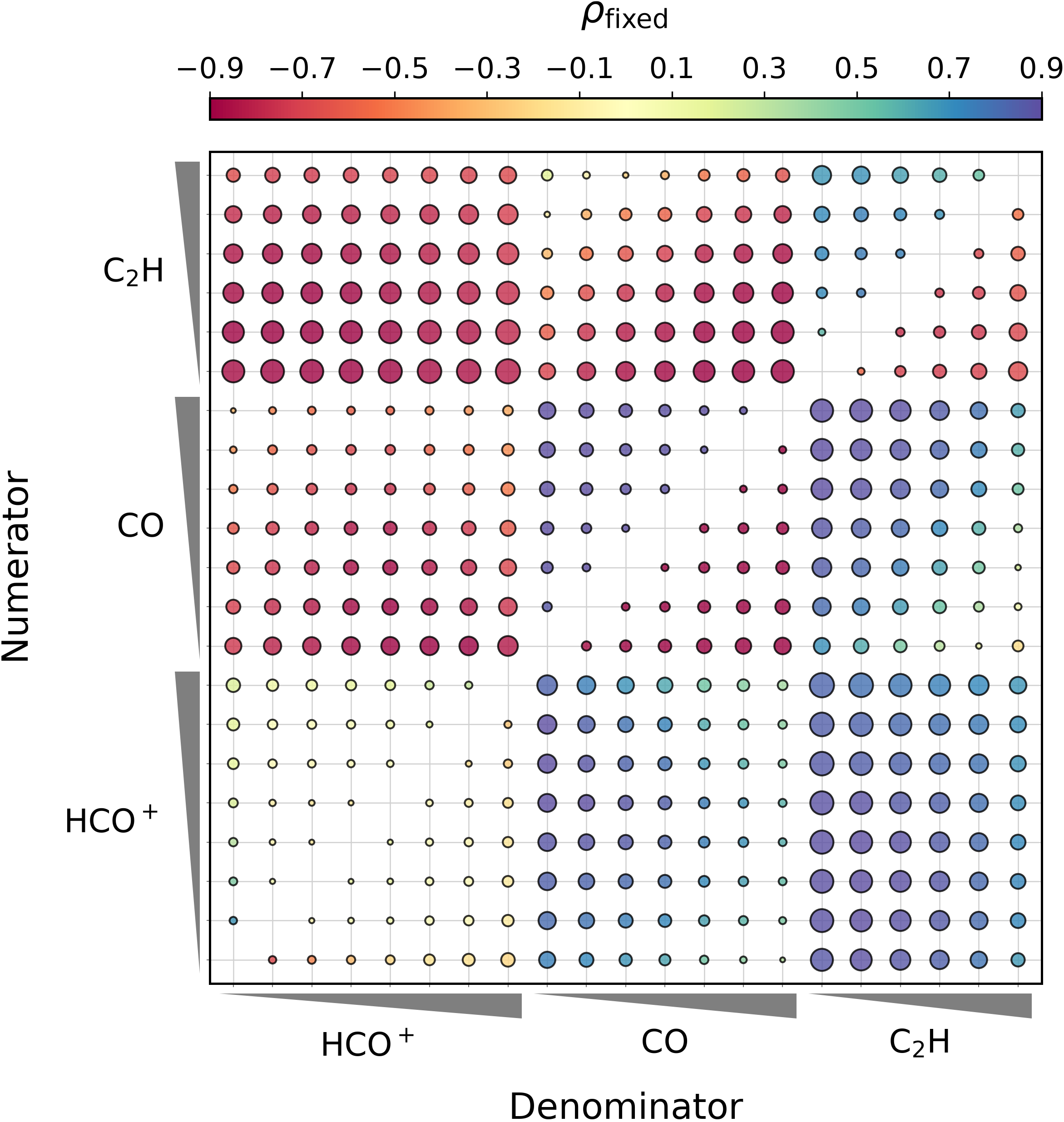}
\caption{Diagnostic power of molecular line flux ratios as tracers of external photoevaporation, shown for a representative subset of molecular species (CO, HCO$^+$, C$_2$H). These species are chosen to illustrate the two classes of diagnostic identified in Section \ref{sec:results_ratios}: excitation-sensitive ratios (e.g. CO high-$J$/low-$J$, HCO$^+$/CO) and chemistry-sensitive ratios with a denominator that responds weakly to the external FUV field (e.g. HCO$^+$/C$_2$H). Each circle represents the ratio of the transition on the $y$-axis (numerator) to that on the $x$-axis (denominator), with colour denoting \rhopartial\ and circle size the enhancement factor $\eta_{\rm FUV}$. Blue indicates that the ratio increases with $F_\text{FUV}$ and red that it decreases. Wedges along the axes indicate increasing $E_\text{u}$ within each species (see Table \ref{table:line_transitions}). The full matrix for all 8 species is shown in Appendix \ref{appendix:full_ratio_matrix}.}
\label{fig_heatmap}
\end{figure}
%-------------------------------------------%

Every ratio in the shortlist correlates positively with $F_{\rm FUV}$. In each case a strongly responding line is placed over a more weakly responding one, so that the ratio grows as the external field increases. The numerators are drawn exclusively from the species identified in Section \ref{sec:fuv_sensitivity} as the most FUV-responsive, including HCO$^+$ across its full rotational ladder, high-$J$ CO, and the [CI] lines. The denominators broadly separate into two categories that trade correlation strength against dynamic range. In the first, the denominator is a low-excitation line tracing essentially the same gas as the numerator, either a low-$J$ line of the same molecule or of a co-spatial species (e.g. CO $J=8-7$/$J=1-0$, CO $J=8-7$/HCN $J=2-1$, HCO$^+$ $J=7-6$/CO $J=1-0$). Because both lines share the same dependence on $M_*$, $r_{\rm d}$ and $\Sigma_{\rm 1au}$, that dependence cancels almost exactly, leaving as a signal the excitation contrast produced by the warm photoevaporative wind. These ratios attain the highest \rhopartial\ (up to $\sim0.91$) and are detectable in nearly every model ($f_{\rm det}\gtrsim0.8$), but their enhancement factors are modest ($\eta_{\rm FUV}\sim5-17$) because the denominator also responds to the FUV field. In the second category, the denominator is a comparatively FUV-insensitive line (see Section \ref{sec:disc_alma_guidance}). Here the change is set almost entirely by the numerator, producing the largest enhancement factors in the shortlist (up to $\eta_{\rm FUV}=89$, e.g. HCO$^+$\,$J=10-9$/CN\,$N=1-0$), at the cost of a lower detection fraction ($f_{\rm det}\sim0.6-0.8$) owing to the faintness of the denominator. The ratios with the greatest dynamic range are therefore less reliably detected, while those detectable in every disk offer more limited contrast. 

Figure \ref{fig_heatmap} illustrates both categories using all ratios formed from CO, HCO$^+$, and C$_2$H, with colour denoting \rhopartial\ and circle size the enhancement factor $\eta_\text{FUV}$. The first category appears in the CO high-$J$/low-$J$ blocks and HCO$^+$/CO blocks, which are strongly correlated with $F_\text{FUV}$, but have modest enhancement factors. The second category appears in the HCO$^+$/C$_2$H block, where the weakly responding C$_2$H denominator produces the largest enhancement factors in the figure. In contrast, ratios between different HCO$^+$ transitions show little dependence on $F_\text{FUV}$ since all respond to the external field in a similar way. The complete matrix of 2256 ratios is presented in Appendix~\ref{appendix:full_ratio_matrix}.

Figure~\ref{fig_ratio_summary} extends this comparison, showing the joint distribution of \rhopartial\ and $\eta_{\rm FUV}$ alongside the detectability of each pair. The figure highlights the trade-off between correlation strength, dynamic range, and detectability, as well as the fraction of ratios that satisfy our selection criteria. Ratios over low-excitation CO attain the highest \rhopartial\ but only modest enhancement factors, while those over CN and C$_2$H reach the largest dynamic ranges at the cost of detectability.

We note that satisfying all three criteria does not necessarily make a ratio the optimal diagnostic for a particular target. The ranking rewards a strong, structure-independent correlation but is largely blind to practical observability beyond the coarse $f_{\rm det}$ cut. Many of the highest-ranked ratios pair widely separated frequencies, often involving Band 9 or Band 10, which demand excellent atmospheric conditions and are therefore considerably more expensive to observe than e.g. a same-band pair such as HCO$^+$\,$J=3-2$/C$_2$H\,$N=3-2$. The detection fraction is similarly a relatively crude filter. A ratio detectable in only a minority of systems may remain an excellent diagnostic within that subset, and even a non-detection of one line yields a useful limit on the ratio. The most appropriate diagnostic therefore depends on the target and the specific science goal. We turn to these practical considerations, and to guidance on applying the ratios to real observations, in Section \ref{sec:disc_alma_guidance}.

%--- FIGURE: Ratio Summary -----------------%
\begin{figure}
\centering
\includegraphics[clip=,width=1\linewidth]{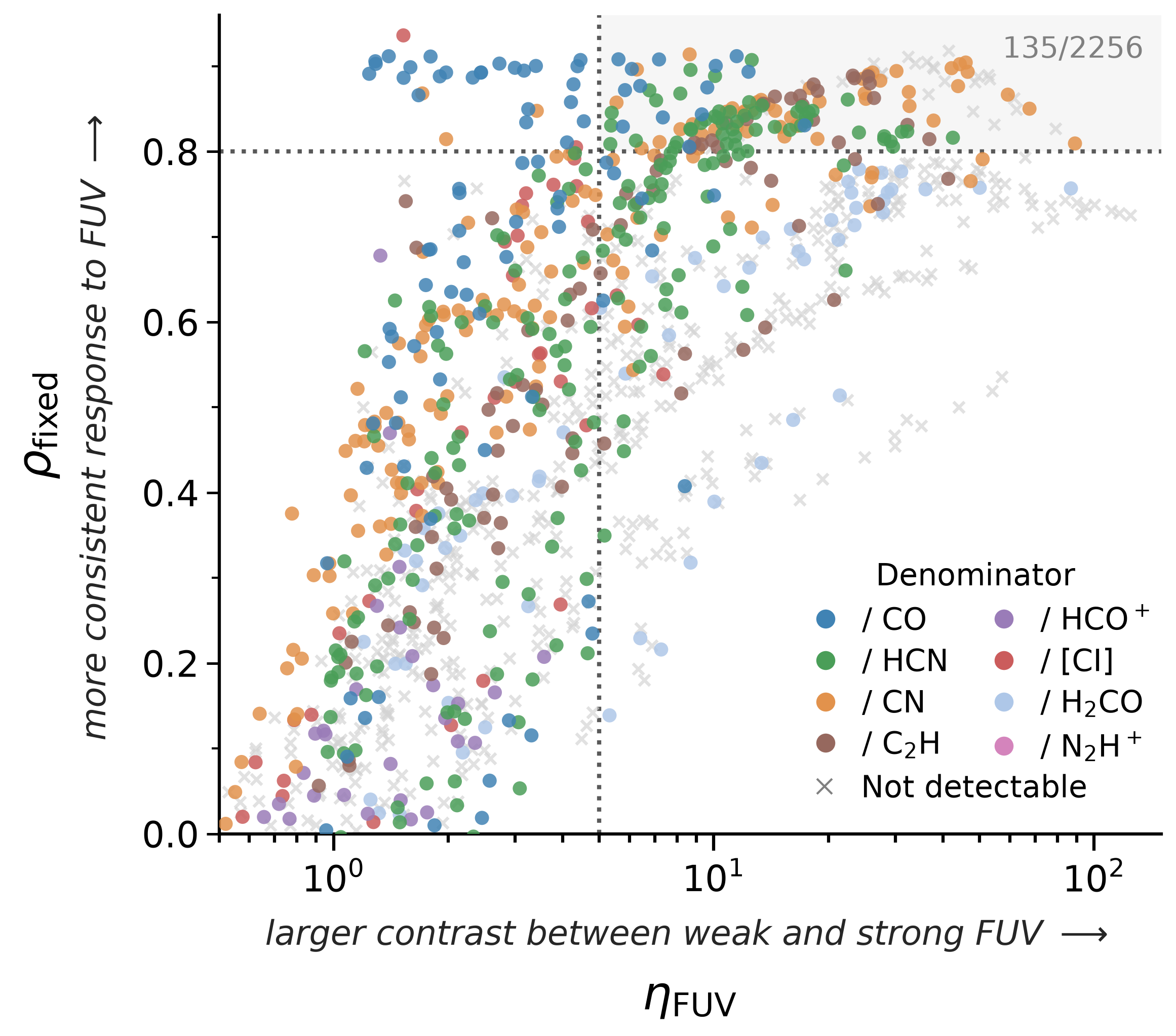}
\caption{Diagnostic performance of all 2256 line flux ratios, shown as the partial Spearman correlation with the external FUV field strength, \rhopartial, against the median FUV enhancement factor, $\eta_\text{FUV}$. Points are coloured by the species forming the denominator of the ratio, and grey crosses mark ratios that fall below our detectability threshold ($f_\text{det} < 0.5$). Dotted lines mark the cuts at \rhopartial\ $=0.8$ and $\eta_\mathrm{FUV} = 5$, and the shaded region indicates where both are satisfied, containing the 135 shortlisted ratios of Table \ref{table:ratio_shortlist}.}
\label{fig_ratio_summary}
\end{figure}
%-------------------------------------------%

%%%%%%%%%%%%%%%%%
% 4. DISCUSSION %
%%%%%%%%%%%%%%%%%
\section{Discussion}
\label{sec:discussion}

% SUBSECTION: CONNECTION TO CHEMISTRY
\subsection{Chemical origins of the line flux trends}
\label{subsec:disc_connection_to_chemistry}

\begin{figure*}
\centering
\includegraphics[clip=,width=1\linewidth]{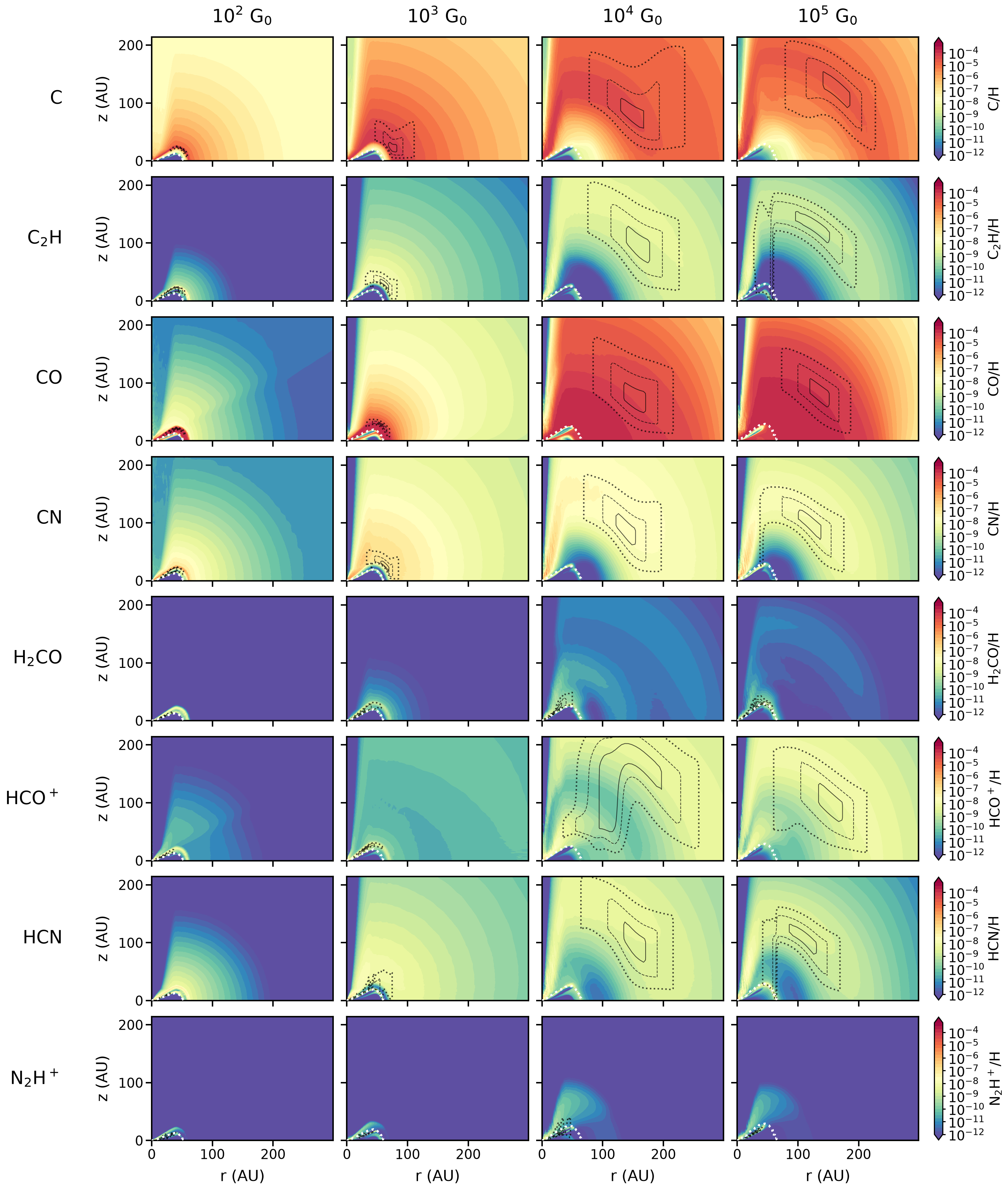}
\caption{Abundance maps for the eight chemical species considered in this work (rows) for a representative model ($M_*=0.6\,M_{\odot}$, $\Sigma_\mathrm{1au}=1000\,{\rm g\,cm^{-2}}$, $r_\mathrm{d}=40$ au) at external FUV field strengths of $10^{2}$--$10^{5}$ G$_0$ (columns). Black contours enclose the central 25\% (solid), 50\% (dashed) and 75\% (dotted) of the emission of the highest-$E_{\rm u}$ transition of each species (Table \ref{table:line_transitions}), computed for a face-on disk. The white dotted line marks the disk--wind boundary, defined as the gas density contour equal to the midplane density where $\tau=1$. With increasing $F_{\rm FUV}$, the emission of [CI], C$_2$H, CO, CN, HCO$^+$ and HCN shifts from the disk surface into the photoevaporative wind, whereas H$_2$CO and N$_2$H$^+$ remain confined to the disk--wind interface.}
\label{fig_abundance_maps}
\end{figure*}

We have shown that most line fluxes increase with $F_{\rm FUV}$ when the stellar and disk properties are held fixed, but that the strength and consistency of this response varies substantially between species and transitions. To explore the origin of these differences, we first consider how the abundance of each species, and the emitting region of its highest-excitation transition, change with $F_{\rm FUV}$ for a representative model ($M_* = 0.6\,M_\odot$, $\Sigma_{\rm 1au} = 1000$\,g\,cm$^{-2}$, $r_{\rm d} = 40$\,au; Figure \ref{fig_abundance_maps}). At the weakest field ($10^2$\,G$_0$), the emission of every species arises from a thin layer close to the disk surface. As the field strengthens, [CI], C$_2$H, CO, CN, HCO$^+$ and HCN become increasingly abundant in the photoevaporative wind, and their emitting regions shift outwards to follow. In contrast, H$_2$CO and N$_2$H$^+$ remain in low abundance in the wind, and their emission stays confined to the disk--wind interface. We now examine the
chemistry underlying these changes for each species in turn.

\textbf{\boldmath HCO$^+$} displays the most consistent intrinsic response to external irradiation of any species in our grid. All transitions have median \rhopartial\ $=1$, with negligible variation between disk structures, showing that their fluxes increase monotonically with \ffuv\ when the stellar and disk properties are held fixed. The lower \rhofuv\ values reflect variations induced by these other parameters rather than an inconsistent response to the external irradiation field. Although the peak HCO$^+$ abundance remains approximately constant (HCO$^+$/H $\sim10^{-8}$), the high abundance layer moves outwards and expands into the photoevaporative wind when \ffuv\ is increased. Within this layer, HCO$^+$ is predominantly produced through FUV-driven chemistry. C$^+$ reacts with OH to form CO$^+$, which is subsequently converted to HCO$^+$ either directly or via HOC$^+$. A secondary minor formation pathway is the reaction of CH$_3^+$ with atomic O. The outer boundary of the high-abundance layer is set by rapid dissociative recombination with electrons, while the inner boundary is controlled by attenuation of the FUV field, which reduces the C$^+$ abundance and suppresses the main formation pathway. Stronger irradiation shifts and broadens this region into the wind, increasing the volume of warm HCO$^+$-emitting gas. Deeper in the disk, where the external FUV is attenuated, HCO$^+$ instead forms predominantly through proton transfer from H$_3^+$ to CO, and its abundance is regulated by ionisation from stellar X-rays and cosmic rays \citep[e.g.][]{cleeves_2014, cleeves_2015}. These are held fixed across our grid, but variations in either would alter this disk component, and could therefore affect HCO$^+$-based ratios at weak external fields, where the disk contribution is largest.

\textbf{CO} behaves similarly to HCO$^+$ in the wind, maintaining a high peak abundance (CO/H $\sim 10^{-4}$) at all field strengths and filling a progressively larger volume as \ffuv\ increases. Every CO transition has median \rhopartial\ $=1$ with negligible variation between disk structures, so the fluxes increase monotonically with the external field whenever the stellar and disk properties are held fixed. The grid-wide correlations are considerably lower (\rhofuv\ $\sim 0.3$--$0.5$, rising with $E_\text{u}$) and fall below the corresponding HCO$^+$ values across the full ladder. This stronger dilution arises because CO remains abundant throughout the disk as well as the wind, so the low-$J$ lines receive a large contribution from gas whose emission is regulated by the disk radius and stellar mass rather than by the external field. Higher-$J$ transitions preferentially trace the warm wind and are correspondingly less diluted. Uniquely among the species considered here, increasing \ffuv\ also enhances the gas-phase CO abundance within the disk itself. This occurs because CO has a relatively low sublimation temperature ($\sim 20$\,K). Although the external FUV is efficiently attenuated at the disk surface even at the highest field strengths, reprocessing of FUV by the photoevaporative wind into infrared radiation can raise dust temperatures within the disk by a few kelvin, sufficient to desorb a substantial fraction of the CO ice reservoir \citep{keyte_haworth_2025}. Nevertheless, the dominant effect of increasing \ffuv\ on the total CO column is still the expansion of the warm wind. Both the desorption of CO ice and the additional emission from the wind may contribute to the weak CO depletion inferred in compact ONC disks by \citet{boyden_eisner_2023}, although testing this would require source-specific models that include the photoevaporative wind.

\textbf{CN} and \textbf{HCN} both respond positively to increasing \ffuv, but the strength and consistency of this response depend on rotational excitation. For HCN, the median \rhopartial\ rises from ${\sim}0.65$ for $J=1-0$ to unity for $J=5-4$ and remains above ${\sim}0.95$ at higher $J$, with little variation between disk structures. CN shows a much stronger dependence on excitation, with \rhopartial\ increasing from ${\sim}0.15$ for $N=1-0$ to ${\sim}0.85$ for $N=7-6$. As for the other species, the grid-wide correlations are lower, reflecting the additional scatter introduced by variations in the stellar and disk properties. These trends are linked to the coupled chemistry of the two species. CN forms predominantly through the photodissociation of HCN and is destroyed primarily through $\mathrm{CN}+\mathrm{H_2}\rightarrow\mathrm{HCN}+\mathrm{H}$, which is also a major HCN formation pathway alongside the dissociative recombination reaction $\mathrm{HCNH^+}+\mathrm{e^-}\rightarrow\mathrm{HCN}+\mathrm{H}$. CN is therefore most abundant in the exposed gas where HCN is photodissociated, and extends farther into the wind, whereas HCN is concentrated in a layer coincident with, or lying beneath, the CN-rich region. As \ffuv\ increases, the high-abundance layers of both species expand into the warm wind. The higher-excitation transitions preferentially trace this warm gas and so brighten consistently with \ffuv. The lower-excitation lines are more easily excited in cooler gas and receive a larger contribution from the disk, particularly at weak fields, which dilutes their response to the wind and produces the rise in \rhopartial\ with $E_\text{u}$ seen for both species.

%--- FIGURE: Abundance Maps ---------------------------------------------%
%------------------------------------------------------------------------%

\textbf{Atomic C} has previously received particular attention as a tracer of the photoevaporative flow interior to the ionisation front, through both the sub-millimetre [CI] fine-structure lines and the optical [CI] 8727\,\AA\ line \citep{haworth_2020, haworth_2022, aru_2024}. In our models, both [CI] transitions show a large gap between the two statistics, with \rhofuv\ $\sim0.2$ but a median \rhopartial\ $\sim0.8$. Their broad central quartile ranges further indicate that, although the [CI] flux generally increases with \ffuv\ at fixed disk structure, the strength and consistency of this response depend sensitively on the underlying stellar and disk properties. Neutral carbon occupies a relatively narrow layer between the CO-rich molecular gas and the C$^+$-dominated outer wind, bounded on the inside by the transition to CO and on the outside by photoionisation of C to C$^+$. As \ffuv\ increases, this layer shifts outwards and extends further into the photoevaporative wind, increasing the volume of warm gas in which atomic carbon is abundant. Since both [CI] transitions have upper-level energies well below the typical gas temperature in this layer ($E_\mathrm{u}\approx24$ and 62\,K), this additional gas is readily excited and contributes to the emission. The position and thickness of the layer are set by the balance between CO photodissociation and C photoionisation, which in turn depends on the density structure of the wind and hence on how rapidly the incident FUV radiation is attenuated. Because the ability of the wind to shield its inner regions varies substantially across the grid, so do the location and volume of the atomic carbon layer. This helps explain why [CI] correlates strongly with \ffuv\ within each disk structure, but only weakly when all models are combined.

\textbf{H$_2$CO} exhibits the largest gap between \rhofuv\ and \rhopartial\ of any species in our grid. Its \rhofuv\ values are among the lowest we measure ($\lesssim 0.3$ for all transitions), yet its median \rhopartial\ is uniformly high ($\sim$0.8-0.9), so the flux rises consistently with \ffuv\ at fixed disk structure while this response is almost entirely masked across the full grid. This reflects the fact that H$_2$CO is the only species whose fluxes are all positively correlated with stellar mass ($\rho_{M_*} \sim 0.2-0.7$; Section \ref{sec:degeneracies}), so that variations in $M_*$ imprint scatter comparable in magnitude to the FUV response itself. H$_2$CO is confined to a thin layer on the disk surface at $10^{2}$ G$_0$ and spreads into the wind at higher \ffuv, but remains relatively diffuse and in low abundance. In our models it is produced predominantly through gas-phase neutral--radical reactions in the warm irradiated gas such as CH$_3$ + O $\rightarrow$ H$_2$CO +H (rather than by grain-surface hydrogenation of CO, which typically dominates in colder disk regions). Its destruction is controlled by proton transfer to the rising HCO$^+$ abundance (HCO$^+$ + H$_2$CO $\rightarrow$ H$_3$CO$^+$ + CO), which counteracts the outward expansion of the H$_2$CO layer and limits the growth of the emitting volume as \ffuv\ increases.

\textbf{C$_2$H} shows a weaker and more structure-dependent response to \ffuv\ than the species discussed above, particularly at low excitation. Its median \rhopartial\ rises from ${\sim}0.2$ for $N=1-0$ to ${\sim}0.7$ for the highest-$N$ transitions, while the central quartile ranges of the low-$N$ lines extend below zero. Thus, although higher-excitation C$_2$H emission generally strengthens with increasing \ffuv, the low-excitation flux can be insensitive to, or even decrease with, the external field for some disk structures. Its \rhofuv\ values are lower still, increasing from approximately zero at $N=1-0$ to ${\sim}0.4$ at $N=7-6$, showing that variations in the underlying stellar and disk properties further dilute this intrinsic trend. Its abundance morphology closely resembles that of atomic C, forming a thin shell near the disk surface that shifts outwards into the photoevaporative wind as \ffuv\ increases. In our models, C$_2$H forms primarily through the dissociative recombination of C$_2$H$_2^+$ and C$_2$H$_3^+$. The production of these parent ions is promoted by FUV-driven ionisation and photochemistry within the hydrocarbon network. By contrast, C$_2$H destruction is dominated almost entirely by photodissociation. The C$_2$H-rich shell is therefore bounded on the inside by the availability of the hydrocarbon ions required for its formation, and on its outside by rapid photodestruction. As \ffuv\ increases, the shell moves outwards and reaches its greatest spatial extent at \ffuv$=10^4$ G$_0$. At \ffuv$=10^5$\,G$_0$, it contracts slightly as the increased photodissociation rate begins to outpace its formation pathways.

\textbf{N$_2$H$^+$} shows the weakest response of any species considered here. Its median \rhopartial\ rises from $\sim$0.15 at $J=1-0$ to $\sim$0.8 at $J=5-4$, but the central quartile ranges of the low-$J$ lines extend well below zero, indicating that for an appreciable fraction of disk structures the flux is insensitive to, or even anti-correlated with, the external field. Its \rhofuv\ values are close to zero across the entire ladder ($\lesssim 0.2$). This behaviour follows from the tight coupling of N$_2$H$^+$ to CO. It forms through H$_3^+$ + N$_2$ $\rightarrow$ N$_2$H$^+$ + H$_2$ and is destroyed rapidly by CO + N$_2$H$^+$ $\rightarrow$ HCO$^+$ + N$_2$, so it survives only where CO is photodissociated while N$_2$ and H$_3^+$ remain available. This confines the species to a narrow layer near the disk--wind interface, which expands only modestly with increasing $F_{\rm FUV}$, while within the disk the FUV-driven desorption of CO ice described above further suppresses the reservoir. The emission is typically the faintest of any species considered here, with median fluxes that fall below our detectability threshold across all lines, although the brightest models exceed it for $J=2-1$ and above (Figure~\ref{fig_line_fluxes_all}). This contrasts with the growing number of N$_2$H$^+$ detections in isolated disks. However, these detections come predominantly from systems with extensive cold outer regions, where CO is frozen out beyond the snowline and no longer destroys N$_2$H$^+$ \citep[e.g.][]{qi_2013, qi_2015, vant_hoff_2017, qi_2019, cataldi_2021_maps10}. Such conditions are largely absent from our models. The disks in our grid are mostly compact and so contain little cold gas beyond the CO snowline, while the wind-driven heating described above desorbs much of the CO ice that remains. The CO-depleted reservoir in which N$_2$H$^+$ is normally abundant is therefore small or absent, and the N$_2$H$^+$ abundance is correspondingly suppressed. We note, however, that N$_2$H$^+$ is also among the species most sensitive to the adopted chemical network, as discussed in Section~\ref{sec:caveats}.

% SUBSECTION: ALMA GUIDANCE 
\subsection{Practical selection of line-ratio diagnostics for ALMA}
\label{sec:disc_alma_guidance}

The ratios listed in Table \ref{table:ratio_shortlist} respond strongly and consistently to the external FUV field across our model grid, but no single ratio is optimal for every target. These ratios have two main observational applications. In surveys of unresolved disks, elevated wind-sensitive ratios can identify systems whose molecular emission is strongly affected by external irradiation. For systems already known or suspected to be externally irradiated, comparison with the model grid can constrain $F_\mathrm{FUV}$, with the precision depending on how well the stellar and disk properties are known. The practical choice of ratio must additionally account for whether the two transitions can be observed in the same spectral setup, how observational and model uncertainties affect the inferred $F_\mathrm{FUV}$, and whether either line is likely to be contaminated by emission from the surrounding environment.

The shortlisted ratios fall broadly into two classes. The first comprises excitation-sensitive ratios in which a high-$E_{\rm u}$ transition is divided by a low-$E_{\rm u}$ transition of the same molecule or of a species with a similar spatial distribution. Examples include CO $J=8-7$ / CO $J=1-0$ and HCO$^+$ $J=7-6$ / CO $J=1-0$. At low and moderate $F_{\rm FUV}$, the low-excitation line receives a substantial contribution from the cooler disk, whereas the high-excitation line preferentially traces the warm photoevaporative wind. At higher $F_{\rm FUV}$, both lines arise largely in the wind, but the high-excitation line is sub-thermally excited at typical wind densities and therefore brightens more steeply as the wind becomes denser and warmer. Their similar dependence on disk size, stellar mass, and surface density produces high \rhopartial\ and, in many cases, detection fractions close to unity. Their main disadvantage is that the large excitation contrast generally corresponds to a large frequency separation, often requiring observations in widely separated receiver bands.

The second class exploits differences in chemical and spatial response rather than excitation alone. The clearest example is HCO$^+$ $J=3-2$ / C$_2$H $N=3-2$. As the external field increases, HCO$^+$ develops a substantially more extended wind contribution, whereas the low-excitation C$_2$H emission responds much more weakly to $F_{\rm FUV}$  (Figure \ref{fig_abundance_maps}). The same principle applies to ratios in which H$_2$CO or a low-frequency CN transition forms the denominator. These ratios can provide large dynamic ranges without requiring a large separation in excitation energy, although their detection fractions are often lower because the more weakly responding denominator is comparatively faint.

For diagnostics with broadly similar predictive power, the possibility of observing both transitions in a single setup can determine the preferred choice. Of the 135 ratios in Table \ref{table:ratio_shortlist}, 13 combine transitions within the same ALMA receiver band. Whether such a pair can be measured simultaneously is determined by the band-dependent intermediate frequency range and the placement of the spectral windows. Using the current ALMA frequency coverage, five of the shortlisted ratios can be accommodated within a single tuning: HCO$^+$/HCN for the matched $J=2$--$1$, $3$--$2$, $4$--$3$, $5$--$4$, and $7$--$6$ transitions (corresponding to Bands 5, 6, 7, 8 and 9). These ratios have \rhopartial\ $\simeq0.81$--$0.84$, $\eta_{\rm FUV}\simeq9.3$--$15.8$, and $f_{\rm det}\simeq0.91$--$1.00$. They therefore provide somewhat less dynamic range than the strongest cross-band diagnostics, but offer a strong combination of detectability and observational efficiency.

Observing both lines in a single tuning also improves the relative calibration of the ratio, since uncertainties on the absolute flux and atmospheric transmission are largely common to both transitions. Ratios constructed from separate observations retain the calibration uncertainty associated with each band and may also differ in weather, elevation, synthesised beam, and $uv$ coverage. Adopting typical flux uncertainties of $5$\% in Bands~1--5, $10$\% in Bands~6--8, and $20$\% in Bands~9--10 \citep{ALMATechnicalHandbook2026}, and treating the two measurements as independent, gives calibration uncertainty contributions of approximately $11$\% for a B3/B6 ratio, $21$\% for a B3/B10 ratio, and up to $28$\% when both observations use Bands~9--10. 

The effect of this uncertainty on the inferred external field depends on the dynamic range of the diagnostic. Ratios with large enhancement factors are relatively robust to observational uncertainty, whereas ratios with more limited dynamic ranges provide weaker constraints. For example, a 20\% uncertainty in the measured ratio corresponds to an uncertainty of roughly a factor of 1.5 in $F_{\rm FUV}$ for the strongest diagnostics, but can increase to a factor of $2$--$3$ for ratios with smaller enhancement factors ($\eta_{\rm FUV}=3$--$5$). These values are only approximate, because the relation between the ratio and $F_{\rm FUV}$ is not necessarily uniform across the grid, and residual dependence on $M_{*}$, $r_{\rm d}$, and $\Sigma_{\rm 1au}$ introduces additional scatter. The choice of line ratio should therefore account for how well the relevant disk and stellar properties of the target are constrained.

Ratios involving Bands~9 and 10 are subject to additional complications. The low precipitable water vapour (PWV) and stable atmospheric phase required at these frequencies occur during only a limited fraction of observing time, with their availability depending on season, time of day, target elevation, and array configuration \citep{ALMATechnicalHandbook2026}. However, this cost is partly offset by the density of useful transitions at high frequency. For example, the CN $N=7-6$, CO $J=7-6$ and [CI] $^3P_2-^3P_1$ transitions span only $16$\,GHz and can be captured across the two sidebands of a single Band 10 tuning. Where a high-frequency observation is undertaken, multiple transitions can be included within a single spectral setup at no additional cost.

Since nearly all of the shortlisted ratios already require two tunings, those tunings can be chosen to deliver several diagnostics rather than one. The numerators in Table \ref{table:ratio_shortlist} are drawn from a small number of FUV-responsive species, whereas the denominators are predominantly low-excitation lines of CO, CN, HCN and C$_2$H. A single tuning can therefore collect several numerators, a second several denominators, and every cross-pairing between them becomes available for the cost of two observations. For example, pairing Bands 7 and 3 allows CO $J=3-2$ and HCO$^+$ $J=4-3$ to be placed in opposite sidebands of a single Band 7 tuning, while CN $N=1-0$ and CO $J=1-0$ share a single sideband in Band 3, yielding three shortlisted ratios. Similar pairings are possible across other bands. The number of lines that can be captured at once is ultimately limited by the sideband separation, but this restriction can be lifted using spectral scan mode, which covers a continuous frequency range with several tunings rather than a single LO setting \citep{ALMATechnicalHandbook2026}. Band 7 is the most line-rich for our purposes, containing five relevant transitions within $17$\,GHz. A scan across this region captures all five, yielding two additional shortlisted ratios that would otherwise require separate tunings.

A final consideration relates to the wider environments in which externally irradiated disks are typically found. For example, many benchmark systems lie within the Orion Nebula Cluster \citep{odell_wen_1994,bally_2000,mann_2014}, where spatially extended molecular emission can contaminate disk spectra or produce absorption against a bright cloud background \citep{bally_2015,boyden_eisner_2020}. Comparable complications are likely in other irradiated regions, such as NGC~2024 and NGC~1977, where disks are likewise found in dense molecular environments or associated with externally driven ionised flows \citep[e.g.][]{graf_2012, kim_2016, haworth_2021}. Low-$J$ CO transitions are especially vulnerable because the ambient cloud is bright at similar excitation temperatures and velocities \citep{boyden_eisner_2023}. Free-free continuum from the ionised photoevaporative envelope can also contribute substantially at long millimetre wavelengths, reducing molecular-line contrast and complicating continuum subtraction \citep{ballering_2023, boyden_eisner_2024}. Chemistry-based ratios that avoid low-$J$ CO are therefore preferable in strongly nebulous regions.

We note that the criteria adopted in Section \ref{sec:results_ratios} exclude a further 28 ratios that satisfy the correlation and enhancement thresholds, but have $f_{\rm det}<0.5$. All contain C$_2$H in the denominator, paired with CO, HCO$^+$, or [CI] in the numerator. These ratios remain strongly correlated with the external field, with \rhopartial\ $=0.80$--$0.92$, and span substantial dynamic ranges of $\eta_\text{FUV}=8.8$--$79.0$, but are limited by the faintness of the low-excitation C$_2$H lines. Ratios using C$_2$H $N=1$--$0$ provide the largest contrasts, including HCO$^+$ $J=10$--$9$ / C$_2$H $N=1$--$0$ with $\eta_\text{FUV}=79.0$, but both transitions are detectable in only $\sim8$\% of the model grid. Using C$_2$H $N=2$--$1$ instead raises the detection fraction to $\simeq0.42$--$0.43$, while retaining enhancement factors of $8.8$--$54.6$. These ratios may therefore remain valuable where sufficiently deep observations are feasible. Conversely, a secure detection of the wind-sensitive numerator together with a non-detection of C$_2$H provides a lower limit on the ratio and may therefore place a useful lower bound on \ffuv.

Finally, we note that the co-observability classification adopted here reflects ALMA's current capabilities. The upcoming Advanced Technology ALMA Correlator, delivered as part of the Wideband Sensitivity Upgrade, will increase the instantaneous bandwidth to $\geq 16$\,GHz per sideband \citep{carpenter_2023_ALMA_WSU}. Many of the ratios that currently require two independent tunings will become accessible within a single spectral setup, decreasing their observational cost, and the balance of recommendations presented here should be revisited accordingly.

% SUBSECTION: CAVEATS AND LIMITATIONS
\subsection{Caveats and limitations}
\label{sec:caveats}

Our models omit external EUV radiation, and therefore do not include an ionisation front. In reality, the ionisation front truncates the photoevaporative flow, and the PDR becomes more compact as the external field increases. Since most of the emission in our strongly irradiated models comes from the wind, this would reduce the emitting volume at high $F_\text{FUV}$ and potentially weaken the trends we identify. The effect should be small over much of our parameter space, where the ionisation front lies far from the disk, but becomes significant for systems in which it sits close to the disk surface \citep[e.g.][]{johnstone_1998}. Our predicted fluxes at the highest field strengths should therefore be treated as upper limits. How much this matters depends on the irradiating star, since the EUV-to-FUV ratio of the incident field rises with effective temperature. Our diagnostics are therefore best suited to regions dominated by cooler massive stars, and complement the optical line ratios of \citet{peake_2025}, which trace the ionised outflow beyond the ionisation front and work best in the extreme environments where our molecular emitting region is most reduced.

The geometry of our models is also simplified. We assume an isotropic external radiation field and a spherically diverging wind, whereas real proplyds are irradiated from a particular direction and develop the familiar cometary morphology \citep[e.g.][]{odell_wen_1994}. Each model also represents a static snapshot of the disk--wind structure. In reality, external photoevaporation progressively truncates the disk, which in turn alters the wind it supplies and hence the wind-dominated emission. Our grid captures this only through the range of disk radii sampled (Section~\ref{sec:degeneracies}), rather than by following the evolution self-consistently. Additionally, all synthetic fluxes presented in the main analysis assume a face-on disk. We assess the sensitivity of our results to disk inclination in Appendix \ref{appendix:inclination}, and find that most of the shortlisted ratios are only weakly affected, though those pairing HCO$^+$ with CO, HCN, or (in a few cases) higher-$N$ CN lose part of their dynamic range when viewed edge-on.

There are also a number of uncertainties associated with the chemical modelling itself. Our models are static, so the chemistry is solved at fixed position and does not follow material as it is advected outwards through the wind. Where the flow timescale is shorter than the relevant chemical timescale, the resulting wind abundances may differ from those predicted here. For example, one-dimensional models that include advection show that molecular ices can survive much farther into the flow than static models predict \citep{keyte_ran_2025}. More generally, thermochemical codes differ from one another in a number of ways, including their chemical networks, heating and cooling prescriptions, treatment of molecular excitation, and can produce noticeably different line fluxes for the same physical structure. Absolute fluxes are therefore the least reliable of our predictions. Ratios are less affected, since these systematic differences act in a similar way on both lines and partially cancel, in the same way that they cancel the dependence on the stellar and disk properties.

This cancellation does not, however, extend to uncertainties that affect only one species, which arise from e.g. our choice of chemical network. Here, we have used a reduced subset of UMIST\,06 \citep{woodall2007}, which includes the principal formation and destruction pathways of the species considered here but omits larger molecules and some reaction channels. Comparisons between chemical networks show that the (sub-)millimetre fluxes of species like HCO$^+$, HCN, and N$_2$H$^+$ in particular can depend on the network adopted, with N$_2$H$^+$ among the most sensitive species to the underlying chemical assumptions \citep[e.g.][]{kamp_2017, wakelam_2019}. C$_2$H is also likely to be affected, since our network lacks some of the larger hydrocarbons that can act as both carbon reservoirs and photodissociative sources of C$_2$H, and its abundance is additionally sensitive to our fixed C/O\,=\,0.5 \citep[e.g.][]{bergin_2016, bosman_2021_maps}. N$_2$H$^+$ does not appear in our shortlist, but HCO$^+$, HCN and C$_2$H do, so the absolute values and detection fractions of the affected ratios should be treated with caution. Our diagnostics, however, are based on relative changes with $F_{\rm FUV}$ computed with the same network across the grid, which are expected to be more robust than absolute fluxes.

Finally, the criteria used to construct the shortlist in Section \ref{sec:results_ratios} are deliberately simple. They identify ratios that respond strongly and consistently to the external field across our grid, but they take no account of the properties of any particular target, nor of what a given observation is intended to measure. As discussed in Section \ref{sec:disc_alma_guidance}, the most appropriate diagnostic depends on the source and the science goal. Table \ref{table:ratio_shortlist} should therefore be treated as a starting point for selecting lines, rather than a definitive ranking.

%%%%%%%%%%%%%%%%%%
% 5. CONCLUSIONS %
%%%%%%%%%%%%%%%%%%
\section{Conclusions}
\label{sec:conclusions}
We have used a large grid of thermochemical models, tailored to the chemistry of externally irradiated protoplanetary disk + wind systems, to identify robust observational diagnostics of external photoevaporation at (sub-)millimetre wavelengths. The grid spans a broad range of stellar masses, disk radii, surface densities, and external FUV field strengths. For each model we synthesised disk-integrated fluxes for 48 transitions of eight species (CO, HCO$^+$, CN, HCN, C$_2$H, N$_2$H$^+$, H$_2$CO, and atomic C), sampling a wide range of excitation energies accessible to ALMA. Our main conclusions are:
\begin{enumerate}
    \item Disk-integrated line fluxes vary significantly across the grid, spanning roughly 13 orders of magnitude between the brightest and faintest values. This variation is driven both by the external FUV field and by the other disk and stellar parameters. For any individual transition the full range typically covers a few orders of magnitude, while the interquartile range spans only $\sim 1-2$ orders of magnitude, indicating that the most extreme fluxes arise from a small subset of models.
    \item We quantify the sensitivity of each transition to the external FUV field using the Spearman rank correlation coefficient, computed in two ways. The grid-wide correlation \rhofuv, evaluated across all models, measures how reliably a line traces the external field in a population of disks with unknown properties. The fixed-structure correlation \rhopartial, evaluated separately for each combination of stellar and disk parameters, measures how the line responds to the external field alone. As expected, \rhopartial\ exceeds \rhofuv\ for every transition, but the size of this gap varies substantially, showing that the underlying FUV response of some species is strongly masked by variations in the other parameters. Both statistics tend to increase with excitation energy, so higher-$E_{\rm u}$ transitions are generally the more sensitive tracers.
    \item HCO$^+$ and CO are the most robust single-line tracers of the external field, with \rhopartial\ $\simeq 1$ and little variation across their rotational ladders. HCN and high-$N$ CN transitions show similarly strong behaviour at moderate-to-high excitation. Atomic carbon ([CI]) and H$_2$CO also respond strongly to the external field when the disk parameters are held fixed, as indicated by their high \rhopartial\ values. However, their low \rhofuv\ values show that this response is readily obscured by variations in other disk properties, making them reliable tracers only when the disk structure is independently constrained. N$_2$H$^+$ and C$_2$H are the least sensitive species, particularly at low excitation, where their line fluxes are largely insensitive to, or may even decrease with, increasing external FUV field strength.
    \item Of the secondary parameters, stellar mass and disk radius exert the strongest influence on the absolute line fluxes. Most transitions correlate negatively with stellar mass ($\rho_{M_*} \sim 0$ to $-0.5$), most strongly at low $E_\text{u}$, because more luminous stars warm a larger fraction of the disk and depopulate the low-$E_\text{u}$ levels that trace cold gas. Disk radius correlates positively with flux ($\rho_{r_\text{d}}$ up to $\sim 0.85$), simply because larger disks present a greater emitting area. Surface density exerts only a weak positive influence ($\rho_{\Sigma_\text{1au}} \sim 0.15$). Because disk radius and stellar mass introduce correlations of comparable strength to the FUV field itself, absolute line fluxes alone do not provide a clean diagnostic of external photoevaporation.
    \item We computed all 2256 pairwise flux ratios and characterised each by three quantities: the fixed-structure correlation with the FUV field (\rhopartial), an enhancement factor $\eta_\text{FUV}$ measuring how strongly the ratio changes between the weakest and strongest fields, and a detection fraction $f_\text{det}$ measuring how often both lines are detectable. Applying cuts on all three yields a shortlist of 135 diagnostic ratios (Table \ref{table:ratio_shortlist}), all correlating positively with the external FUV field. In each, a strongly responding line (typically HCO$^+$, high-$J$ CO, or [CI]) sits over a more weakly responding one. These separate into two classes: those whose numerator and denominator trace the same gas and vary due to excitation conditions, and those whose numerator and denominator trace gas of varying spatial extent due to underlying differences in the chemistry.
    \item Selecting the best ratio for a given target requires practical considerations beyond this ranking. Almost all shortlisted ratios pair transitions in different ALMA bands, so calibration uncertainties do not fully cancel and ratios with large $\eta_\text{FUV}$ are more robust. Bands 9 and 10 demand excellent atmospheric conditions but contain many useful transitions that can be captured in a single setup. More generally, two tunings can be chosen to deliver several diagnostics at once. For example, a Band 7 tuning covering CO $J=3-2$ and HCO$^+$ $J=4-3$, combined with a Band 3 tuning covering CN $N=1-0$ and CO $J=1-0$, yields three shortlisted ratios. The ALMA Wideband Sensitivity Upgrade, which will at least double the instantaneous bandwidth, will bring many more ratios within a single setup, including the line-rich Band 7 window.
\end{enumerate}
Our results show that no single line flux cleanly separates the effects of external irradiation from variations in intrinsic disk properties. Carefully chosen flux ratios, however, trace the increasing contribution of the warm photoevaporative wind across a wide range of disk properties. They can therefore be used to identify unresolved disks whose molecular emission is strongly shaped by external irradiation and, when combined with constraints on the stellar and disk properties, to place limits on $F_\mathrm{FUV}$. The shortlist presented here offers a practical starting point for applying this approach with ALMA in irradiated star-forming regions, particularly where individual disks cannot be spatially resolved. It also follows that, in strongly irradiated regions, unresolved molecular emission cannot be assumed to trace the disk alone, and models that neglect the wind risk drawing the wrong conclusions about the underlying disk.

%%%%%%%%%%%%%%%%%%%%
% ACKNOWLEDGEMENTS %
%%%%%%%%%%%%%%%%%%%%
\section*{Acknowledgements}

LK is funded by UKRI guaranteed funding for a Horizon Europe ERC consolidator grant (EP/Y024710/1). TJH acknowledges UKRI guaranteed funding for a Horizon Europe ERC consolidator grant (EP/Y024710/1) and a Royal Society Dorothy Hodgkin Fellowship. TP is funded as part of the Dorothy Hodgkin Fellowship of TJH. AR has received funding from the Royal Society through a University Research Fellowship grant number URF\textbackslash R1\textbackslash 241791.

%%%%%%%%%%%%%%%%%%%%%%%%%%%%%%%%%%%%%%%%%%%%%%%%%%
\section*{Data Availability}

This work makes use of the open-source Python package \textsc{puffin}, available at \url{https://github.com/lukekeyte/PUFFIN/}. The thermochemical models were computed with \textsc{dali}, which is not publicly available, but the model outputs underlying this article are available from the corresponding author upon reasonable request.



\bibliographystyle{mnras}
\bibliography{bibliography} % if your bibtex file is called example.bib

@article{Bruderer2012,
       author = {{Bruderer}, S. and {van Dishoeck}, E.~F. and {Doty}, S.~D. and {Herczeg}, G.~J.},
        title = "{The warm gas atmosphere of the HD 100546 disk seen by Herschel. Evidence of a gas-rich, carbon-poor atmosphere?}",
      journal = {\aap},
         year = 2012,
        month = may,
       volume = {541},
          eid = {A91},
        pages = {A91},
          doi = {10.1051/0004-6361/201118218},
archivePrefix = {arXiv},
       eprint = {1201.4860},
 primaryClass = {astro-ph.SR},
       adsurl = {https://ui.adsabs.harvard.edu/abs/2012A&A...541A..91B}
}

@article{Bruderer2013,
       author = {{Bruderer}, Simon},
        title = "{Survival of molecular gas in cavities of transition disks. I. CO}",
      journal = {\aap},
         year = 2013,
        month = nov,
       volume = {559},
          eid = {A46},
        pages = {A46},
          doi = {10.1051/0004-6361/201321171},
archivePrefix = {arXiv},
       eprint = {1308.2966},
 primaryClass = {astro-ph.SR},
       adsurl = {https://ui.adsabs.harvard.edu/abs/2013A&A...559A..46B}
}

@article{woodall2007,
       author = {{Woodall}, J. and {Ag{\'u}ndez}, M. and {Markwick-Kemper}, A.~J. and {Millar}, T.~J.},
        title = "{The UMIST database for astrochemistry 2006}",
      journal = {\aap},
         year = 2007,
        month = may,
       volume = {466},
       number = {3},
        pages = {1197-1204},
          doi = {10.1051/0004-6361:20064981},
       adsurl = {https://ui.adsabs.harvard.edu/abs/2007A&A...466.1197W}
}

@ARTICLE{visser_2018,
       author = {{Visser}, Ruud and {Bruderer}, Simon and {Cazzoletti}, Paolo and {Facchini}, Stefano and {Heays}, Alan N. and {van Dishoeck}, Ewine F.},
        title = "{Nitrogen isotope fractionation in protoplanetary disks}",
      journal = {\aap},
         year = 2018,
        month = jul,
       volume = {615},
          eid = {A75},
        pages = {A75},
          doi = {10.1051/0004-6361/201731898},
archivePrefix = {arXiv},
       eprint = {1802.02841},
 primaryClass = {astro-ph.SR},
       adsurl = {https://ui.adsabs.harvard.edu/abs/2018A&A...615A..75V}
}

@ARTICLE{mathis_1977,
       author = {{Mathis}, J.~S. and {Rumpl}, W. and {Nordsieck}, K.~H.},
        title = "{The size distribution of interstellar grains.}",
      journal = {\apj},
         year = 1977,
        month = oct,
       volume = {217},
        pages = {425-433},
          doi = {10.1086/155591},
       adsurl = {https://ui.adsabs.harvard.edu/abs/1977ApJ...217..425M}
}

@ARTICLE{ballering_2021,
       author = {{Ballering}, Nicholas P. and {Cleeves}, L. Ilsedore and {Anderson}, Dana E.},
        title = "{Simulating Observations of Ices in Protoplanetary Disks}",
      journal = {\apj},
         year = 2021,
        month = oct,
       volume = {920},
       number = {2},
          eid = {115},
        pages = {115},
          doi = {10.3847/1538-4357/ac17ed},
archivePrefix = {arXiv},
       eprint = {2105.12169},
 primaryClass = {astro-ph.EP},
       adsurl = {https://ui.adsabs.harvard.edu/abs/2021ApJ...920..115B}
}

@ARTICLE{lada_lada_2003,
       author = {{Lada}, Charles J. and {Lada}, Elizabeth A.},
        title = "{Embedded Clusters in Molecular Clouds}",
      journal = {\araa},
         year = 2003,
        month = jan,
       volume = {41},
        pages = {57-115},
          doi = {10.1146/annurev.astro.41.011802.094844},
archivePrefix = {arXiv},
       eprint = {astro-ph/0301540},
 primaryClass = {astro-ph},
       adsurl = {https://ui.adsabs.harvard.edu/abs/2003ARA&A..41...57L}
}

@ARTICLE{winter_haworth_2022,
       author = {{Winter}, Andrew J. and {Haworth}, Thomas J.},
        title = "{The external photoevaporation of planet-forming discs}",
      journal = {European Physical Journal Plus},
         year = 2022,
        month = oct,
       volume = {137},
       number = {10},
          eid = {1132},
        pages = {1132},
          doi = {10.1140/epjp/s13360-022-03314-1},
archivePrefix = {arXiv},
       eprint = {2206.11910},
 primaryClass = {astro-ph.EP},
       adsurl = {https://ui.adsabs.harvard.edu/abs/2022EPJP..137.1132W}
}

@ARTICLE{ansdell_2016,
       author = {{Ansdell}, M. and {Williams}, J.~P. and {van der Marel}, N. and {Carpenter}, J.~M. and {Guidi}, G. and {Hogerheijde}, M. and {Mathews}, G.~S. and {Manara}, C.~F. and {Miotello}, A. and {Natta}, A. and {Oliveira}, I. and {Tazzari}, M. and {Testi}, L. and {van Dishoeck}, E.~F. and {van Terwisga}, S.~E.},
        title = "{ALMA Survey of Lupus Protoplanetary Disks. I. Dust and Gas Masses}",
      journal = {\apj},
         year = 2016,
        month = sep,
       volume = {828},
       number = {1},
          eid = {46},
        pages = {46},
          doi = {10.3847/0004-637X/828/1/46},
archivePrefix = {arXiv},
       eprint = {1604.05719},
 primaryClass = {astro-ph.EP},
       adsurl = {https://ui.adsabs.harvard.edu/abs/2016ApJ...828...46A}
}

@ARTICLE{isella_2016,
       author = {{Isella}, Andrea and {Guidi}, Greta and {Testi}, Leonardo and {Liu}, Shangfei and {Li}, Hui and {Li}, Shengtai and {Weaver}, Erik and {Boehler}, Yann and {Carperter}, John M. and {De Gregorio-Monsalvo}, Itziar and {Manara}, Carlo F. and {Natta}, Antonella and {P{\'e}rez}, Laura M. and {Ricci}, Luca and {Sargent}, Anneila and {Tazzari}, Marco and {Turner}, Neal},
        title = "{Ringed Structures of the HD 163296 Protoplanetary Disk Revealed by ALMA}",
      journal = {\prl},
         year = 2016,
        month = dec,
       volume = {117},
       number = {25},
          eid = {251101},
        pages = {251101},
          doi = {10.1103/PhysRevLett.117.251101},
       adsurl = {https://ui.adsabs.harvard.edu/abs/2016PhRvL.117y1101I}
}

@ARTICLE{boyden_eisner_2020,
       author = {{Boyden}, Ryan D. and {Eisner}, Josh A.},
        title = "{Protoplanetary Disks in the Orion Nebula Cluster: Gas-disk Morphologies and Kinematics as Seen with ALMA}",
      journal = {\apj},
         year = 2020,
        month = may,
       volume = {894},
       number = {1},
          eid = {74},
        pages = {74},
          doi = {10.3847/1538-4357/ab86b7},
archivePrefix = {arXiv},
       eprint = {2003.12580},
 primaryClass = {astro-ph.EP},
       adsurl = {https://ui.adsabs.harvard.edu/abs/2020ApJ...894...74B}
}

@ARTICLE{adams_2004,
       author = {{Adams}, Fred C. and {Hollenbach}, David and {Laughlin}, Gregory and {Gorti}, Uma},
        title = "{Photoevaporation of Circumstellar Disks Due to External Far-Ultraviolet Radiation in Stellar Aggregates}",
      journal = {\apj},
         year = 2004,
        month = aug,
       volume = {611},
       number = {1},
        pages = {360-379},
          doi = {10.1086/421989},
archivePrefix = {arXiv},
       eprint = {astro-ph/0404383},
 primaryClass = {astro-ph},
       adsurl = {https://ui.adsabs.harvard.edu/abs/2004ApJ...611..360A}
}

@ARTICLE{facchini_2016,
       author = {{Facchini}, Stefano and {Clarke}, Cathie J. and {Bisbas}, Thomas G.},
        title = "{External photoevaporation of protoplanetary discs in sparse stellar groups: the impact of dust growth}",
      journal = {\mnras},
         year = 2016,
        month = apr,
       volume = {457},
       number = {4},
        pages = {3593-3610},
          doi = {10.1093/mnras/stw240},
archivePrefix = {arXiv},
       eprint = {1601.07562},
 primaryClass = {astro-ph.SR},
       adsurl = {https://ui.adsabs.harvard.edu/abs/2016MNRAS.457.3593F}
}

@ARTICLE{odell_wen_hu_1993,
       author = {{O'Dell}, C.~R. and {Wen}, Zheng and {Hu}, Xihai},
        title = "{Discovery of New Objects in the Orion Nebula on HST Images: Shocks, Compact Sources, and Protoplanetary Disks}",
      journal = {\apj},
         year = 1993,
        month = jun,
       volume = {410},
        pages = {696},
          doi = {10.1086/172786},
       adsurl = {https://ui.adsabs.harvard.edu/abs/1993ApJ...410..696O}
}

@ARTICLE{odell_wen_1994,
       author = {{O'Dell}, C.~R. and {Wen}, Zheng},
        title = "{Postrefurbishment Mission Hubble Space Telescope Images of the Core of the Orion Nebula: Proplyds, Herbig-Haro Objects, and Measurements of a Circumstellar Disk}",
      journal = {\apj},
         year = 1994,
        month = nov,
       volume = {436},
        pages = {194},
          doi = {10.1086/174892},
       adsurl = {https://ui.adsabs.harvard.edu/abs/1994ApJ...436..194O}
}

@ARTICLE{bally_1998b,
       author = {{Bally}, John and {Testi}, Leonardo and {Sargent}, Anneila and {Carlstrom}, John},
        title = "{Disk Mass Limits and Lifetimes of Externally Irradiated Young Stellar Objects Embedded in the Orion Nebula}",
      journal = {\aj},
         year = 1998,
        month = aug,
       volume = {116},
       number = {2},
        pages = {854-859},
          doi = {10.1086/300469},
       adsurl = {https://ui.adsabs.harvard.edu/abs/1998AJ....116..854B}
}

@ARTICLE{bally_2000,
       author = {{Bally}, John and {O'Dell}, C.~R. and {McCaughrean}, Mark J.},
        title = "{Disks, Microjets, Windblown Bubbles, and Outflows in the Orion Nebula}",
      journal = {\aj},
         year = 2000,
        month = jun,
       volume = {119},
       number = {6},
        pages = {2919-2959},
          doi = {10.1086/301385},
       adsurl = {https://ui.adsabs.harvard.edu/abs/2000AJ....119.2919B}
}

@ARTICLE{haworth_2018_fried,
       author = {{Haworth}, Thomas J. and {Clarke}, Cathie J. and {Rahman}, Wahidur and {Winter}, Andrew J. and {Facchini}, Stefano},
        title = "{The FRIED grid of mass-loss rates for externally irradiated protoplanetary discs}",
      journal = {\mnras},
         year = 2018,
        month = nov,
       volume = {481},
       number = {1},
        pages = {452-466},
          doi = {10.1093/mnras/sty2323},
archivePrefix = {arXiv},
       eprint = {1808.07484},
 primaryClass = {astro-ph.SR},
       adsurl = {https://ui.adsabs.harvard.edu/abs/2018MNRAS.481..452H}
}

@ARTICLE{haworth_clarke_2019,
       author = {{Haworth}, Thomas J. and {Clarke}, Cathie J.},
        title = "{The first multidimensional view of mass loss from externally FUV irradiated protoplanetary discs}",
      journal = {\mnras},
         year = 2019,
        month = may,
       volume = {485},
       number = {3},
        pages = {3895-3908},
          doi = {10.1093/mnras/stz706},
archivePrefix = {arXiv},
       eprint = {1903.03644},
 primaryClass = {astro-ph.EP},
       adsurl = {https://ui.adsabs.harvard.edu/abs/2019MNRAS.485.3895H}
}

@ARTICLE{mann_2014,
       author = {{Mann}, Rita K. and {Di Francesco}, James and {Johnstone}, Doug and {Andrews}, Sean M. and {Williams}, Jonathan P. and {Bally}, John and {Ricci}, Luca and {Hughes}, A. Meredith and {Matthews}, Brenda C.},
        title = "{ALMA Observations of the Orion Proplyds}",
      journal = {\apj},
         year = 2014,
        month = mar,
       volume = {784},
       number = {1},
          eid = {82},
        pages = {82},
          doi = {10.1088/0004-637X/784/1/82},
archivePrefix = {arXiv},
       eprint = {1403.2026},
 primaryClass = {astro-ph.SR},
       adsurl = {https://ui.adsabs.harvard.edu/abs/2014ApJ...784...82M}
}

@ARTICLE{eisner_2018,
       author = {{Eisner}, J.~A. and {Arce}, H.~G. and {Ballering}, N.~P. and {Bally}, J. and {Andrews}, S.~M. and {Boyden}, R.~D. and {Di Francesco}, J. and {Fang}, M. and {Johnstone}, D. and {Kim}, J.~S. and {Mann}, R.~K. and {Matthews}, B. and {Pascucci}, I. and {Ricci}, L. and {Sheehan}, P.~D. and {Williams}, J.~P.},
        title = "{Protoplanetary Disk Properties in the Orion Nebula Cluster: Initial Results from Deep, High-resolution ALMA Observations}",
      journal = {\apj},
         year = 2018,
        month = jun,
       volume = {860},
       number = {1},
          eid = {77},
        pages = {77},
          doi = {10.3847/1538-4357/aac3e2},
archivePrefix = {arXiv},
       eprint = {1805.03669},
 primaryClass = {astro-ph.SR},
       adsurl = {https://ui.adsabs.harvard.edu/abs/2018ApJ...860...77E}
}

@ARTICLE{van_terwisga_2023,
       author = {{van Terwisga}, S.~E. and {Hacar}, A.},
        title = "{Survey of Orion Disks with ALMA (SODA). II. UV-driven disk mass loss in L1641 and L1647}",
      journal = {\aap},
         year = 2023,
        month = may,
       volume = {673},
          eid = {L2},
        pages = {L2},
          doi = {10.1051/0004-6361/202346135},
archivePrefix = {arXiv},
       eprint = {2304.05777},
 primaryClass = {astro-ph.EP},
       adsurl = {https://ui.adsabs.harvard.edu/abs/2023A&A...673L...2V}
}

@ARTICLE{ballering_2023,
       author = {{Ballering}, Nicholas P. and {Cleeves}, L. Ilsedore and {Haworth}, Thomas J. and {Bally}, John and {Eisner}, Josh A. and {Ginsburg}, Adam and {Boyden}, Ryan D. and {Fang}, Min and {Kim}, Jinyoung Serena},
        title = "{Isolating Dust and Free-Free Emission in ONC Proplyds with ALMA Band 3 Observations}",
      journal = {\apj},
         year = 2023,
        month = sep,
       volume = {954},
       number = {2},
          eid = {127},
        pages = {127},
          doi = {10.3847/1538-4357/ace901},
archivePrefix = {arXiv},
       eprint = {2308.07369},
 primaryClass = {astro-ph.EP},
       adsurl = {https://ui.adsabs.harvard.edu/abs/2023ApJ...954..127B}
}

@ARTICLE{diaz_berrios_2024,
       author = {{D{\'\i}az-Berr{\'\i}os}, Javiera K. and {Guzm{\'a}n}, Viviana V. and {Walsh}, Catherine and {{\"O}berg}, Karin I. and {Cleeves}, L. Ilsedore and {Artur de la Villarmois}, Elizabeth and {Carpenter}, John},
        title = "{Chemistry in Externally FUV-irradiated Disks in the Outskirts of the Orion Nebula Cluster}",
      journal = {\apj},
         year = 2024,
        month = jul,
       volume = {969},
       number = {2},
          eid = {165},
        pages = {165},
          doi = {10.3847/1538-4357/ad4603},
archivePrefix = {arXiv},
       eprint = {2405.00615},
 primaryClass = {astro-ph.EP},
       adsurl = {https://ui.adsabs.harvard.edu/abs/2024ApJ...969..165D}
}

@ARTICLE{goicoechea_2024,
       author = {{Goicoechea}, J.~R. and {Le Bourlot}, J. and {Black}, J.~H. and {Alarc{\'o}n}, F. and {Bergin}, E.~A. and {Bern{\'e}}, O. and {Bron}, E. and {Canin}, A. and {Chapillon}, E. and {Chown}, R. and {Dartois}, E. and {Gerin}, M. and {Habart}, E. and {Haworth}, T.~J. and {Joblin}, C. and {Kannavou}, O. and {Le Petit}, F. and {Onaka}, T. and {Peeters}, E. and {Pety}, J. and {Roueff}, E. and {Sidhu}, A. and {Schroetter}, I. and {Tabone}, B. and {Tielens}, A.~G.~G.~M. and {Trahin}, B. and {Van De Putte}, D. and {Vicente}, S. and {Zannese}, M.},
        title = "{PDRs4All: X. ALMA and JWST detection of neutral carbon in the externally irradiated disk d203-506: Undepleted gas-phase carbon}",
      journal = {\aap},
         year = 2024,
        month = sep,
       volume = {689},
          eid = {L4},
        pages = {L4},
          doi = {10.1051/0004-6361/202450988},
archivePrefix = {arXiv},
       eprint = {2408.06279},
 primaryClass = {astro-ph.GA},
       adsurl = {https://ui.adsabs.harvard.edu/abs/2024A&A...689L...4G}
}

@ARTICLE{walsh_2013,
       author = {{Walsh}, Catherine and {Millar}, T.~J. and {Nomura}, Hideko},
        title = "{Molecular Line Emission from a Protoplanetary Disk Irradiated Externally by a Nearby Massive Star}",
      journal = {\apjl},
         year = 2013,
        month = apr,
       volume = {766},
       number = {2},
          eid = {L23},
        pages = {L23},
          doi = {10.1088/2041-8205/766/2/L23},
archivePrefix = {arXiv},
       eprint = {1303.4903},
 primaryClass = {astro-ph.GA},
       adsurl = {https://ui.adsabs.harvard.edu/abs/2013ApJ...766L..23W}
}

@ARTICLE{ramirez_tannus_2023,
       author = {{Ram{\'\i}rez-Tannus}, Mar{\'\i}a Claudia and {Bik}, Arjan and {Cuijpers}, Lars and {Waters}, Rens and {G{\"o}ppl}, Christiane and {Henning}, Thomas and {Kamp}, Inga and {Preibisch}, Thomas and {Getman}, Konstantin V. and {Chaparro}, Germ{\'a}n and {Cuartas-Restrepo}, Pablo and {de Koter}, Alex and {Feigelson}, Eric D. and {Grant}, Sierra L. and {Haworth}, Thomas J. and {Hern{\'a}ndez}, Sebasti{\'a}n and {Kuhn}, Michael A. and {Perotti}, Giulia and {Povich}, Matthew S. and {Reiter}, Megan and {Roccatagliata}, Veronica and {Sabbi}, Elena and {Tabone}, Beno{\^\i}t and {Winter}, Andrew J. and {McLeod}, Anna F. and {van Boekel}, Roy and {van Terwisga}, Sierk E.},
        title = "{XUE: Molecular Inventory in the Inner Region of an Extremely Irradiated Protoplanetary Disk}",
      journal = {\apjl},
         year = 2023,
        month = dec,
       volume = {958},
       number = {2},
          eid = {L30},
        pages = {L30},
          doi = {10.3847/2041-8213/ad03f8},
archivePrefix = {arXiv},
       eprint = {2310.11074},
 primaryClass = {astro-ph.SR},
       adsurl = {https://ui.adsabs.harvard.edu/abs/2023ApJ...958L..30R}
}

@ARTICLE{boyden_eisner_2023,
       author = {{Boyden}, Ryan D. and {Eisner}, Josh A.},
        title = "{Chemical Modeling of Orion Nebula Cluster Disks: Evidence for Massive, Compact Gas Disks with Interstellar Gas-to-dust Ratios}",
      journal = {\apj},
         year = 2023,
        month = apr,
       volume = {947},
       number = {1},
          eid = {7},
        pages = {7},
          doi = {10.3847/1538-4357/acaf77},
archivePrefix = {arXiv},
       eprint = {2212.12325},
 primaryClass = {astro-ph.EP},
       adsurl = {https://ui.adsabs.harvard.edu/abs/2023ApJ...947....7B}
}

@ARTICLE{johnstone_1998,
       author = {{Johnstone}, Doug and {Hollenbach}, David and {Bally}, John},
        title = "{Photoevaporation of Disks and Clumps by Nearby Massive Stars: Application to Disk Destruction in the Orion Nebula}",
      journal = {\apj},
         year = 1998,
        month = may,
       volume = {499},
       number = {2},
        pages = {758-776},
          doi = {10.1086/305658},
       adsurl = {https://ui.adsabs.harvard.edu/abs/1998ApJ...499..758J}
}

@ARTICLE{ballabio_2023,
       author = {{Ballabio}, Giulia and {Haworth}, Thomas J. and {Henney}, W.~J.},
        title = "{[O I] 6300 {\r{A}} emission as a probe of external photoevaporation of protoplanetary discs}",
      journal = {\mnras},
         year = 2023,
        month = feb,
       volume = {518},
       number = {4},
        pages = {5563-5575},
          doi = {10.1093/mnras/stac3467},
archivePrefix = {arXiv},
       eprint = {2211.13064},
 primaryClass = {astro-ph.EP},
       adsurl = {https://ui.adsabs.harvard.edu/abs/2023MNRAS.518.5563B}
}

@ARTICLE{booth_2024a,
       author = {{Booth}, Alice S. and {Leemker}, Margot and {van Dishoeck}, Ewine F. and {Evans}, Lucy and {Ilee}, John D. and {Kama}, Mihkel and {Keyte}, Luke and {Law}, Charles J. and {van der Marel}, Nienke and {Nomura}, Hideko and {Notsu}, Shota and {{\"O}berg}, Karin and {Temmink}, Milou and {Walsh}, Catherine},
        title = "{An ALMA Molecular Inventory of Warm Herbig Ae Disks. I. Molecular Rings, Asymmetries, and Complexity in the HD 100546 Disk}",
      journal = {\aj},
         year = 2024,
        month = apr,
       volume = {167},
       number = {4},
          eid = {164},
        pages = {164},
          doi = {10.3847/1538-3881/ad2700},
archivePrefix = {arXiv},
       eprint = {2402.04001},
 primaryClass = {astro-ph.EP},
       adsurl = {https://ui.adsabs.harvard.edu/abs/2024AJ....167..164B}
}

@ARTICLE{haworth_2023_fried2,
       author = {{Haworth}, Thomas J. and {Coleman}, Gavin A.~L. and {Qiao}, Lin and {Sellek}, Andrew D. and {Askari}, Kanaar},
        title = "{FRIED v2: a new grid of mass-loss rates for externally irradiated protoplanetary discs}",
      journal = {\mnras},
         year = 2023,
        month = dec,
       volume = {526},
       number = {3},
        pages = {4315-4334},
          doi = {10.1093/mnras/stad3054},
archivePrefix = {arXiv},
       eprint = {2310.03097},
 primaryClass = {astro-ph.EP},
       adsurl = {https://ui.adsabs.harvard.edu/abs/2023MNRAS.526.4315H}
}

@ARTICLE{bosman_2021_maps,
       author = {{Bosman}, Arthur D. and {Alarc{\'o}n}, Felipe and {Bergin}, Edwin A. and {Zhang}, Ke and {van't Hoff}, Merel L.~R. and {{\"O}berg}, Karin I. and {Guzm{\'a}n}, Viviana V. and {Walsh}, Catherine and {Aikawa}, Yuri and {Andrews}, Sean M. and {Bergner}, Jennifer B. and {Booth}, Alice S. and {Cataldi}, Gianni and {Cleeves}, L. Ilsedore and {Czekala}, Ian and {Furuya}, Kenji and {Huang}, Jane and {Ilee}, John D. and {Law}, Charles J. and {Le Gal}, Romane and {Liu}, Yao and {Long}, Feng and {Loomis}, Ryan A. and {M{\'e}nard}, Fran{\c{c}}ois and {Nomura}, Hideko and {Qi}, Chunhua and {Schwarz}, Kamber R. and {Teague}, Richard and {Tsukagoshi}, Takashi and {Yamato}, Yoshihide and {Wilner}, David J.},
        title = "{Molecules with ALMA at Planet-forming Scales (MAPS). VII. Substellar O/H and C/H and Superstellar C/O in Planet-feeding Gas}",
      journal = {\apjs},
         year = 2021,
        month = nov,
       volume = {257},
       number = {1},
          eid = {7},
        pages = {7},
          doi = {10.3847/1538-4365/ac1435},
archivePrefix = {arXiv},
       eprint = {2109.06221},
 primaryClass = {astro-ph.EP},
       adsurl = {https://ui.adsabs.harvard.edu/abs/2021ApJS..257....7B}
}

@ARTICLE{schoier_LAMDA_2005,
       author = {{Sch{\"o}ier}, F.~L. and {van der Tak}, F.~F.~S. and {van Dishoeck}, E.~F. and {Black}, J.~H.},
        title = "{An atomic and molecular database for analysis of submillimetre line observations}",
      journal = {\aap},
         year = 2005,
        month = mar,
       volume = {432},
       number = {1},
        pages = {369-379},
          doi = {10.1051/0004-6361:20041729},
archivePrefix = {arXiv},
       eprint = {astro-ph/0411110},
 primaryClass = {astro-ph},
       adsurl = {https://ui.adsabs.harvard.edu/abs/2005A&A...432..369S}
}

@ARTICLE{aru_2024,
       author = {{Aru}, M. -L. and {Mauc{\'o}}, K. and {Manara}, C.~F. and {Haworth}, T.~J. and {Ballering}, N. and {Boyden}, R. and {Campbell-White}, J. and {Facchini}, S. and {Rosotti}, G.~P. and {Winter}, A. and {Miotello}, A. and {McLeod}, A.~F. and {Robberto}, M. and {Petr-Gotzens}, M.~G. and {Ballabio}, G. and {Vicente}, S. and {Ansdell}, M. and {Cleeves}, L.~I.},
        title = "{A tell-tale tracer for externally irradiated protoplanetary disks: Comparing the [C I] 8727 {\r{A}} line and ALMA observations in proplyds}",
      journal = {\aap},
         year = 2024,
        month = dec,
       volume = {692},
          eid = {A137},
        pages = {A137},
          doi = {10.1051/0004-6361/202451737},
archivePrefix = {arXiv},
       eprint = {2410.21018},
 primaryClass = {astro-ph.SR},
       adsurl = {https://ui.adsabs.harvard.edu/abs/2024A&A...692A.137A}
}

@ARTICLE{eker_2018,
       author = {{Eker}, Z. and {Bak{\i}{\c{s}}}, V. and {Bilir}, S. and {Soydugan}, F. and {Steer}, I. and {Soydugan}, E. and {Bak{\i}{\c{s}}}, H. and {Ali{\c{c}}avu{\c{s}}}, F. and {Aslan}, G. and {Alpsoy}, M.},
        title = "{Interrelated main-sequence mass-luminosity, mass-radius, and mass-effective temperature relations}",
      journal = {\mnras},
         year = 2018,
        month = oct,
       volume = {479},
       number = {4},
        pages = {5491-5511},
          doi = {10.1093/mnras/sty1834},
archivePrefix = {arXiv},
       eprint = {1807.02568},
 primaryClass = {astro-ph.SR},
       adsurl = {https://ui.adsabs.harvard.edu/abs/2018MNRAS.479.5491E}
}

@ARTICLE{keyte_haworth_2025,
       author = {{Keyte}, Luke and {Haworth}, Thomas J.},
        title = "{Impact of photoevaporative winds in chemical models of externally irradiated protoplanetary discs}",
      journal = {\mnras},
         year = 2025,
        month = feb,
       volume = {537},
       number = {1},
        pages = {598-616},
          doi = {10.1093/mnras/staf047},
archivePrefix = {arXiv},
       eprint = {2501.05172},
 primaryClass = {astro-ph.EP},
       adsurl = {https://ui.adsabs.harvard.edu/abs/2025MNRAS.537..598K}
}

@ARTICLE{owen_altaf_2021,
       author = {{Owen}, James E. and {Altaf}, Noumahn},
        title = "{A slim disc approach to external photoevaporation of discs}",
      journal = {\mnras},
         year = 2021,
        month = dec,
       volume = {508},
       number = {2},
        pages = {2493-2504},
          doi = {10.1093/mnras/stab2749},
archivePrefix = {arXiv},
       eprint = {2110.01630},
 primaryClass = {astro-ph.EP},
       adsurl = {https://ui.adsabs.harvard.edu/abs/2021MNRAS.508.2493O}
}

@ARTICLE{paine_2025,
       author = {{Paine}, S. and {Haworth}, T.~J. and {Nelson}, R.~P.},
        title = "{A particle-based approach to dust dynamics in external photoevaporative winds}",
      journal = {\mnras},
         year = 2025,
        month = may,
       volume = {539},
       number = {2},
        pages = {1414-1429},
          doi = {10.1093/mnras/staf582},
archivePrefix = {arXiv},
       eprint = {2504.05388},
 primaryClass = {astro-ph.EP},
       adsurl = {https://ui.adsabs.harvard.edu/abs/2025MNRAS.539.1414P}
}

@ARTICLE{dalessio_2006,
       author = {{D'Alessio}, Paola and {Calvet}, Nuria and {Hartmann}, Lee and {Franco-Hern{\'a}ndez}, Ramiro and {Serv{\'\i}n}, Hermelinda},
        title = "{Effects of Dust Growth and Settling in T Tauri Disks}",
      journal = {\apj},
         year = 2006,
        month = feb,
       volume = {638},
       number = {1},
        pages = {314-335},
          doi = {10.1086/498861},
archivePrefix = {arXiv},
       eprint = {astro-ph/0511564},
 primaryClass = {astro-ph},
       adsurl = {https://ui.adsabs.harvard.edu/abs/2006ApJ...638..314D}
}

@ARTICLE{concha-ramirez2019,
       author = {{Concha-Ram{\'\i}rez}, Francisca and {Wilhelm}, Maite J.~C. and {Portegies Zwart}, Simon and {Haworth}, Thomas J.},
        title = "{External photoevaporation of circumstellar discs constrains the time-scale for planet formation}",
      journal = {\mnras},
         year = 2019,
        month = oct,
       volume = {490},
       number = {4},
        pages = {5678-5690},
          doi = {10.1093/mnras/stz2973},
archivePrefix = {arXiv},
       eprint = {1907.03760},
 primaryClass = {astro-ph.EP},
       adsurl = {https://ui.adsabs.harvard.edu/abs/2019MNRAS.490.5678C}
}

@ARTICLE{ansdell_2017,
       author = {{Ansdell}, M. and {Williams}, J.~P. and {Manara}, C.~F. and {Miotello}, A. and {Facchini}, S. and {van der Marel}, N. and {Testi}, L. and {van Dishoeck}, E.~F.},
        title = "{An ALMA Survey of Protoplanetary Disks in the {\ensuremath{\sigma}} Orionis Cluster}",
      journal = {\aj},
         year = 2017,
        month = may,
       volume = {153},
       number = {5},
          eid = {240},
        pages = {240},
          doi = {10.3847/1538-3881/aa69c0},
archivePrefix = {arXiv},
       eprint = {1703.08546},
 primaryClass = {astro-ph.EP},
       adsurl = {https://ui.adsabs.harvard.edu/abs/2017AJ....153..240A}
}

@ARTICLE{gross_cleeves_2025,
       author = {{Gross}, Rachel E. and {Cleeves}, L. Ilsedore},
        title = "{Modeling the Impact of Moderate External UV Irradiation on Disk Chemistry}",
      journal = {\apj},
         year = 2025,
        month = feb,
       volume = {980},
       number = {2},
          eid = {189},
        pages = {189},
          doi = {10.3847/1538-4357/ada939},
archivePrefix = {arXiv},
       eprint = {2501.06156},
 primaryClass = {astro-ph.EP},
       adsurl = {https://ui.adsabs.harvard.edu/abs/2025ApJ...980..189G}
}

@ARTICLE{calahan_2025,
       author = {{Calahan}, Jenny K. and {{\"O}berg}, Karin and {Booth}, Alice},
        title = "{The Impact of External Radiation on the Inner Disk Chemistry of Planet Formation}",
      journal = {\apj},
         year = 2025,
        month = sep,
       volume = {991},
       number = {1},
          eid = {94},
        pages = {94},
          doi = {10.3847/1538-4357/adfa09},
archivePrefix = {arXiv},
       eprint = {2508.06613},
 primaryClass = {astro-ph.EP},
       adsurl = {https://ui.adsabs.harvard.edu/abs/2025ApJ...991...94C}
}

@ARTICLE{ndugu_2024,
       author = {{Ndugu}, N. and {Bitsch}, B. and {Lienert}, J.~L.},
        title = "{How external photoevaporation changes the chemical composition of the inner disc}",
      journal = {\aap},
         year = 2024,
        month = nov,
       volume = {691},
          eid = {A32},
        pages = {A32},
          doi = {10.1051/0004-6361/202451633},
archivePrefix = {arXiv},
       eprint = {2409.07596},
 primaryClass = {astro-ph.EP},
       adsurl = {https://ui.adsabs.harvard.edu/abs/2024A&A...691A..32N}
}

@ARTICLE{frediani_2025,
       author = {{Frediani}, Jenny and {Bik}, Arjan and {Ram{\'\i}rez-Tannus}, Mar{\'\i}a Claudia and {Waters}, Rens and {Getman}, Konstantin V. and {Feigelson}, Eric D. and {Portilla-Revelo}, Bayron and {Tabone}, Beno{\^\i}t and {Haworth}, Thomas J. and {Winter}, Andrew and {Henning}, Thomas and {Perotti}, Giulia and {Brandeker}, Alexis and {Chaparro}, Germ{\'a}n and {Cuartas-Restrepo}, Pablo and {Hern{\'a}ndez A.}, Sebastian and {Kuhn}, Michael A. and {Preibisch}, Thomas and {Roccatagliata}, Veronica and {van Terwisga}, Sierk E. and {Zeidler}, Peter},
        title = "{XUE: The CO$_{2}$-rich terrestrial planet-forming region of an externally irradiated Herbig disk}",
      journal = {\aap},
         year = 2025,
        month = sep,
       volume = {701},
          eid = {A14},
        pages = {A14},
          doi = {10.1051/0004-6361/202555718},
archivePrefix = {arXiv},
       eprint = {2507.13921},
 primaryClass = {astro-ph.EP},
       adsurl = {https://ui.adsabs.harvard.edu/abs/2025A&A...701A..14F}
}

@ARTICLE{peake_2025,
       author = {{Peake}, Tyger and {Haworth}, Thomas J. and {Aru}, Mari-Liis and {Henney}, William J.},
        title = "{Line ratio identification of external photoevaporation}",
      journal = {\mnras},
         year = 2025,
        month = aug,
       volume = {541},
       number = {4},
        pages = {2917-2933},
          doi = {10.1093/mnras/staf1090},
archivePrefix = {arXiv},
       eprint = {2506.19788},
 primaryClass = {astro-ph.EP},
       adsurl = {https://ui.adsabs.harvard.edu/abs/2025MNRAS.541.2917P}
}

@ARTICLE{mesa_delgado_2016,
       author = {{Mesa-Delgado}, A. and {Zapata}, L. and {Henney}, W.~J. and {Puzia}, T.~H. and {Tsamis}, Y.~G.},
        title = "{Protoplanetary Disks in the Hostile Environment of Carina}",
      journal = {\apjl},
         year = 2016,
        month = jul,
       volume = {825},
       number = {1},
          eid = {L16},
        pages = {L16},
          doi = {10.3847/2041-8205/825/1/L16},
archivePrefix = {arXiv},
       eprint = {1605.08809},
 primaryClass = {astro-ph.EP},
       adsurl = {https://ui.adsabs.harvard.edu/abs/2016ApJ...825L..16M}
}

@ARTICLE{schroetter_2025,
       author = {{Schroetter}, Ilane and {Bern{\'e}}, Olivier and {Bron}, Emeric and {Alarcon}, Felipe and {Amiot}, Paul and {Bergin}, Edwin A. and {Boersma}, Christiaan and {Cami}, Jan and {Coleman}, Gavin A.~L. and {Dartois}, Emmanuel and {Fuente}, Asuncion and {Goicoechea}, Javier R. and {Habart}, Emilie and {Haworth}, Thomas J. and {Joblin}, Christine and {Franck}, Le Petit and {Onaka}, Takashi and {Peeters}, Els and {R{\"o}lling}, Markus and {Tielens}, Alexander G.~G.~M. and {Zannese}, Marion},
        title = "{A solar C/O ratio in planet-forming gas at 1 au in a highly irradiated disk}",
      journal = {Nature Astronomy},
         year = 2025,
        month = sep,
       volume = {9},
        pages = {1326-1336},
          doi = {10.1038/s41550-025-02596-6},
archivePrefix = {arXiv},
       eprint = {2505.22314},
 primaryClass = {astro-ph.GA},
       adsurl = {https://ui.adsabs.harvard.edu/abs/2025NatAs...9.1326S}
}

@ARTICLE{habing_1968,
       author = {{Habing}, H.~J.},
        title = "{The interstellar radiation density between 912 A and 2400 A}",
      journal = {\bain},
         year = 1968,
        month = jan,
       volume = {19},
        pages = {421},
       adsurl = {https://ui.adsabs.harvard.edu/abs/1968BAN....19..421H}
}

@ARTICLE{flaischlen_2021,
       author = {{Flaischlen}, S. and {Preibisch}, T. and {Manara}, C.~F. and {Ercolano}, B.},
        title = "{Testing the models of X-ray driven photoevaporation with accreting stars in the Orion Nebula Cluster}",
      journal = {\aap},
         year = 2021,
        month = apr,
       volume = {648},
          eid = {A121},
        pages = {A121},
          doi = {10.1051/0004-6361/202039746},
archivePrefix = {arXiv},
       eprint = {2103.03039},
 primaryClass = {astro-ph.SR},
       adsurl = {https://ui.adsabs.harvard.edu/abs/2021A&A...648A.121F}
}

@ARTICLE{van_terwisga_2020,
       author = {{van Terwisga}, S.~E. and {van Dishoeck}, E.~F. and {Mann}, R.~K. and {Di Francesco}, J. and {van der Marel}, N. and {Meyer}, M. and {Andrews}, S.~M. and {Carpenter}, J. and {Eisner}, J.~A. and {Manara}, C.~F. and {Williams}, J.~P.},
        title = "{Protoplanetary disk masses in NGC 2024: Evidence for two populations}",
      journal = {\aap},
         year = 2020,
        month = aug,
       volume = {640},
          eid = {A27},
        pages = {A27},
          doi = {10.1051/0004-6361/201937403},
archivePrefix = {arXiv},
       eprint = {2004.13551},
 primaryClass = {astro-ph.SR},
       adsurl = {https://ui.adsabs.harvard.edu/abs/2020A&A...640A..27V}
}

@ARTICLE{kim_2016,
       author = {{Kim}, Jinyoung Serena and {Clarke}, Cathie J. and {Fang}, Min and {Facchini}, Stefano},
        title = "{Proplyds Around a B1 Star: 42 Orionis in NGC 1977}",
      journal = {\apjl},
         year = 2016,
        month = jul,
       volume = {826},
       number = {1},
          eid = {L15},
        pages = {L15},
          doi = {10.3847/2041-8205/826/1/L15},
archivePrefix = {arXiv},
       eprint = {1606.08271},
 primaryClass = {astro-ph.SR},
       adsurl = {https://ui.adsabs.harvard.edu/abs/2016ApJ...826L..15K}
}

@ARTICLE{haworth_2021,
       author = {{Haworth}, Thomas J. and {Kim}, Jinyoung S. and {Winter}, Andrew J. and {Hines}, Dean C. and {Clarke}, Cathie J. and {Sellek}, Andrew D. and {Ballabio}, Giulia and {Stapelfeldt}, Karl R.},
        title = "{Proplyds in the flame nebula NGC 2024}",
      journal = {\mnras},
         year = 2021,
        month = mar,
       volume = {501},
       number = {3},
        pages = {3502-3514},
          doi = {10.1093/mnras/staa3918},
archivePrefix = {arXiv},
       eprint = {2012.09166},
 primaryClass = {astro-ph.SR},
       adsurl = {https://ui.adsabs.harvard.edu/abs/2021MNRAS.501.3502H}
}

@ARTICLE{mauco_2025,
       author = {{Mauc{\'o}}, K. and {Manara}, C.~F. and {Bayo}, A. and {Hern{\'a}ndez}, J. and {Campbell-White}, J. and {Calvet}, N. and {Ballabio}, G. and {Aru}, M.~L. and {Alcal{\'a}}, J.~M. and {Ansdell}, M. and {Brice{\~n}o}, C. and {Facchini}, S. and {Haworth}, T.~J. and {McClure}, M. and {Williams}, J.~P.},
        title = "{A new look at disk winds and external photoevaporation in the {\ensuremath{\sigma}}-Orionis cluster}",
      journal = {\aap},
         year = 2025,
        month = jan,
       volume = {693},
          eid = {A87},
        pages = {A87},
          doi = {10.1051/0004-6361/202452386},
archivePrefix = {arXiv},
       eprint = {2411.19741},
 primaryClass = {astro-ph.SR},
       adsurl = {https://ui.adsabs.harvard.edu/abs/2025A&A...693A..87M}
}

@ARTICLE{keyte_ran_2025,
       author = {{Keyte}, Luke and {Ran}, Jason},
        title = "{SIMBA: a PYTHON-based single-point astrochemical solver and analysis tool}",
      journal = {\mnras},
         year = 2025,
        month = nov,
       volume = {543},
       number = {3},
        pages = {2599-2612},
          doi = {10.1093/mnras/staf1619},
archivePrefix = {arXiv},
       eprint = {2509.12298},
 primaryClass = {astro-ph.IM},
       adsurl = {https://ui.adsabs.harvard.edu/abs/2025MNRAS.543.2599K}
}

@ARTICLE{portilla_revelo_2025,
       author = {{Portilla-Revelo}, Bayron and {Getman}, Konstantin V. and {Ram{\'\i}rez-Tannus}, Mar{\'\i}a Claudia and {Haworth}, Thomas J. and {Waters}, Rens and {Bik}, Arjan and {Feigelson}, Eric D. and {Kamp}, Inga and {van Terwisga}, Sierk E. and {Frediani}, Jenny and {Henning}, Thomas and {Winter}, Andrew J. and {Roccatagliata}, Veronica and {Preibisch}, Thomas and {Sabbi}, E. and {Zeidler}, Peter and {Kuhn}, Michael A.},
        title = "{XUE: Thermochemical Modeling Suggests a Compact and Gas-depleted Structure for a Distant, Irradiated Protoplanetary Disk}",
      journal = {\apj},
         year = 2025,
        month = may,
       volume = {985},
       number = {1},
          eid = {72},
        pages = {72},
          doi = {10.3847/1538-4357/adc91d},
archivePrefix = {arXiv},
       eprint = {2504.00841},
 primaryClass = {astro-ph.EP},
       adsurl = {https://ui.adsabs.harvard.edu/abs/2025ApJ...985...72P}
}

@ARTICLE{keyte_haworth_2026,
       author = {{Keyte}, Luke and {Haworth}, Thomas J.},
        title = "{A parametric model for externally irradiated protoplanetary discs with photoevaporative winds}",
      journal = {\mnras},
         year = 2026,
        month = mar,
       volume = {546},
       number = {4},
          eid = {stag270},
        pages = {stag270},
          doi = {10.1093/mnras/stag270},
archivePrefix = {arXiv},
       eprint = {2602.02011},
 primaryClass = {astro-ph.EP},
       adsurl = {https://ui.adsabs.harvard.edu/abs/2026MNRAS.546ag270K}
}

@ARTICLE{yang_2010_lamda_CO,
       author = {{Yang}, Benhui and {Stancil}, P.~C. and {Balakrishnan}, N. and {Forrey}, R.~C.},
        title = "{Rotational Quenching of CO due to H$_{2}$ Collisions}",
      journal = {\apj},
         year = 2010,
        month = aug,
       volume = {718},
       number = {2},
        pages = {1062-1069},
          doi = {10.1088/0004-637X/718/2/1062},
archivePrefix = {arXiv},
       eprint = {1004.3923},
 primaryClass = {astro-ph.SR},
       adsurl = {https://ui.adsabs.harvard.edu/abs/2010ApJ...718.1062Y}
}

@ARTICLE{jankowski_2005_lamda_CO,
       author = {{Jankowski}, Piotr and {Szalewicz}, Krzysztof},
        title = "{A new ab initio interaction energy surface and high-resolution spectra of the H$_{2}$-CO van der Waals complex}",
      journal = {\jcp},
         year = 2005,
        month = sep,
       volume = {123},
       number = {10},
        pages = {104301-104301},
          doi = {10.1063/1.2008216},
       adsurl = {https://ui.adsabs.harvard.edu/abs/2005JChPh.123j4301J}
}

@ARTICLE{launay_1977_lamda_C,
       author = {{Launay}, J.~M. and {Roueff}, E.},
        title = "{Fine structure excitation of carbon and oxygen by atomic hydrogen impact.}",
      journal = {\aap},
         year = 1977,
        month = apr,
       volume = {56},
       number = {1-2},
        pages = {289-292},
       adsurl = {https://ui.adsabs.harvard.edu/abs/1977A&A....56..289L}
}

@ARTICLE{johnson_1987_lamda_C,
       author = {{Johnson}, C.~T. and {Burke}, P.~G. and {Kingston}, A.~E.},
        title = "{Electron scattering from the fine structure levels within the 1s$^{2}$2s$^{2}$2p$^{2}$ $^{3}$P$^{e}$ ground state of C I}",
      journal = {Journal of Physics B Atomic Molecular Physics},
         year = 1987,
        month = jun,
       volume = {20},
       number = {11},
        pages = {2553-2563},
          doi = {10.1088/0022-3700/20/11/022},
       adsurl = {https://ui.adsabs.harvard.edu/abs/1987JPhB...20.2553J}
}

@ARTICLE{roueff_1990_lamda_C,
       author = {{Roueff}, E. and {Le Bourlot}, J.},
        title = "{Excitation of forbidden C I fine structure transitions by protons}",
      journal = {\aap},
         year = 1990,
        month = sep,
       volume = {236},
       number = {2},
        pages = {515-518},
       adsurl = {https://ui.adsabs.harvard.edu/abs/1990A&A...236..515R}
}

@ARTICLE{staemmler_1991_lamda_C,
       author = {{Staemmler}, V. and {Flower}, D.~R.},
        title = "{Excitation of the C(2p$^{2.}$ $^{3}$P$_{j}$) fine structure states in collisions with He(1s$^{2}$ $^{1}$S$_{0}$)}",
      journal = {Journal of Physics B Atomic Molecular Physics},
         year = 1991,
        month = may,
       volume = {24},
       number = {9},
        pages = {2343-2351},
          doi = {10.1088/0953-4075/24/9/013},
       adsurl = {https://ui.adsabs.harvard.edu/abs/1991JPhB...24.2343S}
}

@ARTICLE{schroder_1991_lamda_C,
       author = {{Schroder}, K. and {Staemmler}, V. and {Smith}, M.~D. and {Flower}, D.~R. and {Jaquet}, R.},
        title = "{Excitation of the fine-structure transitions of C in collisions with ortho- and para-H$_{2}$}",
      journal = {Journal of Physics B Atomic Molecular Physics},
         year = 1991,
        month = may,
       volume = {24},
       number = {10},
        pages = {2487-2502},
          doi = {10.1088/0953-4075/24/10/007},
       adsurl = {https://ui.adsabs.harvard.edu/abs/1991JPhB...24.2487S}
}

@ARTICLE{muller_2005_cdms,
       author = {{M{\"u}ller}, Holger S.~P. and {Schl{\"o}der}, Frank and {Stutzki}, J{\"u}rgen and {Winnewisser}, Gisbert},
        title = "{The Cologne Database for Molecular Spectroscopy, CDMS: a useful tool for astronomers and spectroscopists}",
      journal = {Journal of Molecular Structure},
         year = 2005,
        month = may,
       volume = {742},
       number = {1-3},
        pages = {215-227},
          doi = {10.1016/j.molstruc.2005.01.027},
       adsurl = {https://ui.adsabs.harvard.edu/abs/2005JMoSt.742..215M}
}

@ARTICLE{spielfiedel_2012_lamda_C2H,
       author = {{Spielfiedel}, A. and {Feautrier}, N. and {Najar}, F. and {Ben Abdallah}, D. and {Dayou}, F. and {Senent}, M.~L. and {Lique}, F.},
        title = "{Fine and hyperfine excitation of C$_{2}$H by collisions with He at low temperature}",
      journal = {\mnras},
         year = 2012,
        month = apr,
       volume = {421},
       number = {3},
        pages = {1891-1896},
          doi = {10.1111/j.1365-2966.2011.20225.x},
       adsurl = {https://ui.adsabs.harvard.edu/abs/2012MNRAS.421.1891S}
}

@ARTICLE{dumouchel_2010_lamda_HCN,
       author = {{Dumouchel}, F. and {Faure}, A. and {Lique}, F.},
        title = "{The rotational excitation of HCN and HNC by He: temperature dependence of the collisional rate coefficients}",
      journal = {\mnras},
         year = 2010,
        month = aug,
       volume = {406},
       number = {4},
        pages = {2488-2492},
          doi = {10.1111/j.1365-2966.2010.16826.x},
       adsurl = {https://ui.adsabs.harvard.edu/abs/2010MNRAS.406.2488D}
}

@ARTICLE{faure_2007_lamda_HCN,
       author = {{Faure}, Alexandre and {Varambhia}, Hemal N. and {Stoecklin}, Thierry and {Tennyson}, Jonathan},
        title = "{Electron-impact rotational and hyperfine excitation of HCN, HNC, DCN and DNC}",
      journal = {\mnras},
         year = 2007,
        month = dec,
       volume = {382},
       number = {2},
        pages = {840-848},
          doi = {10.1111/j.1365-2966.2007.12416.x},
archivePrefix = {arXiv},
       eprint = {0709.1904},
 primaryClass = {astro-ph},
       adsurl = {https://ui.adsabs.harvard.edu/abs/2007MNRAS.382..840F}
}

@ARTICLE{denis_alpizar_2020_lamda_HCOp,
       author = {{Denis-Alpizar}, Otoniel and {Stoecklin}, Thierry and {Dutrey}, Anne and {Guilloteau}, St{\'e}phane},
        title = "{Rotational relaxation of HCO$^{+}$ and DCO$^{+}$ by collision with H$_{2}$}",
      journal = {\mnras},
         year = 2020,
        month = oct,
       volume = {497},
       number = {4},
        pages = {4276-4281},
          doi = {10.1093/mnras/staa2308},
       adsurl = {https://ui.adsabs.harvard.edu/abs/2020MNRAS.497.4276D}
}

@ARTICLE{daniel_2005_lamda_N2Hp,
       author = {{Daniel}, F. and {Dubernet}, M.-L. and {Meuwly}, M. and {Cernicharo}, J. and {Pagani}, L.},
        title = "{Collisional excitation rate coefficients of N$_{2}$H$^{+}$ by He}",
      journal = {\mnras},
         year = 2005,
        month = nov,
       volume = {363},
       number = {4},
        pages = {1083-1091},
          doi = {10.1111/j.1365-2966.2005.09542.x},
       adsurl = {https://ui.adsabs.harvard.edu/abs/2005MNRAS.363.1083D}
}

@ARTICLE{klisch_1995_lamda_CN,
       author = {{Klisch}, E. and {Klaus}, T. and {Belov}, S.~P. and {Winnewisser}, G. and {Herbst}, E.},
        title = "{Laboratory rotational spectrum of CN in the 1 THz region.}",
      journal = {\aap},
         year = 1995,
        month = dec,
       volume = {304},
        pages = {L5},
       adsurl = {https://ui.adsabs.harvard.edu/abs/1995A&A...304L...5K}
}

@ARTICLE{thomson_1968_lamda_CN,
       author = {{Thomson}, Ritchie and {Dalby}, F.~W.},
        title = "{Experimental determination of the dipole moments of the X($^{2}$$^{+}$) and B($^{2}$$^{+}$) states of the CN molecule}",
      journal = {Canadian Journal of Physics},
         year = 1968,
        month = dec,
       volume = {46},
       number = {24},
        pages = {2815-2819},
          doi = {10.1139/p68-652},
       adsurl = {https://ui.adsabs.harvard.edu/abs/1968CaJPh..46.2815T}
}

@ARTICLE{lique_2010_lamda_CN,
       author = {{Lique}, Fran{\c{c}}ois and {Spielfiedel}, Annie and {Feautrier}, Nicole and {Schneider}, Ioan F. and {K{\l}os}, Jacek and {Alexander}, Millard H.},
        title = "{Rotational excitation of CN(X $^{2}${\ensuremath{\Sigma}}$^{+}$) by He: Theory and comparison with experiments}",
      journal = {\jcp},
         year = 2010,
        month = jan,
       volume = {132},
       number = {2},
        pages = {024303-024303},
          doi = {10.1063/1.3285811},
       adsurl = {https://ui.adsabs.harvard.edu/abs/2010JChPh.132b4303L}
}

@ARTICLE{allison_1971_lamda_CN,
       author = {{Allison}, A.~C. and {Dalgarno}, A.},
        title = "{Rotational Excitation of CN by Electron Impact}",
      journal = {\aap},
         year = 1971,
        month = jul,
       volume = {13},
        pages = {331},
       adsurl = {https://ui.adsabs.harvard.edu/abs/1971A&A....13..331A}
}

@ARTICLE{wiesenfeld_2013_lamda_H2CO,
       author = {{Wiesenfeld}, L. and {Faure}, A.},
        title = "{Rotational quenching of H$_{2}$CO by molecular hydrogen: cross-sections, rates and pressure broadening}",
      journal = {\mnras},
         year = 2013,
        month = jul,
       volume = {432},
       number = {3},
        pages = {2573-2578},
          doi = {10.1093/mnras/stt616},
archivePrefix = {arXiv},
       eprint = {1304.4804},
 primaryClass = {astro-ph.SR},
       adsurl = {https://ui.adsabs.harvard.edu/abs/2013MNRAS.432.2573W}
}

@manual{ALMATechnicalHandbook2026,
  author = {Cortes, P. C. and Asaki, Y. and Cerrigone, L. and
            Kneissl, R. and Messias, H. and Carpenter, J. and
            Kameno, S. and Loomis, R. and Vila-Vilaro, B. and
            Immer, K. and Plunkett, A. and Law, J. and
            Stoehr, F. and Saini, K. and Hales, A.},
  title = {ALMA Technical Handbook},
  organization = {Atacama Large Millimeter/submillimeter Array},
  number = {ALMA Doc. 13.3},
  edition = {Version 1.0},
  year = {2026},
  doi = {10.5281/zenodo.18793803}
}

@ARTICLE{boyden_eisner_2024,
       author = {{Boyden}, Ryan D. and {Eisner}, Josh A.},
        title = "{Constraining Free─Free Emission and Photoevaporative Mass-loss Rates for Known Proplyds and New VLA─identified Candidate Proplyds in NGC 1977}",
      journal = {\apj},
         year = 2024,
        month = jun,
       volume = {967},
       number = {2},
          eid = {103},
        pages = {103},
          doi = {10.3847/1538-4357/ad3cd5},
archivePrefix = {arXiv},
       eprint = {2404.04437},
 primaryClass = {astro-ph.EP},
       adsurl = {https://ui.adsabs.harvard.edu/abs/2024ApJ...967..103B}
}

@INPROCEEDINGS{carpenter_2023_ALMA_WSU,
       author = {{Carpenter}, John and {Brogan}, Crystal and {Iono}, Daisuke and {Mroczkowski}, Tony},
        title = "{The ALMA Wideband Sensitivity Upgrade}",
    booktitle = {Physics and Chemistry of Star Formation: The Dynamical ISM Across Time and Spatial Scales},
         year = 2023,
       editor = {{Ossenkopf-Okada}, V. and {Schaaf}, R. and {Breloy}, I. and {Stutzki}, J.},
        month = feb,
        pages = {304},
          doi = {10.48550/arXiv.2211.00195},
archivePrefix = {arXiv},
       eprint = {2211.00195},
 primaryClass = {astro-ph.IM},
       adsurl = {https://ui.adsabs.harvard.edu/abs/2023pcsf.conf..304C}
}

@ARTICLE{bally_2015,
       author = {{Bally}, John and {Mann}, Rita K. and {Eisner}, Josh and {Andrews}, Sean M. and {Di Francesco}, James and {Hughes}, Meredith and {Johnstone}, Doug and {Matthews}, Brenda and {Ricci}, Luca and {Williams}, Jonathan P.},
        title = "{ALMA Observations of the Largest Proto-Planetary Disk in the Orion Nebula, 114-426: A CO Silhouette}",
      journal = {\apj},
         year = 2015,
        month = jul,
       volume = {808},
       number = {1},
          eid = {69},
        pages = {69},
          doi = {10.1088/0004-637X/808/1/69},
archivePrefix = {arXiv},
       eprint = {1506.03391},
 primaryClass = {astro-ph.GA},
       adsurl = {https://ui.adsabs.harvard.edu/abs/2015ApJ...808...69B}
}

@ARTICLE{graf_2012,
       author = {{Graf}, U.~U. and {Simon}, R. and {Stutzki}, J. and {Colgan}, S.~W.~J. and {Guan}, X. and {G{\"u}sten}, R. and {Hartogh}, P. and {Honingh}, C.~E. and {H{\"u}bers}, H.-W.},
        title = "{[$^{12}$Cii] and [$^{13}$C ii] 158 {\ensuremath{\mu}}m emission from NGC 2024: Large column densities of ionized carbon}",
      journal = {\aap},
         year = 2012,
        month = jun,
       volume = {542},
          eid = {L16},
        pages = {L16},
          doi = {10.1051/0004-6361/201218930},
archivePrefix = {arXiv},
       eprint = {1203.2012},
 primaryClass = {astro-ph.GA},
       adsurl = {https://ui.adsabs.harvard.edu/abs/2012A&A...542L..16G}
}

@ARTICLE{qi_2013,
       author = {{Qi}, Chunhua and {{\"O}berg}, Karin I. and {Wilner}, David J. and {D'Alessio}, Paola and {Bergin}, Edwin and {Andrews}, Sean M. and {Blake}, Geoffrey A. and {Hogerheijde}, Michiel R. and {van Dishoeck}, Ewine F.},
        title = "{Imaging of the CO Snow Line in a Solar Nebula Analog}",
      journal = {Science},
         year = 2013,
        month = aug,
       volume = {341},
       number = {6146},
        pages = {630-632},
          doi = {10.1126/science.1239560},
archivePrefix = {arXiv},
       eprint = {1307.7439},
 primaryClass = {astro-ph.SR},
       adsurl = {https://ui.adsabs.harvard.edu/abs/2013Sci...341..630Q}
}

@ARTICLE{guilloteau_2016,
       author = {{Guilloteau}, S. and {Reboussin}, L. and {Dutrey}, A. and {Chapillon}, E. and {Wakelam}, V. and {Pi{\'e}tu}, V. and {Di Folco}, E. and {Semenov}, D. and {Henning}, Th.},
        title = "{Chemistry in disks. X. The molecular content of protoplanetary disks in Taurus}",
      journal = {\aap},
         year = 2016,
        month = aug,
       volume = {592},
          eid = {A124},
        pages = {A124},
          doi = {10.1051/0004-6361/201527088},
archivePrefix = {arXiv},
       eprint = {1604.05028},
 primaryClass = {astro-ph.EP},
       adsurl = {https://ui.adsabs.harvard.edu/abs/2016A&A...592A.124G}
}

@ARTICLE{oberg_2021_maps_1,
       author = {{{\"O}berg}, Karin I. and {Guzm{\'a}n}, Viviana V. and {Walsh}, Catherine and {Aikawa}, Yuri and {Bergin}, Edwin A. and {Law}, Charles J. and {Loomis}, Ryan A. and {Alarc{\'o}n}, Felipe and {Andrews}, Sean M. and {Bae}, Jaehan and {Bergner}, Jennifer B. and {Boehler}, Yann and {Booth}, Alice S. and {Bosman}, Arthur D. and {Calahan}, Jenny K. and {Cataldi}, Gianni and {Cleeves}, L. Ilsedore and {Czekala}, Ian and {Furuya}, Kenji and {Huang}, Jane and {Ilee}, John D. and {Kurtovic}, Nicolas T. and {Le Gal}, Romane and {Liu}, Yao and {Long}, Feng and {M{\'e}nard}, Fran{\c{c}}ois and {Nomura}, Hideko and {P{\'e}rez}, Laura M. and {Qi}, Chunhua and {Schwarz}, Kamber R. and {Sierra}, Anibal and {Teague}, Richard and {Tsukagoshi}, Takashi and {Yamato}, Yoshihide and {van't Hoff}, Merel L.~R. and {Waggoner}, Abygail R. and {Wilner}, David J. and {Zhang}, Ke},
        title = "{Molecules with ALMA at Planet-forming Scales (MAPS). I. Program Overview and Highlights}",
      journal = {\apjs},
         year = 2021,
        month = nov,
       volume = {257},
       number = {1},
          eid = {1},
        pages = {1},
          doi = {10.3847/1538-4365/ac1432},
archivePrefix = {arXiv},
       eprint = {2109.06268},
 primaryClass = {astro-ph.EP},
       adsurl = {https://ui.adsabs.harvard.edu/abs/2021ApJS..257....1O}
}

@ARTICLE{bergin_2016,
       author = {{Bergin}, Edwin A. and {Du}, Fujun and {Cleeves}, L. Ilsedore and {Blake}, G.~A. and {Schwarz}, K. and {Visser}, R. and {Zhang}, K.},
        title = "{Hydrocarbon Emission Rings in Protoplanetary Disks Induced by Dust Evolution}",
      journal = {\apj},
         year = 2016,
        month = nov,
       volume = {831},
       number = {1},
          eid = {101},
        pages = {101},
          doi = {10.3847/0004-637X/831/1/101},
archivePrefix = {arXiv},
       eprint = {1609.06337},
 primaryClass = {astro-ph.EP},
       adsurl = {https://ui.adsabs.harvard.edu/abs/2016ApJ...831..101B}
}

@ARTICLE{miotello_2017,
       author = {{Miotello}, A. and {van Dishoeck}, E.~F. and {Williams}, J.~P. and {Ansdell}, M. and {Guidi}, G. and {Hogerheijde}, M. and {Manara}, C.~F. and {Tazzari}, M. and {Testi}, L. and {van der Marel}, N. and {van Terwisga}, S.},
        title = "{Lupus disks with faint CO isotopologues: low gas/dust or high carbon depletion?}",
      journal = {\aap},
         year = 2017,
        month = mar,
       volume = {599},
          eid = {A113},
        pages = {A113},
          doi = {10.1051/0004-6361/201629556},
archivePrefix = {arXiv},
       eprint = {1612.01538},
 primaryClass = {astro-ph.SR},
       adsurl = {https://ui.adsabs.harvard.edu/abs/2017A&A...599A.113M}
}

@ARTICLE{berne_2023,
       author = {{Bern{\'e}}, Olivier and {Martin-Drumel}, Marie-Aline and {Schroetter}, Ilane and {Goicoechea}, Javier R. and {Jacovella}, Ugo and {Gans}, B{\'e}renger and {Dartois}, Emmanuel and {Coudert}, Laurent H. and {Bergin}, Edwin and {Alarcon}, Felipe and {Cami}, Jan and {Roueff}, Evelyne and {Black}, John H. and {Asvany}, Oskar and {Habart}, Emilie and {Peeters}, Els and {Canin}, Amelie and {Trahin}, Boris and {Joblin}, Christine and {Schlemmer}, Stephan and {Thorwirth}, Sven and {Cernicharo}, Jose and {Gerin}, Maryvonne and {Tielens}, Alexander and {Zannese}, Marion and {Abergel}, Alain and {Bernard-Salas}, Jeronimo and {Boersma}, Christiaan and {Bron}, Emeric and {Chown}, Ryan and {Cuadrado}, Sara and {Dicken}, Daniel and {Elyajouri}, Meriem and {Fuente}, Asunci{\'o}n and {Gordon}, Karl D. and {Issa}, Lina and {Kannavou}, Olga and {Khan}, Baria and {Lacinbala}, Ozan and {Languignon}, David and {Le Gal}, Romane and {Maragkoudakis}, Alexandros and {Meshaka}, Raphael and {Okada}, Yoko and {Onaka}, Takashi and {Pasquini}, Sofia and {Pound}, Marc W. and {Robberto}, Massimo and {R{\"o}llig}, Markus and {Schefter}, Bethany and {Schirmer}, Thi{\'e}baut and {Sidhu}, Ameek and {Tabone}, Benoit and {Van De Putte}, Dries and {Vicente}, S{\'\i}lvia and {Wolfire}, Mark G.},
        title = "{Formation of the methyl cation by photochemistry in a protoplanetary disk}",
      journal = {\nat},
         year = 2023,
        month = sep,
       volume = {621},
       number = {7977},
        pages = {56-59},
          doi = {10.1038/s41586-023-06307-x},
archivePrefix = {arXiv},
       eprint = {2401.03296},
 primaryClass = {astro-ph.GA},
       adsurl = {https://ui.adsabs.harvard.edu/abs/2023Natur.621...56B}
}

@ARTICLE{woitke_2016,
       author = {{Woitke}, P. and {Min}, M. and {Pinte}, C. and {Thi}, W.-F. and {Kamp}, I. and {Rab}, C. and {Anthonioz}, F. and {Antonellini}, S. and {Baldovin-Saavedra}, C. and {Carmona}, A. and {Dominik}, C. and {Dionatos}, O. and {Greaves}, J. and {G{\"u}del}, M. and {Ilee}, J.~D. and {Liebhart}, A. and {M{\'e}nard}, F. and {Rigon}, L. and {Waters}, L.~B.~F.~M. and {Aresu}, G. and {Meijerink}, R. and {Spaans}, M.},
        title = "{Consistent dust and gas models for protoplanetary disks. I. Disk shape, dust settling, opacities, and PAHs}",
      journal = {\aap},
         year = 2016,
        month = feb,
       volume = {586},
          eid = {A103},
        pages = {A103},
          doi = {10.1051/0004-6361/201526538},
archivePrefix = {arXiv},
       eprint = {1511.03431},
 primaryClass = {astro-ph.EP},
       adsurl = {https://ui.adsabs.harvard.edu/abs/2016A&A...586A.103W}
}

@ARTICLE{kamp_2017,
       author = {{Kamp}, I. and {Thi}, W.-F. and {Woitke}, P. and {Rab}, C. and {Bouma}, S. and {M{\'e}nard}, F.},
        title = "{Consistent dust and gas models for protoplanetary disks. II. Chemical networks and rates}",
      journal = {\aap},
         year = 2017,
        month = nov,
       volume = {607},
          eid = {A41},
        pages = {A41},
          doi = {10.1051/0004-6361/201730388},
archivePrefix = {arXiv},
       eprint = {1707.07281},
 primaryClass = {astro-ph.SR},
       adsurl = {https://ui.adsabs.harvard.edu/abs/2017A&A...607A..41K}
}

@ARTICLE{aresu_2011,
       author = {{Aresu}, G. and {Kamp}, I. and {Meijerink}, R. and {Woitke}, P. and {Thi}, W.-F. and {Spaans}, M.},
        title = "{X-ray impact on the protoplanetary disks around T Tauri stars}",
      journal = {\aap},
         year = 2011,
        month = feb,
       volume = {526},
          eid = {A163},
        pages = {A163},
          doi = {10.1051/0004-6361/201015449},
archivePrefix = {arXiv},
       eprint = {1011.5623},
 primaryClass = {astro-ph.SR},
       adsurl = {https://ui.adsabs.harvard.edu/abs/2011A&A...526A.163A}
}

@ARTICLE{walsh_2015,
       author = {{Walsh}, Catherine and {Nomura}, Hideko and {van Dishoeck}, Ewine},
        title = "{The molecular composition of the planet-forming regions of protoplanetary disks across the luminosity regime}",
      journal = {\aap},
         year = 2015,
        month = oct,
       volume = {582},
          eid = {A88},
        pages = {A88},
          doi = {10.1051/0004-6361/201526751},
archivePrefix = {arXiv},
       eprint = {1507.08544},
 primaryClass = {astro-ph.EP},
       adsurl = {https://ui.adsabs.harvard.edu/abs/2015A&A...582A..88W}
}

@ARTICLE{wakelam_2019,
       author = {{Wakelam}, V. and {Chapillon}, E. and {Dutrey}, A. and {Guilloteau}, S. and {Iqbal}, W. and {Coutens}, A. and {Majumdar}, L.},
        title = "{Protoplanetary discs: sensitivity of the chemical composition to various model parameters}",
      journal = {\mnras},
         year = 2019,
        month = apr,
       volume = {484},
       number = {2},
        pages = {1563-1573},
          doi = {10.1093/mnras/stz068},
archivePrefix = {arXiv},
       eprint = {1901.10191},
 primaryClass = {astro-ph.GA},
       adsurl = {https://ui.adsabs.harvard.edu/abs/2019MNRAS.484.1563W}
}

@ARTICLE{gavino_2021,
       author = {{Gavino}, S. and {Dutrey}, A. and {Wakelam}, V. and {Guilloteau}, S. and {Kobus}, J. and {Wolf}, S. and {Iqbal}, W. and {Di Folco}, E. and {Chapillon}, E. and {Pi{\'e}tu}, V.},
        title = "{Impact of size-dependent grain temperature on gas-grain chemistry in protoplanetary disks: The case of low-mass star disks}",
      journal = {\aap},
         year = 2021,
        month = oct,
       volume = {654},
          eid = {A65},
        pages = {A65},
          doi = {10.1051/0004-6361/202038788},
archivePrefix = {arXiv},
       eprint = {2106.05888},
 primaryClass = {astro-ph.GA},
       adsurl = {https://ui.adsabs.harvard.edu/abs/2021A&A...654A..65G}
}

@ARTICLE{cleeves_2013,
       author = {{Cleeves}, L. Ilsedore and {Adams}, Fred C. and {Bergin}, Edwin A.},
        title = "{Exclusion of Cosmic Rays in Protoplanetary Disks: Stellar and Magnetic Effects}",
      journal = {\apj},
         year = 2013,
        month = jul,
       volume = {772},
       number = {1},
          eid = {5},
        pages = {5},
          doi = {10.1088/0004-637X/772/1/5},
archivePrefix = {arXiv},
       eprint = {1306.0902},
 primaryClass = {astro-ph.SR},
       adsurl = {https://ui.adsabs.harvard.edu/abs/2013ApJ...772....5C}
}

@ARTICLE{agundez_2018,
       author = {{Ag{\'u}ndez}, Marcelino and {Roueff}, Evelyne and {Le Petit}, Franck and {Le Bourlot}, Jacques},
        title = "{The chemistry of disks around T Tauri and Herbig Ae/Be stars}",
      journal = {\aap},
         year = 2018,
        month = aug,
       volume = {616},
          eid = {A19},
        pages = {A19},
          doi = {10.1051/0004-6361/201732518},
archivePrefix = {arXiv},
       eprint = {1803.09450},
 primaryClass = {astro-ph.GA},
       adsurl = {https://ui.adsabs.harvard.edu/abs/2018A&A...616A..19A}
}

@ARTICLE{vasyunin_2011,
       author = {{Vasyunin}, A.~I. and {Wiebe}, D.~S. and {Birnstiel}, T. and {Zhukovska}, S. and {Henning}, T. and {Dullemond}, C.~P.},
        title = "{Impact of Grain Evolution on the Chemical Structure of Protoplanetary Disks}",
      journal = {\apj},
         year = 2011,
        month = feb,
       volume = {727},
       number = {2},
          eid = {76},
        pages = {76},
          doi = {10.1088/0004-637X/727/2/76},
archivePrefix = {arXiv},
       eprint = {1011.4420},
 primaryClass = {astro-ph.GA},
       adsurl = {https://ui.adsabs.harvard.edu/abs/2011ApJ...727...76V}
}

@ARTICLE{bergner_2021_maps_xi,
       author = {{Bergner}, Jennifer B. and {{\"O}berg}, Karin I. and {Guzm{\'a}n}, Viviana V. and {Law}, Charles J. and {Loomis}, Ryan A. and {Cataldi}, Gianni and {Bosman}, Arthur D. and {Aikawa}, Yuri and {Andrews}, Sean M. and {Bergin}, Edwin A. and {Booth}, Alice S. and {Cleeves}, L. Ilsedore and {Czekala}, Ian and {Huang}, Jane and {Ilee}, John D. and {Le Gal}, Romane and {Long}, Feng and {Nomura}, Hideko and {M{\'e}nard}, Fran{\c{c}}ois and {Qi}, Chunhua and {Schwarz}, Kamber R. and {Teague}, Richard and {Tsukagoshi}, Takashi and {Walsh}, Catherine and {Wilner}, David J. and {Yamato}, Yoshihide},
        title = "{Molecules with ALMA at Planet-forming Scales (MAPS). XI. CN and HCN as Tracers of Photochemistry in Disks}",
      journal = {\apjs},
         year = 2021,
        month = nov,
       volume = {257},
       number = {1},
          eid = {11},
        pages = {11},
          doi = {10.3847/1538-4365/ac143a},
archivePrefix = {arXiv},
       eprint = {2109.06694},
 primaryClass = {astro-ph.SR},
       adsurl = {https://ui.adsabs.harvard.edu/abs/2021ApJS..257...11B}
}

@ARTICLE{wolfire_2022,
       author = {{Wolfire}, Mark G. and {Vallini}, Livia and {Chevance}, M{\'e}lanie},
        title = "{Photodissociation and X-Ray-Dominated Regions}",
      journal = {\araa},
         year = 2022,
        month = aug,
       volume = {60},
        pages = {247-318},
          doi = {10.1146/annurev-astro-052920-010254},
archivePrefix = {arXiv},
       eprint = {2202.05867},
 primaryClass = {astro-ph.GA},
       adsurl = {https://ui.adsabs.harvard.edu/abs/2022ARA&A..60..247W}
}

@ARTICLE{kaufman_1999,
       author = {{Kaufman}, Michael J. and {Wolfire}, Mark G. and {Hollenbach}, David J. and {Luhman}, Michael L.},
        title = "{Far-Infrared and Submillimeter Emission from Galactic and Extragalactic Photodissociation Regions}",
      journal = {\apj},
         year = 1999,
        month = dec,
       volume = {527},
       number = {2},
        pages = {795-813},
          doi = {10.1086/308102},
archivePrefix = {arXiv},
       eprint = {astro-ph/9907255},
 primaryClass = {astro-ph},
       adsurl = {https://ui.adsabs.harvard.edu/abs/1999ApJ...527..795K}
}

@ARTICLE{chapillon_2012,
       author = {{Chapillon}, E. and {Guilloteau}, S. and {Dutrey}, A. and {Pi{\'e}tu}, V. and {Gu{\'e}lin}, M.},
        title = "{Chemistry in disks. VI. CN and HCN in protoplanetary disks}",
      journal = {\aap},
         year = 2012,
        month = jan,
       volume = {537},
          eid = {A60},
        pages = {A60},
          doi = {10.1051/0004-6361/201116762},
archivePrefix = {arXiv},
       eprint = {1109.5595},
 primaryClass = {astro-ph.GA},
       adsurl = {https://ui.adsabs.harvard.edu/abs/2012A&A...537A..60C}
}

@ARTICLE{coleman_haworth_2022,
       author = {{Coleman}, Gavin A.~L. and {Haworth}, Thomas J.},
        title = "{Dispersal of protoplanetary discs: how stellar properties and the local environment determine the pathway of evolution}",
      journal = {\mnras},
         year = 2022,
        month = aug,
       volume = {514},
       number = {2},
        pages = {2315-2332},
          doi = {10.1093/mnras/stac1513},
archivePrefix = {arXiv},
       eprint = {2204.02303},
 primaryClass = {astro-ph.EP},
       adsurl = {https://ui.adsabs.harvard.edu/abs/2022MNRAS.514.2315C}
}

@ARTICLE{nicholson_2019,
       author = {{Nicholson}, Rhana B. and {Parker}, Richard J. and {Church}, Ross P. and {Davies}, Melvyn B. and {Fearon}, Niamh M. and {Walton}, Sam R.~J.},
        title = "{Rapid destruction of protoplanetary discs due to external photoevaporation in star-forming regions}",
      journal = {\mnras},
         year = 2019,
        month = jun,
       volume = {485},
       number = {4},
        pages = {4893-4905},
          doi = {10.1093/mnras/stz606},
archivePrefix = {arXiv},
       eprint = {1902.11094},
 primaryClass = {astro-ph.GA},
       adsurl = {https://ui.adsabs.harvard.edu/abs/2019MNRAS.485.4893N}
}

@ARTICLE{winter_2020,
       author = {{Winter}, Andrew J. and {Kruijssen}, J.~M. Diederik and {Chevance}, M{\'e}lanie and {Keller}, Benjamin W. and {Longmore}, Steven N.},
        title = "{Prevalent externally driven protoplanetary disc dispersal as a function of the galactic environment}",
      journal = {\mnras},
         year = 2020,
        month = jan,
       volume = {491},
       number = {1},
        pages = {903-922},
          doi = {10.1093/mnras/stz2747},
archivePrefix = {arXiv},
       eprint = {1907.04602},
 primaryClass = {astro-ph.EP},
       adsurl = {https://ui.adsabs.harvard.edu/abs/2020MNRAS.491..903W}
}

@ARTICLE{long_2017,
       author = {{Long}, Feng and {Herczeg}, Gregory J. and {Pascucci}, Ilaria and {Drabek-Maunder}, Emily and {Mohanty}, Subhanjoy and {Testi}, Leonardo and {Apai}, Daniel and {Hendler}, Nathan and {Henning}, Thomas and {Manara}, Carlo F. and {Mulders}, Gijs D.},
        title = "{An ALMA Survey of CO Isotopologue Emission from Protoplanetary Disks in Chamaeleon I}",
      journal = {\apj},
         year = 2017,
        month = aug,
       volume = {844},
       number = {2},
          eid = {99},
        pages = {99},
          doi = {10.3847/1538-4357/aa78fc},
archivePrefix = {arXiv},
       eprint = {1706.03320},
 primaryClass = {astro-ph.SR},
       adsurl = {https://ui.adsabs.harvard.edu/abs/2017ApJ...844...99L}
}

@ARTICLE{long_2019,
       author = {{Long}, Feng and {Herczeg}, Gregory J. and {Harsono}, Daniel and {Pinilla}, Paola and {Tazzari}, Marco and {Manara}, Carlo F. and {Pascucci}, Ilaria and {Cabrit}, Sylvie and {Nisini}, Brunella and {Johnstone}, Doug and {Edwards}, Suzan and {Salyk}, Colette and {Menard}, Francois and {Lodato}, Giuseppe and {Boehler}, Yann and {Mace}, Gregory N. and {Liu}, Yao and {Mulders}, Gijs D. and {Hendler}, Nathanial and {Ragusa}, Enrico and {Fischer}, William J. and {Banzatti}, Andrea and {Rigliaco}, Elisabetta and {van de Plas}, Gerrit and {Dipierro}, Giovanni and {Gully-Santiago}, Michael and {Lopez-Valdivia}, Ricardo},
        title = "{Compact Disks in a High-resolution ALMA Survey of Dust Structures in the Taurus Molecular Cloud}",
      journal = {\apj},
         year = 2019,
        month = sep,
       volume = {882},
       number = {1},
          eid = {49},
        pages = {49},
          doi = {10.3847/1538-4357/ab2d2d},
archivePrefix = {arXiv},
       eprint = {1906.10809},
 primaryClass = {astro-ph.SR},
       adsurl = {https://ui.adsabs.harvard.edu/abs/2019ApJ...882...49L}
}

@ARTICLE{zhang_2025_agepro_i,
       author = {{Zhang}, Ke and {P{\'e}rez}, Laura M. and {Pascucci}, Ilaria and {Pinilla}, Paola and {Cieza}, Lucas A. and {Carpenter}, John and {Trapman}, Leon and {Deng}, Dingshan and {Agurto-Gangas}, Carolina and {Sierra}, Anibal and {Kurtovic}, Nicol{\'a}s T. and {Ruiz-Rodriguez}, Dary A. and {Vioque}, Miguel and {Miley}, James and {Tabone}, Beno{\^\i}t and {Gonz{\'a}lez-Ruilova}, Camilo and {Anania}, Rossella and {Rosotti}, Giovanni P. and {TorresVillanueva}, Estephani and {Hogerheijde}, Michiel R. and {Schwarz}, Kamber and {Kuznetsova}, Aleksandra},
        title = "{The ALMA Survey of Gas Evolution of PROtoplanetary Disks (AGE-PRO). I. Program Overview and Summary of First Results}",
      journal = {\apj},
         year = 2025,
        month = aug,
       volume = {989},
       number = {1},
          eid = {1},
        pages = {1},
          doi = {10.3847/1538-4357/addebe},
archivePrefix = {arXiv},
       eprint = {2506.10719},
 primaryClass = {astro-ph.EP},
       adsurl = {https://ui.adsabs.harvard.edu/abs/2025ApJ...989....1Z}
}

@ARTICLE{planet_formation_collab_2025,
       author = {{Allen}, Megan and {Anania}, Rossella and {Andersen}, Morten and {Aru}, Mari-Liis and {Ballabio}, Giulia and {Ballering}, Nicholas P. and {Beccari}, Giacomo and {Bern{\'e}}, Olivier and {Bik}, Arjan and {Boyden}, Ryan and {Coleman}, Gavin and {D{\'\i}az-Berrios}, Javiera and {Eatson}, Joseph W. and {Frediani}, Jenny and {Forbrich}, Jan and {Gkimisi}, Katia and {Goicoechea}, Javier R. and {Gupta}, Saumya and {Guarcello}, Mario G. and {Haworth}, Thomas J. and {Henney}, William J. and {Isella}, Andrea and {Itrich}, Dominika and {Keyte}, Luke and {Kim}, Jinyoung Serena and {Kuhn}, Michael and {Le Petit}, Frank and {Luo}, Lilian and {Manara}, Carlo and {Mauco}, Karina and {Meshaka}, Rapha{\"e}l and {Millstone}, Samuel and {Owen}, James E. and {Paine}, S{\'e}bastien and {Parker}, Richard and {Peake}, Tyger and {Peatt}, Megan and {Pinilla}, Paola and {Qiao}, Lin and {Ram{\'\i}rez-Tannus}, Mar{\'\i}a Claudia and {Ramsay}, Suzanne and {Reiter}, Megan and {Rogers}, Ciar{\'a}n and {Rosotti}, Giovanni and {Schroetter}, Ilane and {Sellek}, Andrew and {Testi}, Leonardo and {van Terwisga}, Sierk and {Vicente}, Silvia and {Walsh}, Catherine and {Winter}, Andrew and {Wright}, Nicholas J. and {Zeidler}, Peter},
        title = "{The past, present and future of observations of externally irradiated disks}",
      journal = {The Open Journal of Astrophysics},
         year = 2025,
        month = may,
       volume = {8},
          eid = {54},
        pages = {54},
          doi = {10.33232/001c.137538},
archivePrefix = {arXiv},
       eprint = {2502.12255},
 primaryClass = {astro-ph.SR},
       adsurl = {https://ui.adsabs.harvard.edu/abs/2025OJAp....8E..54A}
}

@ARTICLE{guzman_2015,
       author = {{Guzm{\'a}n}, V.~V. and {{\"O}berg}, K.~I. and {Loomis}, R. and {Qi}, C.},
        title = "{Cyanide Photochemistry and Nitrogen Fractionation in the MWC 480 Disk}",
      journal = {\apj},
         year = 2015,
        month = nov,
       volume = {814},
       number = {1},
          eid = {53},
        pages = {53},
          doi = {10.1088/0004-637X/814/1/53},
archivePrefix = {arXiv},
       eprint = {1511.03313},
 primaryClass = {astro-ph.SR},
       adsurl = {https://ui.adsabs.harvard.edu/abs/2015ApJ...814...53G}
}

@ARTICLE{fuente_1993,
       author = {{Fuente}, A. and {Martin-Pintado}, J. and {Cernicharo}, J. and {Bachiller}, R.},
        title = "{A chemical study of the photodissociation region NGC 7023.}",
      journal = {\aap},
         year = 1993,
        month = sep,
       volume = {276},
        pages = {473-488},
       adsurl = {https://ui.adsabs.harvard.edu/abs/1993A&A...276..473F}
}

@ARTICLE{boger_sternberg_2005,
       author = {{Boger}, Gai I. and {Sternberg}, Amiel},
        title = "{CN and HCN in Dense Interstellar Clouds}",
      journal = {\apj},
         year = 2005,
        month = oct,
       volume = {632},
       number = {1},
        pages = {302-315},
          doi = {10.1086/432864},
archivePrefix = {arXiv},
       eprint = {astro-ph/0506535},
 primaryClass = {astro-ph},
       adsurl = {https://ui.adsabs.harvard.edu/abs/2005ApJ...632..302B}
}

@ARTICLE{fatuzzo_adams_2008,
       author = {{Fatuzzo}, Marco and {Adams}, Fred C.},
        title = "{UV Radiation Fields Produced by Young Embedded Star Clusters}",
      journal = {\apj},
         year = 2008,
        month = mar,
       volume = {675},
       number = {2},
        pages = {1361-1374},
          doi = {10.1086/527469},
archivePrefix = {arXiv},
       eprint = {0712.3487},
 primaryClass = {astro-ph},
       adsurl = {https://ui.adsabs.harvard.edu/abs/2008ApJ...675.1361F}
}

@ARTICLE{champion_2017,
       author = {{Champion}, J. and {Bern{\'e}}, O. and {Vicente}, S. and {Kamp}, I. and {Le Petit}, F. and {Gusdorf}, A. and {Joblin}, C. and {Goicoechea}, J.~R.},
        title = "{Herschel survey and modelling of externally-illuminated photoevaporating protoplanetary disks}",
      journal = {\aap},
         year = 2017,
        month = aug,
       volume = {604},
          eid = {A69},
        pages = {A69},
          doi = {10.1051/0004-6361/201629404},
archivePrefix = {arXiv},
       eprint = {1702.00251},
 primaryClass = {astro-ph.EP},
       adsurl = {https://ui.adsabs.harvard.edu/abs/2017A&A...604A..69C}
}

@ARTICLE{guarcello_2023,
       author = {{Guarcello}, M.~G. and {Drake}, J.~J. and {Wright}, N.~J. and {Albacete-Colombo}, J.~F. and {Clarke}, C. and {Ercolano}, B. and {Flaccomio}, E. and {Kashyap}, V. and {Micela}, G. and {Naylor}, T. and {Schneider}, N. and {Sciortino}, S. and {Vink}, J.~S.},
        title = "{Photoevaporation and Close Encounters: How the Environment around Cygnus OB2 Affects the Evolution of Protoplanetary Disks}",
      journal = {\apjs},
         year = 2023,
        month = nov,
       volume = {269},
       number = {1},
          eid = {13},
        pages = {13},
          doi = {10.3847/1538-4365/acdd67},
       adsurl = {https://ui.adsabs.harvard.edu/abs/2023ApJS..269...13G}
}

@ARTICLE{balog_2007,
       author = {{Balog}, Zoltan and {Muzerolle}, James and {Rieke}, G.~H. and {Su}, Kate Y.~L. and {Young}, Eric T. and {Megeath}, S. Tom},
        title = "{Spitzer/IRAC-MIPS Survey of NGC 2244: Protostellar Disk Survival in the Vicinity of Hot Stars}",
      journal = {\apj},
         year = 2007,
        month = may,
       volume = {660},
       number = {2},
        pages = {1532-1540},
          doi = {10.1086/513311},
archivePrefix = {arXiv},
       eprint = {astro-ph/0701741},
 primaryClass = {astro-ph},
       adsurl = {https://ui.adsabs.harvard.edu/abs/2007ApJ...660.1532B}
}

@ARTICLE{haworth_2020,
       author = {{Haworth}, Thomas J. and {Owen}, James E.},
        title = "{The observational anatomy of externally photoevaporating planet-forming discs - I. Atomic carbon}",
      journal = {\mnras},
         year = 2020,
        month = mar,
       volume = {492},
       number = {4},
        pages = {5030-5040},
          doi = {10.1093/mnras/staa151},
archivePrefix = {arXiv},
       eprint = {2001.05004},
 primaryClass = {astro-ph.SR},
       adsurl = {https://ui.adsabs.harvard.edu/abs/2020MNRAS.492.5030H}
}

@ARTICLE{haworth_2022,
       author = {{Haworth}, Thomas J. and {Kim}, Jinyoung S. and {Qiao}, Lin and {Winter}, Andrew J. and {Williams}, Jonathan P. and {Clarke}, Cathie J. and {Owen}, James E. and {Facchini}, Stefano and {Ansdell}, Megan and {Kama}, Mikhel and {Ballabio}, Giulia},
        title = "{An APEX search for carbon emission from NGC 1977 proplyds}",
      journal = {\mnras},
         year = 2022,
        month = may,
       volume = {512},
       number = {2},
        pages = {2594-2603},
          doi = {10.1093/mnras/stac656},
archivePrefix = {arXiv},
       eprint = {2203.03928},
 primaryClass = {astro-ph.EP},
       adsurl = {https://ui.adsabs.harvard.edu/abs/2022MNRAS.512.2594H}
}

@ARTICLE{frediani_2026,
       author = {{Frediani}, Jenny and {Getman}, Konstantin V. and {Woitke}, Peter and {Portilla-Revelo}, Bayron and {Feigelson}, Eric D. and {Rab}, Christian and {Bik}, Arjan and {Ram{\'\i}rez-Tannus}, Mar{\'\i}a Claudia and {Waters}, Rens and {Henning}, Thomas and {Kamp}, Inga and {Chaparro}, Germ{\'a}n and {Hern{\'a}ndez}, Sebasti{\'a}n and {Rodr{\'\i}guez}, Sim{\'o}n and {Cuartas-Restrepo}, Pablo and {Winter}, Andrew and {Haworth}, Thomas J. and {Preibisch}, Thomas and {van Terwisga}, Sierk E. and {Roccatagliata}, Veronica and {Brandeker}, Alexis},
        title = "{XUE. ProDiMo models of internally and externally irradiated planet-forming disks around 0.3-4.0 solar mass stars (The IRIS project I)}",
      journal = {arXiv e-prints},
         year = 2026,
        month = sep,
          eid = {arXiv:2609.08621},
        pages = {arXiv:2609.08621},
          doi = {10.48550/arXiv.2609.08621},
archivePrefix = {arXiv},
       eprint = {2609.08621},
 primaryClass = {astro-ph.EP},
       adsurl = {https://ui.adsabs.harvard.edu/abs/2026arXiv260908621F}
}

@ARTICLE{france_2014,
       author = {{France}, Kevin and {Schindhelm}, Rebecca and {Bergin}, Edwin A. and {Roueff}, Evelyne and {Abgrall}, Herv{\'e}},
        title = "{High-resolution Ultraviolet Radiation Fields of Classical T Tauri Stars}",
      journal = {\apj},
         year = 2014,
        month = apr,
       volume = {784},
       number = {2},
          eid = {127},
        pages = {127},
          doi = {10.1088/0004-637X/784/2/127},
archivePrefix = {arXiv},
       eprint = {1402.6341},
 primaryClass = {astro-ph.SR},
       adsurl = {https://ui.adsabs.harvard.edu/abs/2014ApJ...784..127F}
}

@ARTICLE{bohlin_1978,
       author = {{Bohlin}, R.~C. and {Savage}, B.~D. and {Drake}, J.~F.},
        title = "{A survey of interstellar H I from Lalpha absorption measurements. II.}",
      journal = {\apj},
         year = 1978,
        month = aug,
       volume = {224},
        pages = {132-142},
          doi = {10.1086/156357},
       adsurl = {https://ui.adsabs.harvard.edu/abs/1978ApJ...224..132B}
}

@ARTICLE{qi_2015,
       author = {{Qi}, Chunhua and {{\"O}berg}, Karin I. and {Andrews}, Sean M. and {Wilner}, David J. and {Bergin}, Edwin A. and {Hughes}, A. Meredith and {Hogerheijde}, Michiel and {D'Alessio}, Paola},
        title = "{Chemical Imaging of the CO Snow Line in the HD 163296 Disk}",
      journal = {\apj},
         year = 2015,
        month = nov,
       volume = {813},
       number = {2},
          eid = {128},
        pages = {128},
          doi = {10.1088/0004-637X/813/2/128},
archivePrefix = {arXiv},
       eprint = {1510.00968},
 primaryClass = {astro-ph.SR},
       adsurl = {https://ui.adsabs.harvard.edu/abs/2015ApJ...813..128Q}
}

@ARTICLE{cataldi_2021_maps10,
       author = {{Cataldi}, Gianni and {Yamato}, Yoshihide and {Aikawa}, Yuri and {Bergner}, Jennifer B. and {Furuya}, Kenji and {Guzm{\'a}n}, Viviana V. and {Huang}, Jane and {Loomis}, Ryan A. and {Qi}, Chunhua and {Andrews}, Sean M. and {Bergin}, Edwin A. and {Booth}, Alice S. and {Bosman}, Arthur D. and {Cleeves}, L. Ilsedore and {Czekala}, Ian and {Ilee}, John D. and {Law}, Charles J. and {Le Gal}, Romane and {Liu}, Yao and {Long}, Feng and {M{\'e}nard}, Fran{\c{c}}ois and {Nomura}, Hideko and {{\"O}berg}, Karin I. and {Schwarz}, Kamber R. and {Teague}, Richard and {Tsukagoshi}, Takashi and {Walsh}, Catherine and {Wilner}, David J. and {Zhang}, Ke},
        title = "{Molecules with ALMA at Planet-forming Scales (MAPS). X. Studying Deuteration at High Angular Resolution toward Protoplanetary Disks}",
      journal = {\apjs},
         year = 2021,
        month = nov,
       volume = {257},
       number = {1},
          eid = {10},
        pages = {10},
          doi = {10.3847/1538-4365/ac143d},
archivePrefix = {arXiv},
       eprint = {2109.06462},
 primaryClass = {astro-ph.SR},
       adsurl = {https://ui.adsabs.harvard.edu/abs/2021ApJS..257...10C}
}

@ARTICLE{qi_2019,
       author = {{Qi}, Chunhua and {{\"O}berg}, Karin I. and {Espaillat}, Catherine C. and {Robinson}, Connor E. and {Andrews}, Sean M. and {Wilner}, David J. and {Blake}, Geoffrey A. and {Bergin}, Edwin A. and {Cleeves}, L. Ilsedore},
        title = "{Probing CO and N$_{2}$ Snow Surfaces in Protoplanetary Disks with N$_{2}$H$^{+}$ Emission}",
      journal = {\apj},
         year = 2019,
        month = sep,
       volume = {882},
       number = {2},
          eid = {160},
        pages = {160},
          doi = {10.3847/1538-4357/ab35d3},
archivePrefix = {arXiv},
       eprint = {1907.10647},
 primaryClass = {astro-ph.SR},
       adsurl = {https://ui.adsabs.harvard.edu/abs/2019ApJ...882..160Q}
}

@ARTICLE{vant_hoff_2017,
       author = {{van 't Hoff}, M.~L.~R. and {Walsh}, C. and {Kama}, M. and {Facchini}, S. and {van Dishoeck}, E.~F.},
        title = "{Robustness of N$_{2}$H$^{+}$ as tracer of the CO snowline}",
      journal = {\aap},
         year = 2017,
        month = mar,
       volume = {599},
          eid = {A101},
        pages = {A101},
          doi = {10.1051/0004-6361/201629452},
archivePrefix = {arXiv},
       eprint = {1610.06788},
 primaryClass = {astro-ph.SR},
       adsurl = {https://ui.adsabs.harvard.edu/abs/2017A&A...599A.101V}
}

@ARTICLE{cleeves_2014,
       author = {{Cleeves}, L. Ilsedore and {Bergin}, Edwin A. and {Adams}, Fred C.},
        title = "{Exclusion of Cosmic Rays in Protoplanetary Disks. II. Chemical Gradients and Observational Signatures}",
      journal = {\apj},
         year = 2014,
        month = oct,
       volume = {794},
       number = {2},
          eid = {123},
        pages = {123},
          doi = {10.1088/0004-637X/794/2/123},
archivePrefix = {arXiv},
       eprint = {1408.2835},
 primaryClass = {astro-ph.SR},
       adsurl = {https://ui.adsabs.harvard.edu/abs/2014ApJ...794..123C}
}

@ARTICLE{cleeves_2015,
       author = {{Cleeves}, L. Ilsedore and {Bergin}, Edwin A. and {Qi}, Chunhua and {Adams}, Fred C. and {{\"O}berg}, Karin I.},
        title = "{Constraining the X-Ray and Cosmic-Ray Ionization Chemistry of the TW Hya Protoplanetary Disk: Evidence for a Sub-interstellar Cosmic-Ray Rate}",
      journal = {\apj},
         year = 2015,
        month = feb,
       volume = {799},
       number = {2},
          eid = {204},
        pages = {204},
          doi = {10.1088/0004-637X/799/2/204},
archivePrefix = {arXiv},
       eprint = {1412.1491},
 primaryClass = {astro-ph.SR},
       adsurl = {https://ui.adsabs.harvard.edu/abs/2015ApJ...799..204C}
}

@article{kendall_1942,
    author = {Kendall, M. G.},
    title = {Partial Rank Correlation},
    journal = {Biometrika},
    volume = {32},
    number = {3-4},
    pages = {277-283},
    year = {1942},
    month = {04},
    issn = {0006-3444},
    doi = {10.1093/biomet/32.3-4.277},
    url = {https://doi.org/10.1093/biomet/32.3-4.277},
    eprint = {https://academic.oup.com/biomet/article-pdf/32/3-4/277/514664/32-3-4-277.pdf},
}

%%%%%%%%%%%%%%%%%%%%%%%%%%%%%%%%%%%%%%%%%%%%%%%%%%
%%%%%%%%%%%%%%%%% APPENDICES %%%%%%%%%%%%%%%%%%%%%
%%%%%%%%%%%%%%%%%%%%%%%%%%%%%%%%%%%%%%%%%%%%%%%%%%

\appendix

\section{Synthetic atomic and molecular transitions}

\begin{table*}
\centering
\caption{Molecular and atomic transitions used in this work. Spectroscopic and collisional rate data were obtained from the Leiden Atomic and Molecular Database \citep[LAMDA;][]{schoier_LAMDA_2005}, which compiles data from the following sources for the species considered here: \citet{thomson_1968_lamda_CN, allison_1971_lamda_CN, launay_1977_lamda_C, johnson_1987_lamda_C, roueff_1990_lamda_C, staemmler_1991_lamda_C, schroder_1991_lamda_C, klisch_1995_lamda_CN, jankowski_2005_lamda_CO, muller_2005_cdms, daniel_2005_lamda_N2Hp, faure_2007_lamda_HCN, dumouchel_2010_lamda_HCN, yang_2010_lamda_CO, lique_2010_lamda_CN, spielfiedel_2012_lamda_C2H, wiesenfeld_2013_lamda_H2CO, denis_alpizar_2020_lamda_HCOp}.}\label{table:line_transitions}
\begin{tabular}{llllll}
\hline
\hline
Molecule & Transition & Freq (GHz) & $E_\text{u}$ (K) & $A_{\mathrm{ul}}$ (s$^{-1}$) & $g_\mathrm{u}$  \\
\hline
$[$CI$]$ & $^3P_1 - ^3P_0$ & 492.161 & 23.6 & $7.88\times10^{-8}$ & 3 \\
$[$CI$]$ & $^3P_2 - ^3P_1$ & 809.342 & 62.5 & $2.65\times10^{-7}$ & 5 \\
C$_2$H & $N=1-0$,\; $J=3/2-1/2$,\; $F=2-1$ & 87.317 & 4.2 & $1.53\times10^{-6}$ & 5 \\
C$_2$H & $N=2-1$,\; $J=5/2-3/2$,\; $F=3-2$ & 174.663 & 12.6 & $1.47\times10^{-5}$ & 7 \\
C$_2$H & $N=3-2$,\; $J=5/2-3/2$,\; $F=3-2$ & 262.065 & 25.2 & $4.89\times10^{-5}$ & 7 \\
C$_2$H & $N=4-3$,\; $J=9/2-7/2$,\; $F=5-4$ & 349.338 & 41.9 & $1.31\times10^{-4}$ & 11 \\
C$_2$H & $N=5-4$,\; $J=11/2-9/2$,\; $F=6-5$ & 436.661 & 62.9 & $2.61\times10^{-4}$ & 13 \\
C$_2$H & $N=7-6$,\; $J=15/2-13/2$,\; $F=8-7$ & 611.267 & 117.4 & $7.36\times10^{-4}$ & 17 \\
CO & $J=1-0$ & 115.271 & 5.5 & $7.15\times10^{-8}$ & 3 \\
CO & $J=2-1$ & 230.538 & 16.6 & $6.86\times10^{-7}$ & 5 \\
CO & $J=3-2$ & 345.796 & 33.2 & $2.48\times10^{-6}$ & 7 \\
CO & $J=4-3$ & 461.041 & 55.3 & $6.08\times10^{-6}$ & 9 \\
CO & $J=6-5$ & 691.473 & 116.2 & $2.12\times10^{-5}$ & 13 \\
CO & $J=7-6$ & 806.652 & 154.9 & $3.39\times10^{-5}$ & 15 \\
CO & $J=8-7$ & 921.800 & 199.1 & $5.09\times10^{-5}$ & 17 \\
CN & $N=1-0$,\; $J=3/2-1/2$ & 113.495 & 5.5 & $1.19\times10^{-5}$ & 4 \\
CN & $N=2-1$,\; $J=5/2-3/2$ & 226.876 & 16.3 & $1.14\times10^{-4}$ & 6 \\
CN & $N=3-2$,\; $J=5/2-3/2$ & 340.031 & 32.6 & $3.85\times10^{-4}$ & 6 \\
CN & $N=4-3$,\; $J=9/2-7/2$ & 453.607 & 54.4 & $1.02\times10^{-3}$ & 10 \\
CN & $N=6-5$,\; $J=11/2-9/2$ & 680.047 & 114.2 & $3.50\times10^{-3}$ & 12 \\
CN & $N=7-6$,\; $J=13/2-11/2$ & 793.336 & 152.3 & $5.64\times10^{-3}$ & 14 \\
o-H$_2$CO & $J_\mathrm{K_a,K_c} = 2_{1,1} - 1_{1,0}$ & 150.498 & 22.6 & $6.47\times10^{-5}$ & 5 \\
o-H$_2$CO & $J_\mathrm{K_a,K_c} = 3_{1,2} - 2_{1,1}$ & 225.698 & 33.4 & $2.77\times10^{-4}$ & 7 \\
o-H$_2$CO & $J_\mathrm{K_a,K_c} = 5_{3,2} - 4_{3,1}$ & 364.289 & 158.4 & $8.90\times10^{-4}$ & 11 \\
o-H$_2$CO & $J_\mathrm{K_a,K_c} = 7_{1,7} - 6_{1,6}$ & 491.968 & 106.3 & $3.44\times10^{-3}$ & 15 \\
o-H$_2$CO & $J_\mathrm{K_a,K_c} = 9_{1,8} - 8_{1,7}$ & 674.810 & 174.0 & $9.09\times10^{-3}$ & 19 \\
o-H$_2$CO & $J_\mathrm{K_a,K_c} = 11_{1,10} - 10_{1,9}$ & 823.083 & 249.4 & $1.67\times10^{-2}$ & 23 \\
HCN & $J=1-0$ & 88.632 & 4.3 & $2.41\times10^{-5}$ & 3 \\
HCN & $J=2-1$ & 177.261 & 12.8 & $2.31\times10^{-4}$ & 5 \\
HCN & $J=3-2$ & 265.886 & 25.5 & $8.36\times10^{-4}$ & 7 \\
HCN & $J=4-3$ & 354.505 & 42.5 & $2.05\times10^{-3}$ & 9 \\
HCN & $J=5-4$ & 443.116 & 63.8 & $4.10\times10^{-3}$ & 11 \\
HCN & $J=7-6$ & 620.304 & 119.1 & $1.16\times10^{-2}$ & 15 \\
HCN & $J=8-7$ & 708.877 & 153.1 & $1.74\times10^{-2}$ & 17 \\
HCN & $J=10-9$ & 885.971 & 233.9 & $3.43\times10^{-2}$ & 21 \\
HCO$^+$ & $J=1-0$ & 89.189 & 4.3 & $4.25\times10^{-5}$ & 3 \\
HCO$^+$ & $J=2-1$ & 178.375 & 12.8 & $4.08\times10^{-4}$ & 5 \\
HCO$^+$ & $J=3-2$ & 267.558 & 25.7 & $1.48\times10^{-3}$ & 7 \\
HCO$^+$ & $J=4-3$ & 356.734 & 42.8 & $3.63\times10^{-3}$ & 9 \\
HCO$^+$ & $J=5-4$ & 445.903 & 64.2 & $7.24\times10^{-3}$ & 11 \\
HCO$^+$ & $J=7-6$ & 624.209 & 119.8 & $2.04\times10^{-2}$ & 15 \\
HCO$^+$ & $J=8-7$ & 713.342 & 154.1 & $3.07\times10^{-2}$ & 17 \\
HCO$^+$ & $J=10-9$ & 891.558 & 235.4 & $6.07\times10^{-2}$ & 21 \\
N$_2$H$^+$ & $J=1-0$,\; $F_1=1-1$,\; $F=0-1$ & 93.172 & 4.5 & $3.90\times10^{-5}$ & 1 \\
N$_2$H$^+$ & $J=2-1$,\; $F_1=3-2$,\; $F=4-3$ & 186.345 & 13.4 & $3.74\times10^{-4}$ & 9 \\
N$_2$H$^+$ & $J=3-2$,\; $F_1=4-3$,\; $F=5-4$ & 279.512 & 26.8 & $1.35\times10^{-3}$ & 11 \\
N$_2$H$^+$ & $J=4-3$,\; $F_1=5-4$,\; $F=6-5$ & 372.673 & 44.7 & $3.32\times10^{-3}$ & 13 \\
N$_2$H$^+$ & $J=5-4$,\; $F_1=6-5$,\; $F=7-6$ & 465.825 & 67.1 & $6.64\times10^{-3}$ & 15 \\

\hline
\end{tabular}
\end{table*}

\section{DALI model parameters}

%------------------------------------------------------------------------%
% TABLE: Initial Abundances
\begin{table}
\caption{Initial abundances for all chemical models presented in Section \ref{sec:chemistry}, adapted from \citet{ballering_2021}. These values represent a typical protoplanetary disk with Solar-like C/O=0.5 and composition inherited from the parent molecular cloud.}        
\label{table:initial_abundances}      
\centering
\begin{tabular}{l l}     
\hline\hline       
Species & Abundance (X/H)  \\ 
\hline                    
H$_2$    & $5.00 \times 10^{-1}$    \\
He       & $9.00 \times 10^{-2}$    \\
H$_2$O   & $1.00 \times 10^{-4}$    \\
CO       & $1.00 \times 10^{-4}$    \\
N$_2$    & $3.51 \times 10^{-5}$    \\
C$^+$    & $1.00 \times 10^{-5}$    \\
NH$_3$   & $4.80 \times 10^{-6}$    \\
HCN      & $2.00 \times 10^{-8}$    \\
H$_3^+$  & $1.00 \times 10^{-8}$    \\
S$^+$    & $9.00 \times 10^{-9}$    \\
HCO$^+$  & $9.00 \times 10^{-9}$    \\
C$_2$H   & $8.00 \times 10^{-9}$    \\
Mg$^+$   & $1.00 \times 10^{-11}$   \\
Si$^+$   & $1.00 \times 10^{-11}$   \\
Fe$^+$   & $1.00 \times 10^{-11}$   \\
\hline                  
\end{tabular}
\end{table}
%------------------------------------------------------------------------%

\section{Flux ratio shortlist}

\onecolumn
\begin{small}

\renewcommand{\arraystretch}{1.2} 

\begin{longtable}{lllll}
\caption{Ranked shortlist of line flux ratio diagnostics of the external FUV field. \rhopartial is the partial Spearman correlation with $F_\mathrm{FUV}$, controlling for $M_*$, $r_\mathrm{d}$, and $\Sigma_\mathrm{1au}$. A ratio and its reciprocal contain identical information, with the sign of \rhopartial reversed, so we only retain ratios for which \rhopartial is positive. $\eta_\mathrm{FUV}$ is the median change from $10^2$ to $10^5\,\mathrm{G}_0$ at fixed structure. $f_\mathrm{det}$ is the fraction of models in which both lines exceed the $3\sigma$ detection threshold defined in Section~\ref{sec:individual}. Note that hyperfine transition labels have been shortened for clarity. Full transition details are presented in Table~\ref{table:line_transitions}.}\label{table:ratio_shortlist}\\
\hline\hline
Line flux ratio & ALMA bands & \rhopartial & $\eta_\text{FUV}$ & $f_{\mathrm{det}}$ \\
\hline
\endfirsthead

\hline\hline
Line flux ratio & ALMA bands & \rhopartial & $\eta_\text{FUV}$ & $f_{\mathrm{det}}$ \\
\hline
\endhead

\hline
\endfoot
$[$CI$]$ $^3$P$_2$-$^3$P$_1$ \;/\; CN N=1–0 & B10\;/\;B3 & 0.91 & 8.6 & 0.80 \\
HCO$^+$ J=7-6 \;/\; CO J=1-0 & B9\;/\;B3 & 0.91 & 11.5 & 1.00 \\
CO J=7-6 \;/\; CO J=1-0 & B10\;/\;B3 & 0.91 & 5.6 & 1.00 \\
CO J=8-7 \;/\; CO J=1-0 & B10\;/\;B3 & 0.91 & 7.2 & 1.00 \\
CO J=8-7 \;/\; HCN J=2-1 & B10\;/\;B5 & 0.91 & 12.6 & 0.99 \\
HCO$^+$ J=4-3 \;/\; CN N=1–0 & B7\;/\;B3 & 0.90 & 46.0 & 0.80 \\
HCO$^+$ J=3-2 \;/\; CN N=1–0 & B6\;/\;B3 & 0.90 & 44.3 & 0.80 \\
HCO$^+$ J=5-4 \;/\; CO J=1-0 & B8\;/\;B3 & 0.90 & 10.1 & 1.00 \\
HCO$^+$ J=2-1 \;/\; CN N=1–0 & B5\;/\;B3 & 0.90 & 42.2 & 0.80 \\
HCO$^+$ J=5-4 \;/\; CO J=2-1 & B8\;/\;B6 & 0.90 & 6.1 & 1.00 \\
$[$CI$]$ $^3$P$_1$-$^3$P$_0$ \;/\; CN N=1–0 & B8\;/\;B3 & 0.90 & 6.3 & 0.80 \\
CO J=8-7 \;/\; HCN J=3-2 & B10\;/\;B6 & 0.90 & 8.7 & 1.00 \\
HCO$^+$ J=1-0 \;/\; CN N=1–0 & B3\;/\;B3 & 0.89 & 30.2 & 0.60 \\
HCO$^+$ J=8-7 \;/\; CO J=1-0 & B9\;/\;B3 & 0.89 & 12.3 & 1.00 \\
HCO$^+$ J=5-4 \;/\; CN N=1–0 & B8\;/\;B3 & 0.89 & 46.5 & 0.80 \\
HCO$^+$ J=4-3 \;/\; CN N=2–1 & B7\;/\;B6 & 0.89 & 26.1 & 0.98 \\
HCO$^+$ J=3-2 \;/\; CN N=2–1 & B6\;/\;B6 & 0.89 & 25.5 & 0.98 \\
CO J=8-7 \;/\; C$_2$H N=3–2 & B10\;/\;B6 & 0.89 & 23.2 & 0.58 \\
CO J=7-6 \;/\; HCN J=2-1 & B10\;/\;B5 & 0.89 & 10.1 & 0.99 \\
HCO$^+$ J=3-2 \;/\; C$_2$H N=3–2 & B6\;/\;B6 & 0.89 & 25.4 & 0.58 \\
HCO$^+$ J=2-1 \;/\; C$_2$H N=3–2 & B5\;/\;B6 & 0.89 & 23.4 & 0.58 \\
HCO$^+$ J=2-1 \;/\; CN N=2–1 & B5\;/\;B6 & 0.89 & 24.6 & 0.91 \\
HCO$^+$ J=5-4 \;/\; CN N=2–1 & B8\;/\;B6 & 0.88 & 27.4 & 0.98 \\
HCO$^+$ J=4-3 \;/\; C$_2$H N=3–2 & B7\;/\;B6 & 0.88 & 25.7 & 0.58 \\
CO J=7-6 \;/\; C$_2$H N=3–2 & B10\;/\;B6 & 0.88 & 18.3 & 0.58 \\
HCO$^+$ J=7-6 \;/\; CO J=2-1 & B9\;/\;B6 & 0.88 & 6.4 & 1.00 \\
CO J=8-7 \;/\; CN N=1–0 & B10\;/\;B3 & 0.88 & 43.9 & 0.80 \\
HCO$^+$ J=4-3 \;/\; CO J=1-0 & B7\;/\;B3 & 0.88 & 9.6 & 1.00 \\
HCO$^+$ J=1-0 \;/\; CN N=2–1 & B3\;/\;B6 & 0.87 & 17.9 & 0.63 \\
HCO$^+$ J=4-3 \;/\; CO J=2-1 & B7\;/\;B6 & 0.87 & 5.8 & 1.00 \\
CO J=7-6 \;/\; HCN J=3-2 & B10\;/\;B6 & 0.87 & 6.8 & 1.00 \\
HCO$^+$ J=1-0 \;/\; C$_2$H N=3–2 & B3\;/\;B6 & 0.87 & 18.7 & 0.53 \\
CO J=7-6 \;/\; CN N=1–0 & B10\;/\;B3 & 0.87 & 32.6 & 0.80 \\
CO J=8-7 \;/\; CN N=2–1 & B10\;/\;B6 & 0.87 & 24.9 & 0.98 \\
CO J=6-5 \;/\; HCN J=2-1 & B9\;/\;B5 & 0.87 & 8.2 & 0.99 \\
HCO$^+$ J=7-6 \;/\; CN N=1–0 & B9\;/\;B3 & 0.87 & 59.3 & 0.80 \\
HCO$^+$ J=3-2 \;/\; C$_2$H N=4–3 & B6\;/\;B7 & 0.87 & 16.8 & 0.65 \\
CO J=6-5 \;/\; C$_2$H N=3–2 & B9\;/\;B6 & 0.86 & 14.5 & 0.58 \\
HCO$^+$ J=5-4 \;/\; C$_2$H N=3–2 & B8\;/\;B6 & 0.86 & 26.4 & 0.58 \\
HCO$^+$ J=2-1 \;/\; C$_2$H N=4–3 & B5\;/\;B7 & 0.86 & 16.0 & 0.65 \\
CO J=6-5 \;/\; CN N=1–0 & B9\;/\;B3 & 0.86 & 25.1 & 0.80 \\
CO J=8-7 \;/\; HCN J=1-0 & B10\;/\;B3 & 0.86 & 22.5 & 0.87 \\
HCO$^+$ J=4-3 \;/\; CN N=3–2 & B7\;/\;B7 & 0.86 & 13.3 & 0.99 \\
CO J=8-7 \;/\; HCN J=4-3 & B10\;/\;B7 & 0.86 & 7.0 & 1.00 \\
CO J=7-6 \;/\; CN N=2–1 & B10\;/\;B6 & 0.86 & 19.0 & 0.98 \\
HCO$^+$ J=3-2 \;/\; CN N=3–2 & B6\;/\;B7 & 0.86 & 13.0 & 0.99 \\
HCO$^+$ J=7-6 \;/\; HCN J=3-2 & B9\;/\;B6 & 0.86 & 12.4 & 1.00 \\
$[$CI$]$ $^3$P$_2$-$^3$P$_1$ \;/\; CN N=2–1 & B10\;/\;B6 & 0.86 & 5.5 & 0.88 \\
CO J=8-7 \;/\; C$_2$H N=4–3 & B10\;/\;B7 & 0.86 & 13.5 & 0.65 \\
HCO$^+$ J=5-4 \;/\; CN N=3–2 & B8\;/\;B7 & 0.86 & 13.9 & 0.99 \\
HCO$^+$ J=4-3 \;/\; C$_2$H N=4–3 & B7\;/\;B7 & 0.85 & 17.1 & 0.65 \\
HCO$^+$ J=8-7 \;/\; HCN J=3-2 & B9\;/\;B6 & 0.85 & 13.4 & 1.00 \\
HCO$^+$ J=7-6 \;/\; CN N=2–1 & B9\;/\;B6 & 0.85 & 32.7 & 0.98 \\
HCO$^+$ J=7-6 \;/\; HCN J=2-1 & B9\;/\;B5 & 0.85 & 17.0 & 0.99 \\
CO J=8-7 \;/\; CN N=3–2 & B10\;/\;B7 & 0.85 & 11.2 & 0.99 \\
HCO$^+$ J=8-7 \;/\; CN N=1–0 & B9\;/\;B3 & 0.85 & 67.6 & 0.80 \\
HCO$^+$ J=2-1 \;/\; CN N=3–2 & B5\;/\;B7 & 0.85 & 12.6 & 0.91 \\
CO J=6-5 \;/\; CN N=2–1 & B9\;/\;B6 & 0.85 & 14.5 & 0.98 \\
HCO$^+$ J=8-7 \;/\; HCN J=2-1 & B9\;/\;B5 & 0.85 & 18.2 & 0.99 \\
HCO$^+$ J=4-3 \;/\; CN N=4–3 & B7\;/\;B8 & 0.85 & 12.0 & 1.00 \\
HCO$^+$ J=7-6 \;/\; HCN J=4-3 & B9\;/\;B7 & 0.85 & 10.3 & 1.00 \\
CO J=8-7 \;/\; CN N=4–3 & B10\;/\;B8 & 0.85 & 10.8 & 1.00 \\
HCO$^+$ J=5-4 \;/\; HCN J=2-1 & B8\;/\;B5 & 0.85 & 16.6 & 0.99 \\
HCO$^+$ J=8-7 \;/\; HCN J=4-3 & B9\;/\;B7 & 0.85 & 11.0 & 1.00 \\
HCO$^+$ J=5-4 \;/\; CN N=4–3 & B8\;/\;B8 & 0.85 & 12.1 & 1.00 \\
CO J=7-6 \;/\; HCN J=1-0 & B10\;/\;B3 & 0.85 & 17.8 & 0.87 \\
CO J=6-5 \;/\; HCN J=3-2 & B9\;/\;B6 & 0.85 & 5.4 & 1.00 \\
HCO$^+$ J=5-4 \;/\; HCN J=3-2 & B8\;/\;B6 & 0.85 & 12.8 & 1.00 \\
HCO$^+$ J=3-2 \;/\; CO J=1-0 & B6\;/\;B3 & 0.84 & 9.3 & 1.00 \\
HCO$^+$ J=4-3 \;/\; HCN J=3-2 & B7\;/\;B6 & 0.84 & 12.0 & 1.00 \\
HCO$^+$ J=3-2 \;/\; CN N=4–3 & B6\;/\;B8 & 0.84 & 11.9 & 1.00 \\
HCO$^+$ J=5-4 \;/\; HCN J=4-3 & B8\;/\;B7 & 0.84 & 10.1 & 1.00 \\
HCO$^+$ J=4-3 \;/\; HCN J=2-1 & B7\;/\;B5 & 0.84 & 17.4 & 0.99 \\
CO J=4-3 \;/\; CN N=1–0 & B8\;/\;B3 & 0.84 & 16.0 & 0.80 \\
HCO$^+$ J=8-7 \;/\; CO J=2-1 & B9\;/\;B6 & 0.84 & 7.3 & 1.00 \\
HCO$^+$ J=8-7 \;/\; HCN J=5-4 & B9\;/\;B8 & 0.84 & 10.3 & 0.99 \\
HCO$^+$ J=1-0 \;/\; C$_2$H N=4–3 & B3\;/\;B7 & 0.84 & 12.2 & 0.58 \\
HCO$^+$ J=4-3 \;/\; HCN J=4-3 & B7\;/\;B7 & 0.84 & 9.5 & 1.00 \\
HCO$^+$ J=5-4 \;/\; C$_2$H N=4–3 & B8\;/\;B7 & 0.84 & 18.2 & 0.65 \\
HCO$^+$ J=8-7 \;/\; CN N=2–1 & B9\;/\;B6 & 0.84 & 37.8 & 0.98 \\
HCO$^+$ J=3-2 \;/\; HCN J=2-1 & B6\;/\;B5 & 0.84 & 17.5 & 0.99 \\
HCO$^+$ J=3-2 \;/\; HCN J=3-2 & B6\;/\;B6 & 0.84 & 11.5 & 1.00 \\
HCO$^+$ J=5-4 \;/\; HCN J=5-4 & B8\;/\;B8 & 0.83 & 9.3 & 0.99 \\
CO J=7-6 \;/\; C$_2$H N=4–3 & B10\;/\;B7 & 0.83 & 10.8 & 0.65 \\
HCO$^+$ J=7-6 \;/\; HCN J=5-4 & B9\;/\;B8 & 0.83 & 10.5 & 0.99 \\
CO J=7-6 \;/\; CN N=3–2 & B10\;/\;B7 & 0.83 & 8.8 & 0.99 \\
HCO$^+$ J=7-6 \;/\; C$_2$H N=3–2 & B9\;/\;B6 & 0.83 & 32.3 & 0.58 \\
HCO$^+$ J=10-9 \;/\; CO J=1-0 & B10\;/\;B3 & 0.83 & 17.3 & 0.96 \\
HCO$^+$ J=10-9 \;/\; HCN J=3-2 & B10\;/\;B6 & 0.83 & 16.9 & 0.96 \\
CO J=7-6 \;/\; HCN J=4-3 & B10\;/\;B7 & 0.83 & 5.4 & 1.00 \\
HCO$^+$ J=7-6 \;/\; CN N=3–2 & B9\;/\;B7 & 0.83 & 16.5 & 0.99 \\
HCO$^+$ J=3-2 \;/\; CO J=2-1 & B6\;/\;B6 & 0.83 & 5.8 & 1.00 \\
HCO$^+$ J=2-1 \;/\; CN N=4–3 & B5\;/\;B8 & 0.83 & 11.5 & 0.91 \\
HCO$^+$ J=7-6 \;/\; CN N=4–3 & B9\;/\;B8 & 0.83 & 14.9 & 1.00 \\
CO J=6-5 \;/\; HCN J=1-0 & B9\;/\;B3 & 0.83 & 14.5 & 0.87 \\
HCO$^+$ J=4-3 \;/\; HCN J=5-4 & B7\;/\;B8 & 0.83 & 8.7 & 0.99 \\
CO J=7-6 \;/\; CN N=4–3 & B10\;/\;B8 & 0.83 & 8.2 & 1.00 \\
HCO$^+$ J=10-9 \;/\; HCN J=5-4 & B10\;/\;B8 & 0.83 & 11.2 & 0.96 \\
HCO$^+$ J=10-9 \;/\; HCN J=4-3 & B10\;/\;B7 & 0.83 & 12.9 & 0.96 \\
HCO$^+$ J=2-1 \;/\; HCN J=2-1 & B5\;/\;B5 & 0.83 & 15.8 & 0.91 \\
HCO$^+$ J=3-2 \;/\; HCN J=4-3 & B6\;/\;B7 & 0.83 & 8.7 & 1.00 \\
HCO$^+$ J=1-0 \;/\; CN N=3–2 & B3\;/\;B7 & 0.83 & 10.1 & 0.63 \\
HCO$^+$ J=8-7 \;/\; HCN J=1-0 & B9\;/\;B3 & 0.82 & 32.2 & 0.87 \\
CO J=3-2 \;/\; CN N=1–0 & B7\;/\;B3 & 0.82 & 12.2 & 0.80 \\
HCO$^+$ J=7-6 \;/\; HCN J=1-0 & B9\;/\;B3 & 0.82 & 31.2 & 0.87 \\
HCO$^+$ J=10-9 \;/\; HCN J=2-1 & B10\;/\;B5 & 0.82 & 24.1 & 0.96 \\
HCO$^+$ J=2-1 \;/\; HCN J=3-2 & B5\;/\;B6 & 0.82 & 10.5 & 0.91 \\
CO J=4-3 \;/\; CN N=2–1 & B8\;/\;B6 & 0.82 & 9.3 & 0.98 \\
HCO$^+$ J=5-4 \;/\; HCN J=1-0 & B8\;/\;B3 & 0.82 & 28.1 & 0.87 \\
HCO$^+$ J=10-9 \;/\; HCN J=7-6 & B10\;/\;B9 & 0.82 & 11.4 & 0.96 \\
HCO$^+$ J=8-7 \;/\; CN N=4–3 & B9\;/\;B8 & 0.82 & 15.3 & 1.00 \\
HCO$^+$ J=10-9 \;/\; HCN J=1-0 & B10\;/\;B3 & 0.82 & 42.5 & 0.86 \\
HCO$^+$ J=8-7 \;/\; CN N=3–2 & B9\;/\;B7 & 0.81 & 18.7 & 0.99 \\
HCO$^+$ J=4-3 \;/\; HCN J=1-0 & B7\;/\;B3 & 0.81 & 28.1 & 0.87 \\
HCO$^+$ J=8-7 \;/\; C$_2$H N=3–2 & B9\;/\;B6 & 0.81 & 36.9 & 0.58 \\
HCO$^+$ J=3-2 \;/\; C$_2$H N=5–4 & B6\;/\;B8 & 0.81 & 9.9 & 0.62 \\
CO J=8-7 \;/\; HCN J=5-4 & B10\;/\;B8 & 0.81 & 6.2 & 0.99 \\
HCO$^+$ J=3-2 \;/\; HCN J=1-0 & B6\;/\;B3 & 0.81 & 29.2 & 0.87 \\
CO J=6-5 \;/\; CN N=3–2 & B9\;/\;B7 & 0.81 & 7.2 & 0.99 \\
HCO$^+$ J=7-6 \;/\; C$_2$H N=4–3 & B9\;/\;B7 & 0.81 & 21.3 & 0.65 \\
HCO$^+$ J=3-2 \;/\; HCN J=5-4 & B6\;/\;B8 & 0.81 & 8.0 & 0.99 \\
CO J=4-3 \;/\; C$_2$H N=3–2 & B8\;/\;B6 & 0.81 & 8.9 & 0.58 \\
HCO$^+$ J=10-9 \;/\; CN N=1–0 & B10\;/\;B3 & 0.81 & 89.0 & 0.79 \\
HCO$^+$ J=7-6 \;/\; HCN J=7-6 & B9\;/\;B9 & 0.81 & 10.8 & 0.98 \\
CO J=4-3 \;/\; HCN J=2-1 & B8\;/\;B5 & 0.81 & 5.3 & 0.99 \\
HCO$^+$ J=2-1 \;/\; C$_2$H N=5–4 & B5\;/\;B8 & 0.81 & 9.3 & 0.62 \\
HCO$^+$ J=8-7 \;/\; HCN J=7-6 & B9\;/\;B9 & 0.81 & 11.4 & 0.98 \\
HCO$^+$ J=2-1 \;/\; HCN J=1-0 & B5\;/\;B3 & 0.81 & 29.6 & 0.84 \\
HCO$^+$ J=2-1 \;/\; CO J=1-0 & B5\;/\;B3 & 0.81 & 8.6 & 0.91 \\
HCO$^+$ J=2-1 \;/\; HCN J=4-3 & B5\;/\;B7 & 0.81 & 7.9 & 0.91 \\
HCO$^+$ J=4-3 \;/\; C$_2$H N=5–4 & B7\;/\;B8 & 0.81 & 10.2 & 0.62 \\
CO J=6-5 \;/\; C$_2$H N=4–3 & B9\;/\;B7 & 0.80 & 8.7 & 0.65 \\
CO J=6-5 \;/\; CN N=4–3 & B9\;/\;B8 & 0.80 & 6.5 & 1.00 \\
HCO$^+$ J=1-0 \;/\; CN N=4–3 & B3\;/\;B8 & 0.80 & 9.1 & 0.63 \\
HCO$^+$ J=10-9 \;/\; HCN J=8-7 & B10\;/\;B9 & 0.80 & 12.2 & 0.95 \\
\end{longtable}
\renewcommand{\arraystretch}{1} % restore default
\end{small}

\twocolumn % Switch back to two-column mode

\clearpage

\section{Sensitivity to disk inclination}
\label{appendix:inclination}

%--- FIGURE: Abundance Maps ---------------------------------------------%
\begin{figure*}
\centering
\includegraphics[clip=,width=1\linewidth]{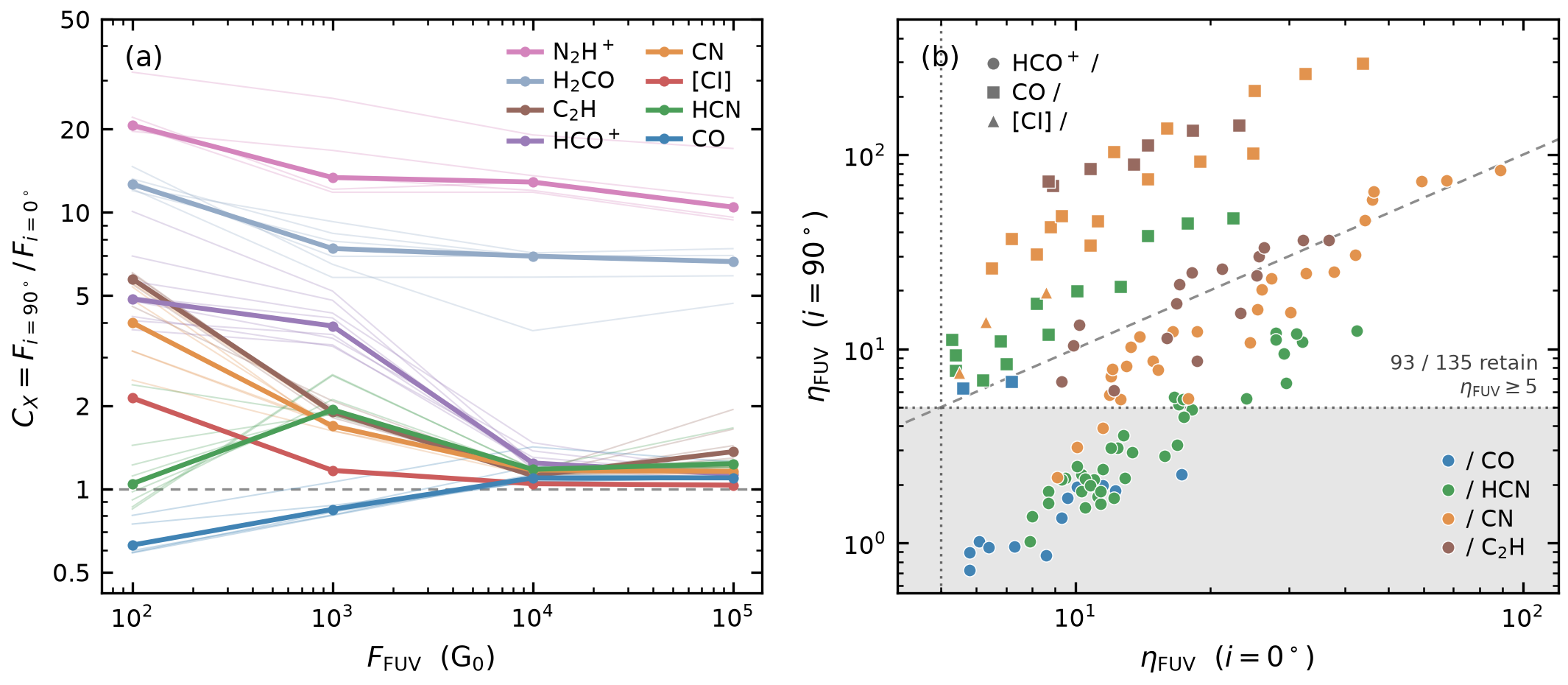}
\caption{Effect of disk inclination on the line fluxes and flux ratio diagnostics, based on the representative model of Figure \ref{fig_abundance_maps} recomputed edge-on. \textit{Left:} Inclination correction factor $C_X = F_{i=90^\circ}/F_{i=0^\circ}$ against external FUV field strength. Faint lines show individual transitions and heavy lines the median for each species. The dashed line marks $C_X=1$, where the flux is independent of viewing geometry. \textit{Right:} FUV enhancement factor $\eta_{\rm FUV}$ at $i=0^\circ$ compared with $i=90^\circ$, for the 135 ratios shortlisted in Table \ref{table:ratio_shortlist}. Colour denotes the denominator species and shape the numerator. The dashed line is the 1:1 relation and the shaded region marks $\eta_{\rm FUV}<5$, below which a ratio no longer satisfies the selection criteria of Section \ref{sec:results_ratios}.}
\label{fig_inclination_test}
\end{figure*}
%------------------------------------------------------------------------%

All fluxes presented in this work assume a face-on geometry ($i=0^\circ$). Recomputing the full grid at multiple inclinations is computationally prohibitive, so we instead assess the sensitivity of our results using the representative model of Figure \ref{fig_abundance_maps} ($M_*=0.6\,M_\odot$, $\Sigma_{\rm 1au}=1000\,\mathrm{g\,cm^{-2}}$, $r_{\rm d}=40$~au), with all fluxes recomputed edge-on ($i=90^\circ$) at each of $F_{\rm FUV}=10^2$, $10^3$, $10^4$ and $10^5$~G$_0$. 

For each transition we define an inclination correction factor $C_X = F_{i=90^\circ}/F_{i=0^\circ}$, where $F_{i=90^\circ}$ is the disk-integrated flux at $i=90^\circ$ and $F_{i=0^\circ}$ is the disk-integrated flux at $i=0^\circ$. Figure \ref{fig_inclination_test} (panel a) shows $C_X$ as a function of $F_{\rm FUV}$. The behaviour varies according to whether emission from a given species predominantly traces the disk or the photoevaporative wind. At $10^2$~G$_0$ the wind is negligible and the emission from all species arises from thin layers near the disk surface. In this scenario, the column density along the line of sight is much larger when the disk is edge-on compared to face-on. For the edge-on disk,  median fluxes therefore increase by factors of $4.9$, $5.8$ and $4.0$ for HCO$^+$, C$_2$H and CN respectively, while optically thick CO instead decreases to $C_{\rm CO}=0.63$ as the projected emitting area is reduced. As $F_{\rm FUV}$ increases, these species begin to fill a large, spherical wind, and the correction converges towards unity, reaching $C_X = 1.0-1.4$ for CO, HCN, [CI], CN, HCO$^+$ and C$_2$H at $10^4$~G$_0$ and above. The exceptions are H$_2$CO and N$_2$H$^+$, which retain $C_X \sim 7-13$ and $10-21$ respectively at all external FUV field strengths. Neither expands into the wind, remaining confined to a narrow layer close to the disk--wind interface (as described in Section \ref{subsec:disc_connection_to_chemistry}).

Because the correction is largest at weaker FUV fields and negligible at strong ones, its principal effect on the diagnostics is to compress or extend the dynamic range $\eta_{\rm FUV}$, rather than to remove the FUV dependence itself. Figure \ref{fig_inclination_test} (panel b) compares $\eta_{\rm FUV}$ at the two inclinations for the ratios shortlisted in Table \ref{table:ratio_shortlist}. Of the 135 ratios, 93 retain $\eta_{\rm FUV}\geq5$ when viewed edge-on. Just as ratios were used in Section \ref{sec:results_ratios} to cancel the dependence on the stellar and disk properties, ratios whose numerator and denominator respond similarly to inclination also leave $\eta_{\rm FUV}$ essentially unchanged. This is the case for same-species excitation ratios, HCO$^+$/C$_2$H, and HCO$^+$/CN $N=1-0$. Of the 42 ratios that fall below the threshold, 39 pair HCO$^+$ with CO or HCN, for which the edge-on brightening of HCO$^+$ at low $F_{\rm FUV}$ raises the weak-field end of the ratio and reduces $\eta_{\rm FUV}$ by factors of $\sim 5-7$. This includes the matched-$J$ HCO$^+$/HCN pairs identified in Section \ref{sec:disc_alma_guidance}. The remaining three pair low-$J$ HCO$^+$ with CN $N=3-2$ or $4-3$, where the reduction is smaller but their face-on enhancement factors are already modest ($\eta_{\rm FUV} \approx 9-12$). Conversely, ratios of high-$J$ CO over CN or C$_2$H gain dynamic range when edge-on, since their numerator and denominator respond in opposite directions. This is not necessarily an advantage, however, since it reflects a strong sensitivity to inclination, making them poor choices where the geometry is unknown. For a target with no constraint on its inclination at all, the corrections in Figure \ref{fig_inclination_test} (panel a) imply a systematic uncertainty of $\sim1$~dex on the inferred field strength. In practice this is rarely the situation, since inclination is routinely measured from continuum imaging, and can therefore be applied as a known correction rather than left as a degeneracy.

Two caveats apply to this. First, this test uses a single disk structure. Second, the photoevaporative wind in \textsc{puffin} is spherically diverging by construction, and a spherical emitting volume looks the same from every viewing angle, so the convergence of $C_X$ towards unity at high $F_{\rm FUV}$ is partly imposed by the model. Real flows are asymmetric and, at high $F_{\rm FUV}$, truncated by an ionisation front. Their emission would depend on the star-disk-observer orientation rather than on inclination alone. Our corrections are therefore minimum estimates, and are most reliable at $10^2-10^3$~G$_0$, where the emission arises from the disk itself.

\section{Full ratio matrix}
\label{appendix:full_ratio_matrix}

\begin{figure*}
\centering
\includegraphics[clip=,width=1\linewidth]{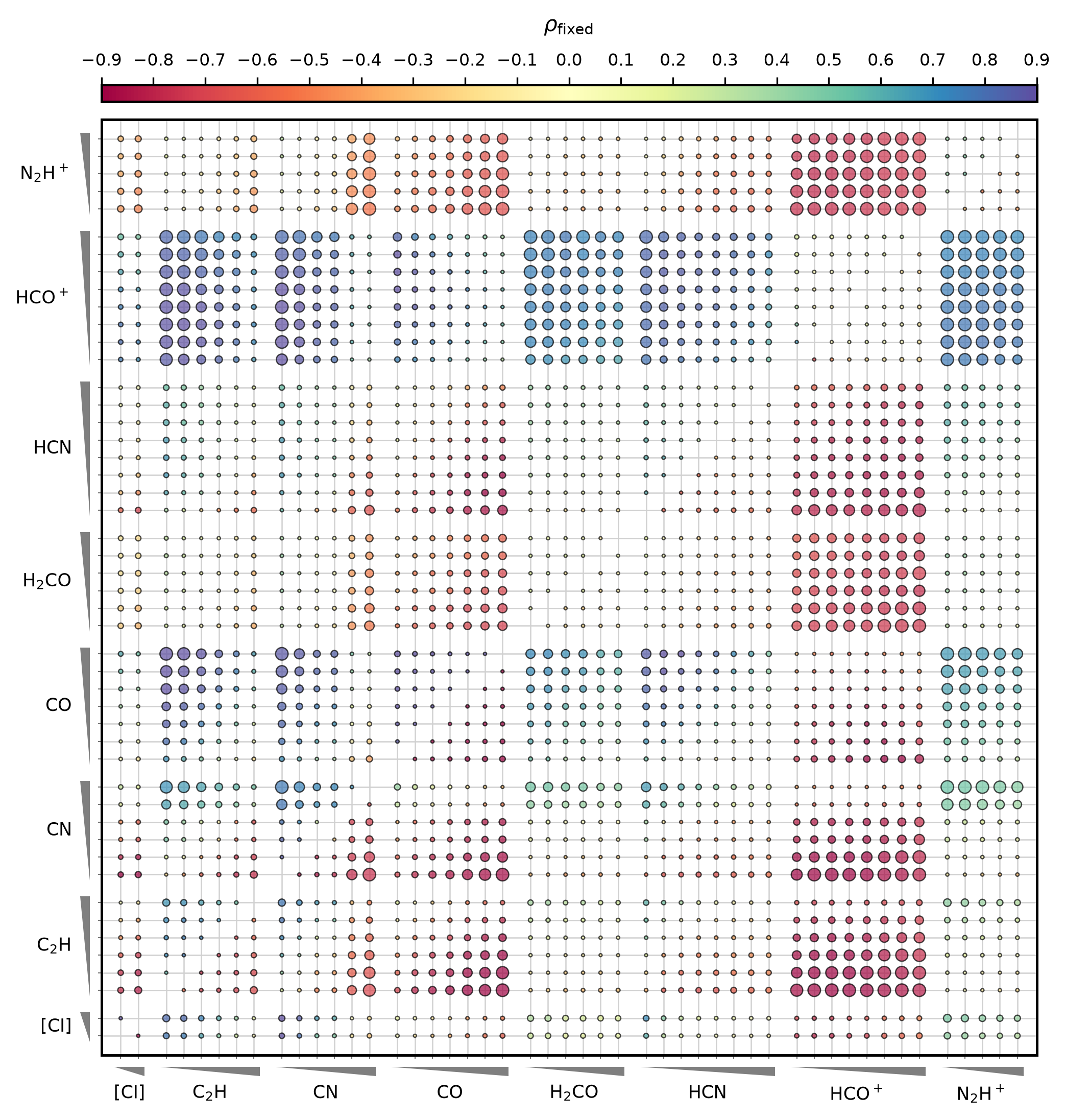}
\caption{Full matrix of the 2256 line flux ratios formed from the 48 transitions considered in this work, grouped by species. Each circle represents the ratio of the transition on the $y$-axis (numerator) to that on the $x$-axis (denominator). Self-ratios along the diagonal are omitted. Colour denotes \rhopartial, the correlation of each ratio with the external FUV field when the stellar and disk properties are held fixed, with blue indicating ratios that increase with \ffuv\ and red those that decrease. Circle size indicates the enhancement factor $\eta_{\rm FUV}$. Reciprocal ratios appear as mirror images across the diagonal, with the same circle size and the opposite sign of \rhopartial. Wedges along each axis indicate increasing $E_{\rm u}$ within each species (see Table \ref{table:line_transitions} for full details of the individual transitions).}
\label{fig_heatmap_full}
\end{figure*}

Figure \ref{fig_heatmap_full} shows the diagnostic performance of all 2256 line flux ratios considered in this work, extending the subset shown in Figure \ref{fig_heatmap} to all eight species.

%--- FIGURE: Heatmap -----------------------%
%-------------------------------------------%

%%%%%%%%%%%%%%%%%%%%%%%%%%%%%%%%%%%%%%%%%%%%%%%%%%

% Don't change these lines
\bsp	% typesetting comment
\label{lastpage}
\end{document}